\documentclass[fleqn,usenatbib]{mnras}

\usepackage[T1]{fontenc}
\usepackage{ae,aecompl}

\pdfoutput=1
\usepackage{graphicx}	
\usepackage{amsmath}	
\usepackage{amssymb}	
\usepackage{enumitem}
\usepackage{xcolor}
\usepackage{caption}
\usepackage{subcaption}
\usepackage{verbatim}
\usepackage{float}
\usepackage{longtable}

\usepackage{mathptmx}
\usepackage{txfonts}

\title[A Gemini--NIFS view of the merger remnant NGC~34]{A Gemini--NIFS view of the merger remnant NGC~34}

\author[J. C. Motter et al.]{
J. C. Motter,$^{1}$\thanks{E-mail: juliana.motter@ufrgs.br (JCM)}
R. Riffel,$^{1}$\thanks{E-mail: riffel@ufrgs.br (RR)}
T. V. Ricci,$^{2}$\thanks{E-mail: tiago.ricci@uffs.edu.br (TVR)}
R. A. Riffel,$^{3}$
T. Storchi-Bergmann,$^{1}$
\newauthor
M. G. Pastoriza,$^{1}$
A. Rodriguez-Ardila,$^{4}$
D. Ruschel-Dutra,$^{5}$
L. G. Dahmer-Hahn,$^{4,6}$
\newauthor
N. Z. Dametto,$^{7}$
M. R. Diniz$^{3}$
\\
$^{1}$Departamento de Astronomia, Universidade Federal do Rio Grande do Sul, Porto Alegre, RS, 91501-970, Brazil\\
$^{2}$Universidade Federal da Fronteira Sul, Campus Cerro Largo, RS, 97900-000, Brazil\\
$^{3}$Departamento de F\'{i}sica, Centro de Ci\^{e}ncias Naturais e Exatas, Universidade Federal de Santa Maria, Santa Maria, RS, 97105-900, Brazil\\
$^{4}$Laborat\'{o}rio Nacional de Astrof\'{i}sica, Itajub\'{a}, MG, 37500-000, Brazil\\
$^{5}$Departamento de F\'{i}sica, Centro de Ci\^{e}ncias F\'{i}sicas e Matem\'{a}ticas, Universidade Federal de Santa Catarina, Florian\'{o}polis, SC, 88040-900, Brazil\\
$^{6}$Shanghai Astronomical Observatory, Chinese Academy of Sciences, 80 Nandan Road, Shanghai 200030, China\\
$^{7}$Centro de Astronom\'{i}a (CITEVA), Universidad de Antofagasta, Antofagasta,  02800, Chile
}

\date{Accepted XXX. Received YYY; in original form ZZZ}

\pubyear{2021}

\begin{document}
\label{firstpage}
\pagerange{\pageref{firstpage}--\pageref{lastpage}}
\maketitle

\begin{abstract}
The merger remnant NGC~34 is a local luminous infrared galaxy (LIRG) hosting a nuclear starburst and a hard X-ray source associated with a putative, obscured Seyfert~2 nucleus. In this work, we use adaptive optics assisted near infrared (NIR) integral field unit observations of this galaxy to map the distribution and kinematics of the ionized and molecular gas in its inner $\mathrm{1.2\,kpc \times 1.2\,kpc}$, with a spatial resolution of 70~pc. The molecular and ionized gas kinematics is consistent with a disc with projected major axis along a mean PA~=~$\mathrm{-9^{\circ}.2 \pm 0^{\circ}.9}$. Our main findings are that NGC~34 hosts an AGN and that the nuclear starburst is distributed in a circumnuclear star-formation ring with inner and outer radii of $\approx$~60 and 180~pc, respectively, as revealed by maps of the $\mathrm{[\ion{Fe}{ii}] / Pa\beta}$ and $\mathrm{H_{2} / Br\gamma}$ emission-line ratios, and corroborated by PCA Tomography analysis. The spatially resolved NIR diagnostic diagram of NGC~34 also identifies a circumnuclear structure dominated by processes related to the stellar radiation field and a nuclear region where $[\ion{Fe}{ii}]$ and H$_2$ emissions are enhanced relative to the hydrogen recombination lines. We estimate that the nuclear X-ray source can account for the central H$_2$ enhancement and conclude that $[\ion{Fe}{ii}]$ and H$_2$ emissions are due to a combination of photo-ionization by young stars, excitation by X-rays produced by the AGN and shocks. These emission lines show nuclear, broad, blue-shifted components that can be interpreted as nuclear outflows driven by the AGN.
\end{abstract}





\section{Introduction} \label{sec:Intro}

Local Universe galaxies display a wide range of luminosities, sizes, gas and stellar distribution and kinematics, that are the outcome of a process that lasts $\sim$13.8~Gyr. In the hierarchical scenario of galaxy formation and evolution, early-type galaxies form as a result of the merger of spiral galaxies whose discs are disrupted as the merging evolves (e.g.: \citealt{1978MNRAS.183..341W,2008ApJS..175..356H,2017MNRAS.467.3934D}). In cases where the interacting galaxies are gas-rich (`wet-merger'), the migration of vast amounts of gas towards the innermost regions of the merging system can lead to the onset of starbursts and/or active galactic nuclei (AGN) phenomena. However, this poses several challenges for the understanding of the fate of the interacting gas during these galaxies encounters; at the same time that gas is required either to form stars or to feed the AGN, it can also be expelled from the merger remnant by galactic winds related with these processes. Moreover, it is well known that star formation (SF) on galactic scales may be quenched, suppressed or triggered as a result of the AGN feedback \citep{2017NatAs...1E.165H}, thus raising the question if the AGN feeding happens simultaneously with the SF \citep{2008ApJ...681...73K}, follows it during a post-starburst phase \citep{2007ApJ...671.1388D,2009MNRAS.400..273R} or lack any association with recent SF \citep{2007ASPC..373..643S}.  

Luminous and Ultra Luminous Infrared Galaxies (LIRGs and ULIRGs, respectively) emit the bulk of their energy in the infrared (IR). For LIRGs, IR luminosities are in the range $11.0 \leq \log(L_{IR}/L_{\odot}) \leq 12.0$, while for ULIRGs, $\log(L_{IR}/L_{\odot}) \geq 12.0$ \citep{2003AJ....126.1607S}. These objects harbor a strong nuclear starburst and/or an AGN \citep{1996ARAA..34..749S,2006asup.book..285L}, and most of local ULIRGs, and LIRGs with $L_{IR} > 10^{11.5}L_{\odot}$ are often associated with gas-rich merging systems (\citealt{2014ApJ...791...63H}, for a recent review on the properties of LIRGs, see \citealt{2021A&ARv..29....2P}). Therefore, these objects have become increasingly important for galaxy evolution studies since they provide a laboratory for understanding the starburst/AGN feedback processes that shape the evolution of a merger remnant. The main power source of U(LIRGs) is the heating of dust by a starburst and/or an AGN, but the relative contribution of each process to the observed IR luminosities is still under debate. This happens mainly due to difficulties in carrying out high resolution optical/IR studies of these objects as a consequence of the heavy dust obscuration towards their nuclear regions and to the fact that they are more numerous at $z \geq 1$ \citep{1996ARAA..34..749S,2006asup.book..285L}. Therefore, it is clear that in order to map in detail their stellar and gas properties, it is necessary to search for local prototypes of the (U)LIRGs population. Observations indicate that, at least for local LIRGs, the bulk of their IR luminosites is dominated by starburst emission \citep{2012ApJ...744....2A}, and that the AGN contribution to the observed IR luminosities increases with the IR luminosity of the system, and may dominate the energy budget for objects with $L_{IR} > 10^{12.5}L_{\odot}$ \citep{2010MNRAS.405.2505N,2009ApJS..182..628V}. 

The galaxy NGC~34 (NGC~17, Mrk~938), $z=0.0196$ \citep{2006AJ....131..185R} and at a distance of $\mathrm{\approx 82\,Mpc }$ ($\mathrm{H_0} = 67.4\,\mathrm{km\,s^{-1}\,Mpc^{-1}}$), is a local representative of the LIRGs class ($\log(L_{IR}/L_{\odot}) = 11.61$, \citealt{1992AA...255...87C}). It hosts a dominant central starburst (e.g.: \citealt{2006AA...457...61R,2007AJ....133.2132S,2008MNRAS.388..803R,2014MNRAS.443.1754D}) with inferred star formation rates (SFR) in the range 50--90~$\mathrm{M_{\odot} yr^{-1}}$  \citep{2005AA...434..149V,2004AA...421..115P}. However, evidence on AGN activity has been extensively debated. Based on near-infrared (NIR) studies in the range 0.8--2.4~$\mathrm{\mu m}$, \citet{2006AA...457...61R} found that NGC~34 displays a poor emission-line spectrum and that the continuum emission is dominated by stellar absorption features. Therefore, they classified NGC~34 as a starburst galaxy. In the optical domain, \citet{1996ApJS..102..309M} carried out an imaging survey of the $\mathrm{[\ion{O}{iii}]\lambda 5007}$ and $\mathrm{H\alpha + [\ion{N}{ii}]\lambda\lambda 6548,6543}$ emission lines for a sample of early-type Seyfert galaxies and found that NGC~34 not only is a week $[\ion{O}{iii}]$ emitter compared to most of the Seyferts in their sample, but also that $\mathrm{H\alpha}$ emission is strong over the entire galaxy, indicating that the gas ionization is not related to any AGN. However, \citet{2010ApJ...709..884Y} and \citet{2011MNRAS.414.3084B} used Baldwin, Phillips and Terlevich (BPT) diagnostic diagrams \citep{1981PASP...93....5B,2001ApJS..132...37K,2006MNRAS.372..961K} to confirm that NGC~34 hosts a Seyfert~2 nucleus. Despite the highly controversial nature of the NGC~34 nuclear spectrum, X-ray observations provide compelling evidence for the presence of an obscured AGN in its central regions (e.g.: \citealt{2005AA...444..119G,2011MNRAS.413.1206B,2012MNRAS.423..185E}). \citet{2012MNRAS.423..185E} found that the NGC~34 X-ray luminosity in the 2--10~keV energy range ($\mathrm{L_X = 1.4^{+0.3}_{-0.2} \times 10^{42} erg\,s^{-1})}$ is too high to be solely due to the nuclear starburst, and is, therefore, dominated by an AGN.

The detailed study of the optical morphological properties of NGC~34 presented by \citet{2007AJ....133.2132S} shows that this galaxy features a red nucleus, with a nuclear starburst confined to a radius $\leq$1~kpc; a young and blue central exponential disc at a position angle (PA) of -9$^{\circ}$, a system of young massive star clusters and a pair of unequal tidal tails indicative of the merger of two former gas-rich disc galaxies. In addition, \citet{2007AJ....133.2132S} also reported that blueshifted \ion{Na}{i}~D lines reveal that the inner regions of NGC~34 drive a strong outflow of cool, neutral gas with a mean velocity of $\mathrm{-620 \pm 60\,km\,s^{-1}}$ that could be due to the concentrated starburst and/or the hidden AGN.

The galaxy NGC~34 has also been probed in the radio and submillimetre regimes. \citet{2010AJ....140.1965F} used 21~cm Very Large Array (VLA) observations to map the radio continuum emission and the \ion{H}{i} distribution and kinematics. They found that the radio continuum emission structure consists of an extended nuclear component that is dominated by the central starburst, and an outer, extra--nuclear diffuse extended component in the shape of two radio lobes that could be evidence for past AGN activity or due to a starburst-driven super-wind. The authors detected a broad \ion{H}{i} absorption profile with both blue-shifted and red-shifted velocities that could be explained by the presence of a circumnuclear disc (CND) of neutral and molecular gas. The existence of the CND was later confirmed by \citet{2014AJ....147...74F}. They detected a rotating CO disc of 2.1~kpc in diameter using Combined Array for Research in Millimeter-wave Astronomy (CARMA) observations of the CO(1-0) transition at 115~GHz. An even more compact molecular rotating disc with a size of 200~pc was later detected by \citet{2014ApJ...787...48X} using Atacama Large Millimeter Array (ALMA) observations of the CO(6-5) emission line (rest-frame frequency = 691.473~GHz). 

Finally, \citet{2018MNRAS.474.3640M} used archival \textit{Herschel} and ALMA observations of multiple CO transitions along with X-ray data from \textit{NuSTAR} and \textit{XMM-Newton} to perform the modelling of the CO Spectral Line Energy Distribution (SLED) - that is, the modelling of the luminosities of the CO lines as a function of their upper rotational levels. They found that a combination of a cold and diffuse photo-dissociation region (PDR) to account for the low-J transitions and a warmer and denser X-ray dominated region (XDR) to account for the high-J transitions was necessary to properly fit the CO line luminosities, and concluded that AGN contribution is significant in heating the molecular gas in NGC~34.

In this work, we use adaptive optics (AO) assisted NIR integral field unit (IFU) observations of the galaxy NGC~34 to map the distribution and kinematics of the ionized and molecular gas distributed in the inner $\mathrm{1.2\,kpc \times 1.2\,kpc}$. Our main goal is to investigate the nature of the NGC~34 NIR emission line spectrum in order to put tighter constraints on the presence of an AGN in its centre. Starburst phenomena, as in the case of NGC~34, are known for residing in dusty systems, therefore NIR observations are ideally suited for probing the central regions of these objects since they can pierce through highly obscured regions. 

This paper is structured as follows. In Sec.~\ref{sec:Obs} we present the observations and data reduction procedures, in Sec.~\ref{sec:DataAnalysis} we describe our data analysis. Our results are shown in Sec.~\ref{sec:Results} followed by the discussion in Sec.~\ref{sec:Disc}. We summarise our findings in Sec.~\ref{sec:Conclusions}. At the redshift of $z=0.0196$, 1$\arcsec$ corresponds to 411~pc for the adopted cosmology: $\mathrm{H_0} = 67.4\,\mathrm{km\,s^{-1}\,Mpc^{-1}}$, $\Omega_\mathrm{M} = 0.31$ and $\Omega_{\Lambda} = 0.69$ \citep{2020A&A...641A...6P}. The NGC~34 systemic velocity is $v_{sys} = 5881\,\mathrm{km \, s^{-1}}$.

\section{Observations and Data Reduction} \label{sec:Obs}

Data for this work were obtained with the Near-Infrared Integral Field Spectrograph (NIFS) of the Gemini North Telescope under the Gemini Science Programme GN-2011B-Q-71 (PI: Rog\'{e}rio Riffel) using the ALTitude conjugate Adaptive optics for the InfraRed (ALTAIR) system. The NIFS instrument provides a field of view (FoV) of $\mathrm{3.0\arcsec \times 3.0\arcsec}$ that is sampled by spatial pixels (spaxels) with dimensions of  $\mathrm{0.103\arcsec \times 0.043\arcsec}$ as a result of its optical and geometrical properties. 

Observations of NGC~34 were carried out in 2011 September 09 and 24 in the \textit{K}$_l$ (1.99--2.40$\mu m$) and \textit{J} (1.15--1.33$\mu m$) bands, respectively. Eight on target (T) and four sky (S) exposures of 350~s were obtained in each band following the sequence TST. Observations of standard stars for telluric absorption removal and flux calibration, as well as flat-field, darks, Ronchi-flat and arc-lamp (Ar and ArXe for the \textit{J} and \textit{K}$_l$ bands, respectively) were obtained for the data reduction process. The spectral resolution for both bands is 20~$\mathrm{km\,s^{-1}}$ as measured by the full width at half maximum (FWHM) of the Ar and ArXe lines. 

The data reduction was performed using the {\sc iraf} (Image Reduction and Analysis Facility) environment along with standard reduction scripts made available by the Gemini team. The data reduction process includes the trimming of the images, sky subtraction, flat-field, bad pixel and spatial distortion corrections, wavelength calibration, telluric absorption removal and flux calibration by fitting a blackbody function to the spectrum of the telluric standard star. Individual reduced data cubes were constructed with spaxels of $\mathrm{0.05\arcsec \times 0.05\arcsec}$ and were median combined in each band to a single data cube.

Fully calibrated and redshift corrected data cubes were obtained following the treatment procedures for Gemini--NIFS data cubes presented by \citet{2014MNRAS.438.2597M}, including the re-sampling of the image to spaxels with dimensions $\mathrm{0.021\arcsec \times 0.021\arcsec}$. High-frequency spatial noise was removed using a Butterworth filter with a cut-off frequency of 0.22$F_{NY}$ (where $F_{NY}$ is the Nyquist frequency) for the \textit{J}-band cube and 0.33$F_{NY}$ for the \textit{K}-band cube, both with a filter order $n = 2$. Then, we removed low-frequency instrumental fingerprints using Principal Component Analysis Tomography \citep{2009MNRAS.395...64S}.

Our final data cubes have FoV's of $\approx \mathrm{3.0\arcsec \times 3.0\arcsec}$ corresponding to $\mathrm{1.2\,kpc \times 1.2\,kpc}$ at the galaxy, and spectral ranges of 1.128--1.310$\mathrm{\mu m}$ and 2.081--2.370$\mathrm{\mu m}$ in the \textit{J} and \textit{K} bands, respectively. The spatial resolution corresponding to the FWHM of the brightness profile of the telluric standard star is 0.17$\arcsec$ ($\approx \mathrm{70\,pc}$) for both the \textit{J} and \textit{K} bands.

\begin{figure*}
\begin{center}

	\begin{minipage}{\textwidth}
	\centering
	    \includegraphics[width=\textwidth]{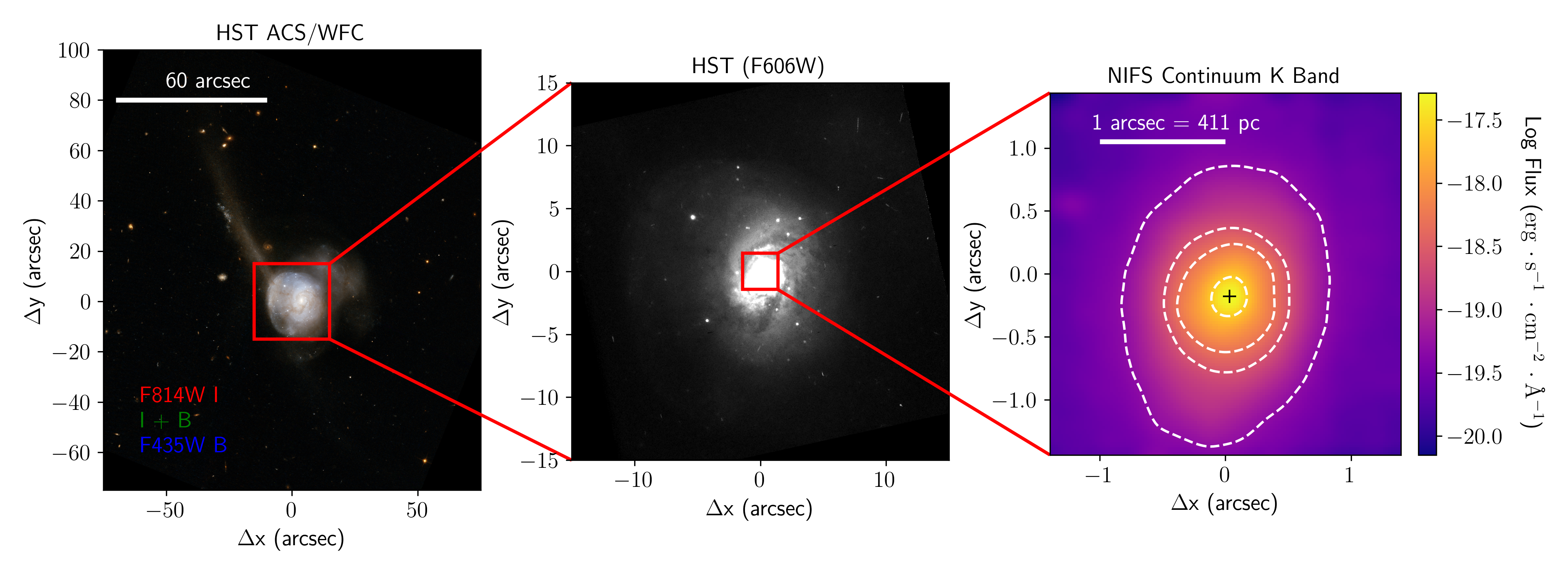}
	\begin{minipage}{.49\textwidth}
		\includegraphics[width=\textwidth]{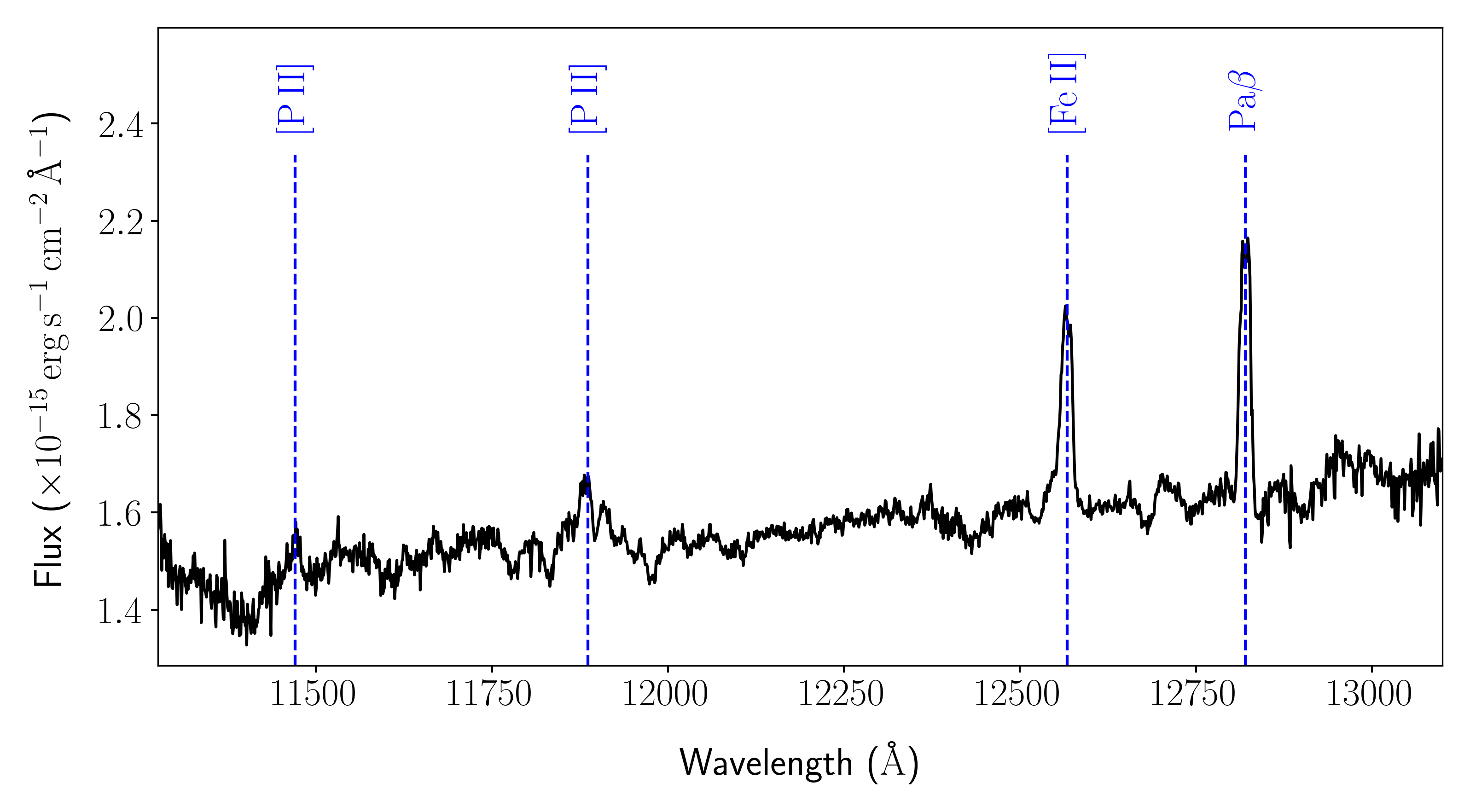}
	\end{minipage}
	\quad
	\begin{minipage}{.49\textwidth}
		\includegraphics[width=\textwidth]{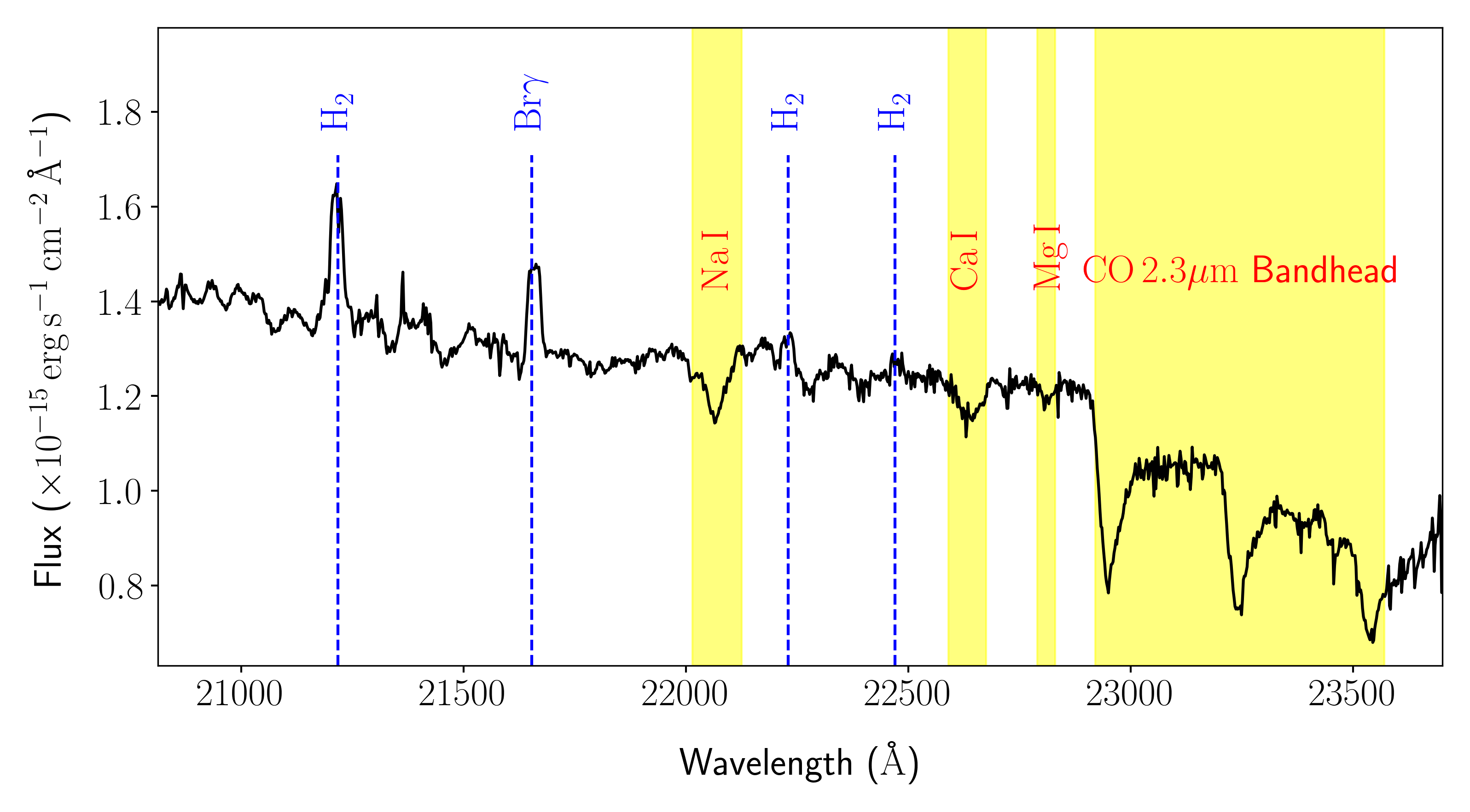}
	\end{minipage}
	\end{minipage}
	
\caption{Top-left: NGC~34 seen by the \textit{HST} ACS/WFC instrument (Image credits: NASA, ESA, the Hubble Heritage (STScI/AURA)-ESA/Hubble Collaboration, and A. Evans (University of Virginia, Charlottesville/NRAO/Stony Brook University)). Top-middle: Optical image of NGC~34 obtained by the \textit{HST} WFPC2 with the filter \textit{F606W} \citep{1998ApJS..117...25M} with the NIFS FoV indicated by the red square. Top-right: average \textit{K}-band NIFS continuum image. The black cross marks the nucleus of the galaxy. The color bar shows the fluxes in units of $\mathrm{ergs\cdot s^{-1} \cdot cm^{-2} \cdot \AA^{-1}}$. White, dashed contours are drawn at 1, 5, 10 and 50 per cent of the continuum peak. Bottom: integrated NIFS \textit{J} (left) and \textit{K} (right) band spectra over the whole FoV. The blue vertical dashed lines show the main NIR emission lines. The yellow shaded areas show the absorption features.
 }
 \label{fig:HST_NIFS}
\end{center}
\end{figure*}

In the top-left panel of Fig.~\ref{fig:HST_NIFS}, we show an image of NGC~34 obtained by the Advance Camera for Surveys/Wide Field Camera (ACS/WFC) on board of the \textit{Hubble Space Telescope} (\textit{HST}) with the filters \textit{F814W} and \textit{F435W}. We show in the middle panel of Fig.~\ref{fig:HST_NIFS} an \textit{HST} Wide Field and Planetary Camera 2 (WFPC2) image of NGC~34 obtained with the filter \textit{F606W} \citep{1998ApJS..117...25M} along with a \textit{K}-band NIFS continuum image obtained from the average flux in the 2.180--2.195$\mathrm{\mu m}$ wavelength range (top-right). Integrated NIFS \textit{J} and \textit{K} bands spectra over the entire FoV are shown in the bottom panels of Fig.~\ref{fig:HST_NIFS}. In the \textit{J} band, the main NIR emission lines that can be seen in our spectra are $\mathrm{[\ion{P}{ii}]\,\lambda 11470}$\AA\ , $\mathrm{[\ion{P}{ii}]\,\lambda 11886}$\AA\ , $\mathrm{[\ion{Fe}{ii}]\,\lambda 12570}$\AA\ and $\mathrm{Pa\beta}$. In the \textit{K} band, the $\mathrm{H_2\,\lambda 21218}$\AA\ , $\mathrm{Br\gamma}$, $\mathrm{H_2\,\lambda 22230}$\AA\ and $\mathrm{H_2\,\lambda 22470}$\AA\ emission lines can be seen, in addition to the \ion{Na}{i}~2.20$\mathrm{\mu m}$, \ion{Ca}{i}~2.26$\mathrm{\mu m}$, \ion{Mg}{i}~2.28$\mathrm{\mu m}$ and CO~2.3$\mathrm{\mu m}$ absorption features. 

\section{Data Analysis} \label{sec:DataAnalysis}

\subsection{Spectral synthesis} \label{sec:SpecSyn}

In order to study the nature of the gas emission lines present in the observed spectra of active/starburst galaxies, first, one needs to obtain spectra that are free from contamination caused by other components of the galaxy, such as the stellar population, dust content and AGN continuum (accretion disc). This is usually done through the application of the spectral synthesis technique, which requires a base of theoretical and/or empirical stellar spectra models to account for the other galaxy components and a code that quantifies the contribution of all these elements to the observed spectra. In this work, we used the NASA InfraRed Telescope Facility (IRTF) Spectral Library \citep{2005ApJ...623.1115C,2009ApJS..185..289R} and the Penalized Pixel-Fitting ({\sc ppxf}, \citealt{Cappellari2017}) code to obtain pure emission line (gas) spectra for the galaxy NGC~34 at the \textit{J} and \textit{K} bands. 

\begin{figure}
\centering
	\includegraphics[width=\linewidth]{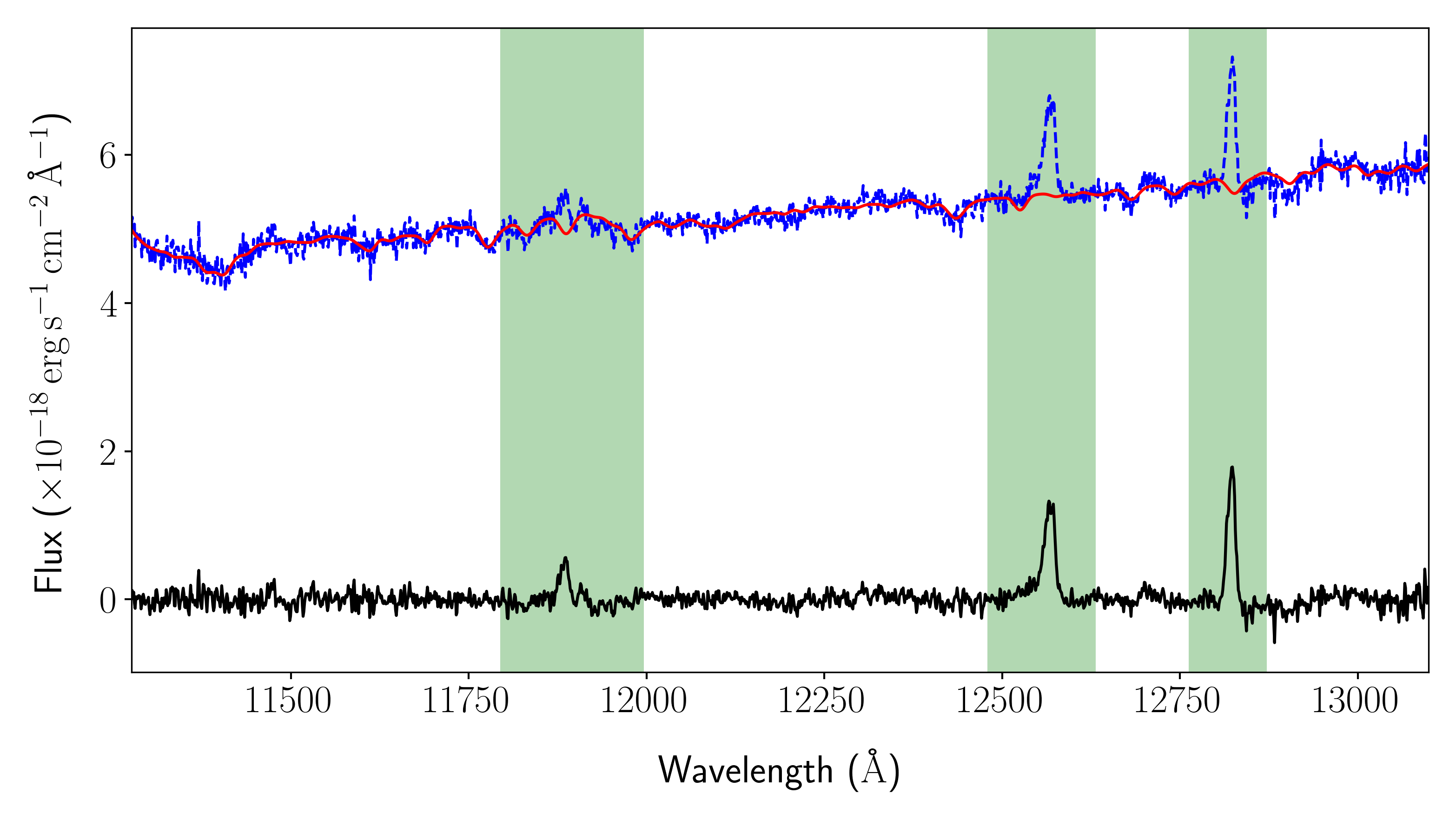}
	\includegraphics[width=\linewidth]{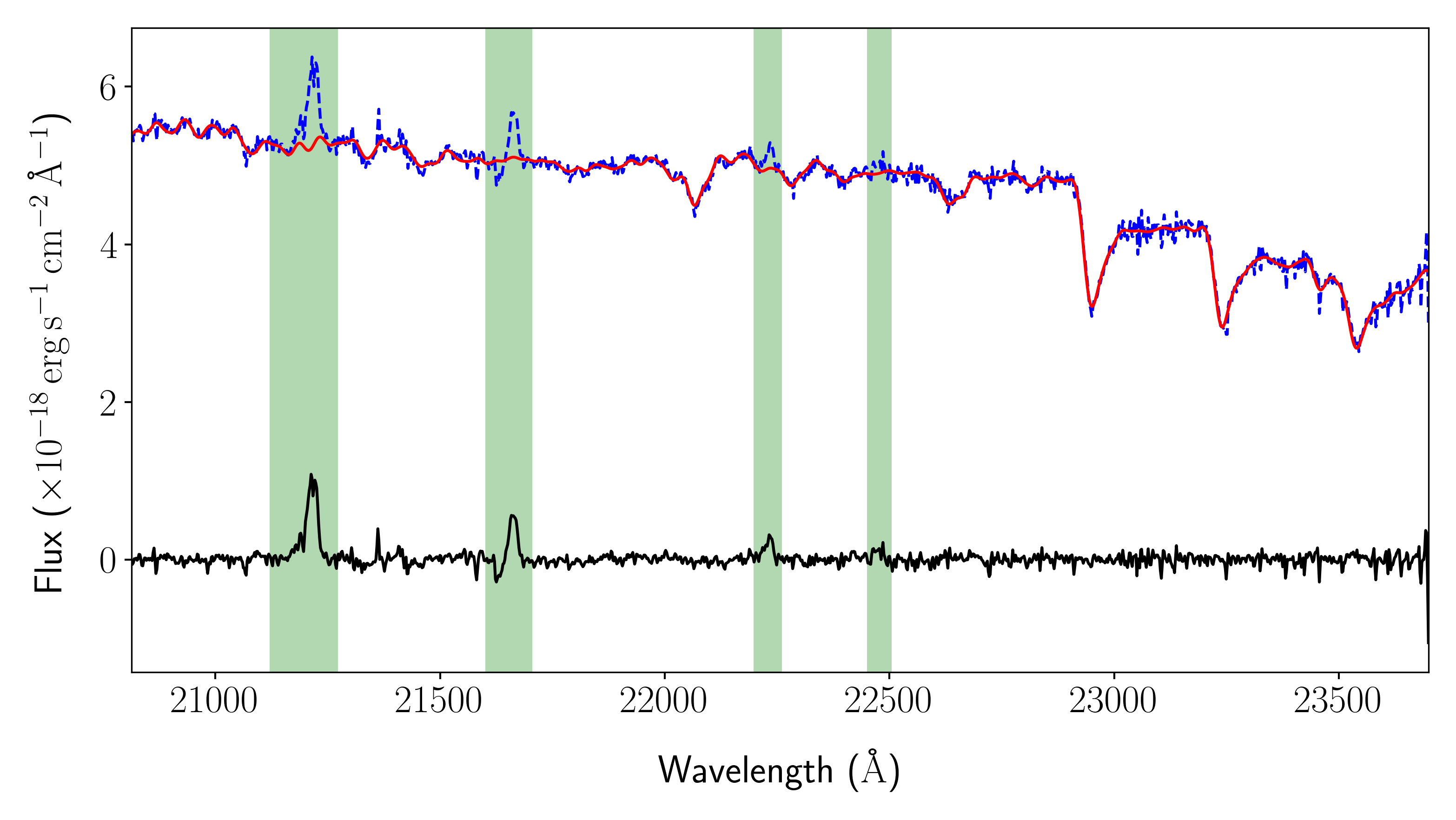}
	\caption{Examples of the pPXF fitting results for \textit{J} (top) and \textit{K} (bottom) bands at the spaxel corresponding to the nucleus of the galaxy. Blue dashed, solid red and solid black lines indicate the observed, fitted stellar model and residual (gas) spectra, respectively. The light green, vertical, shaded areas indicate masking intervals for the fit.}
   \label{fig:ppxf}
\end{figure}

The IRTF Spectral Library ($R=2000$) contains observed spectra in the 0.8--5$\mu \mathrm{m}$ wavelength range for 210 cool stars, with mostly near-solar metallicities, and spectral types between F and M and luminosity classes between I and V, in addition to some AGB, carbon, and S stars \citep[see][for a similar application]{2015MNRAS.450.3069R}. We used this set of observed spectra as an input template for the {\sc ppxf} package to fit the continuum of the NGC~34 observed spectra in every spaxel of the data cubes. The fitting was performed separately for the \textit{J} and \textit{K} bands and the emission lines were masked before the fit. The stellar features were fitted using Gauss-Hermite profiles and we used only multiplicative polynomials to prevent changes in the line strength of the absorption features in the templates as recommended by \citet{Cappellari2017}. Examples of the gas spectra obtained following this procedure are shown in Fig.~\ref{fig:ppxf} for the \textit{J} and \textit{K} bands at the spaxel corresponding to the nucleus of the galaxy.

\subsection{Emission-line fitting} \label{sec:LineFit}

We fitted the profiles of the emission lines using the package {\sc ifscube} \citep{daniel_ruschel_dutra_2020_4065550}, which is a {\sc python} based package designed to carry out analysis of data cubes of integral field spectroscopy. {\sc ifscube} provides a set of tasks that allows the simultaneous fitting by multiple Gaussian curves or Gauss-Hermite series emission line profiles in velocity space. It also provides a Monte-Carlo implementation to estimate fitting uncertainties pixel-by-pixel.

We used Gaussian functions to fit separately, in individual {\sc ifscube} runs, each one of the emission lines seen in Fig.~\ref{fig:HST_NIFS}, thus obtaining their amplitudes, radial velocities, velocity dispersion ($\sigma$) and integrated fluxes. Pixels where the peak of a given emission line compared to the standard deviation of the adjacent continuum was smaller than five were masked during the fit. We also included a one degree polynomial to account for the contribution of the underlying residual continuum to the observed emission. The $\mathrm{[\ion{P}{ii}]\,\lambda 11886}$\AA\ , $\mathrm{Pa\beta}$, $\mathrm{Br\gamma}$, $\mathrm{H_2\,\lambda 22230}$\AA\ and $\mathrm{H_2\,\lambda 22470}$\AA\ emission lines were well fitted by a single Gaussian component. In the cases of the $\mathrm{[\ion{Fe}{ii}]\,\lambda 12570}$\AA\ and $\mathrm{H_2\,\lambda 21218}$\AA\ emission lines, we started by fitting them with a single Gaussian, however this procedure was not able to account for the presence of broad, blue wings in a region of radius $\approx 0.5 \arcsec$ around the nucleus of the galaxy, therefore we included a second Gaussian to fit their profiles. We fixed the radial velocities of these components at $-420$ and $-250$~$\mathrm{km\,s^{-1}}$ and their velocity dispersions at 350 and 215~$\mathrm{km\,s^{-1}}$ for the $\mathrm{[\ion{Fe}{ii}]\,\lambda 12570}$\AA\ and $\mathrm{H_2\,\lambda 21218}$\AA\ emission lines, respectively, as determined from the fitting of the nuclear spectra, integrated in an aperture of 0.2$\arcsec \times$0.2$\arcsec$, also using the package {\sc ifscube}. However, since these components are fitted simultaneously with their main narrow counterparts, a further threshold needs to be applied to claim their detection (see Sec.~\ref{sec:FluxRes}). We note that we did not detect broad, blue-shifted components associated with the other H$_2$ emission features. This could be due to the fact the H$_2$ emissions at $\mathrm{\lambda 22230}$\AA\ and $\mathrm{\lambda 22470}$\AA\ are much weaker compared to the emission at $\mathrm{\lambda 21218}$\AA\ or a result of different excitation mechanisms of the H$_2$ molecule, as will be discussed in Sec.~\ref{sec:FeIIH2Nature}.

We point out that while fitting the $\mathrm{[\ion{P}{ii}]\,\lambda 11886}$\AA\ emission line, we noticed a neighbouring emission feature at the wavelength 11910~\AA\ corresponding to a forbidden transition of $\mathrm{[\ion{Ni}{ii}]}$. For this reason, we fitted these lines simultaneously, although we kept their kinematics independent. The $\mathrm{[\ion{Ni}{ii}]\,\lambda 11910}$\AA\ emission is confined to a region of radius smaller than 0.3$\arcsec$ in the centre of o the FoV and it does not affect the fitting of the $\mathrm{[\ion{P}{ii}]\,\lambda 11886}$\AA\ profile. 

\begin{figure*}
\begin{center}

	\begin{minipage}{\textwidth}
	\centering
	\begin{minipage}{.33\textwidth}
		\includegraphics[width=\textwidth]{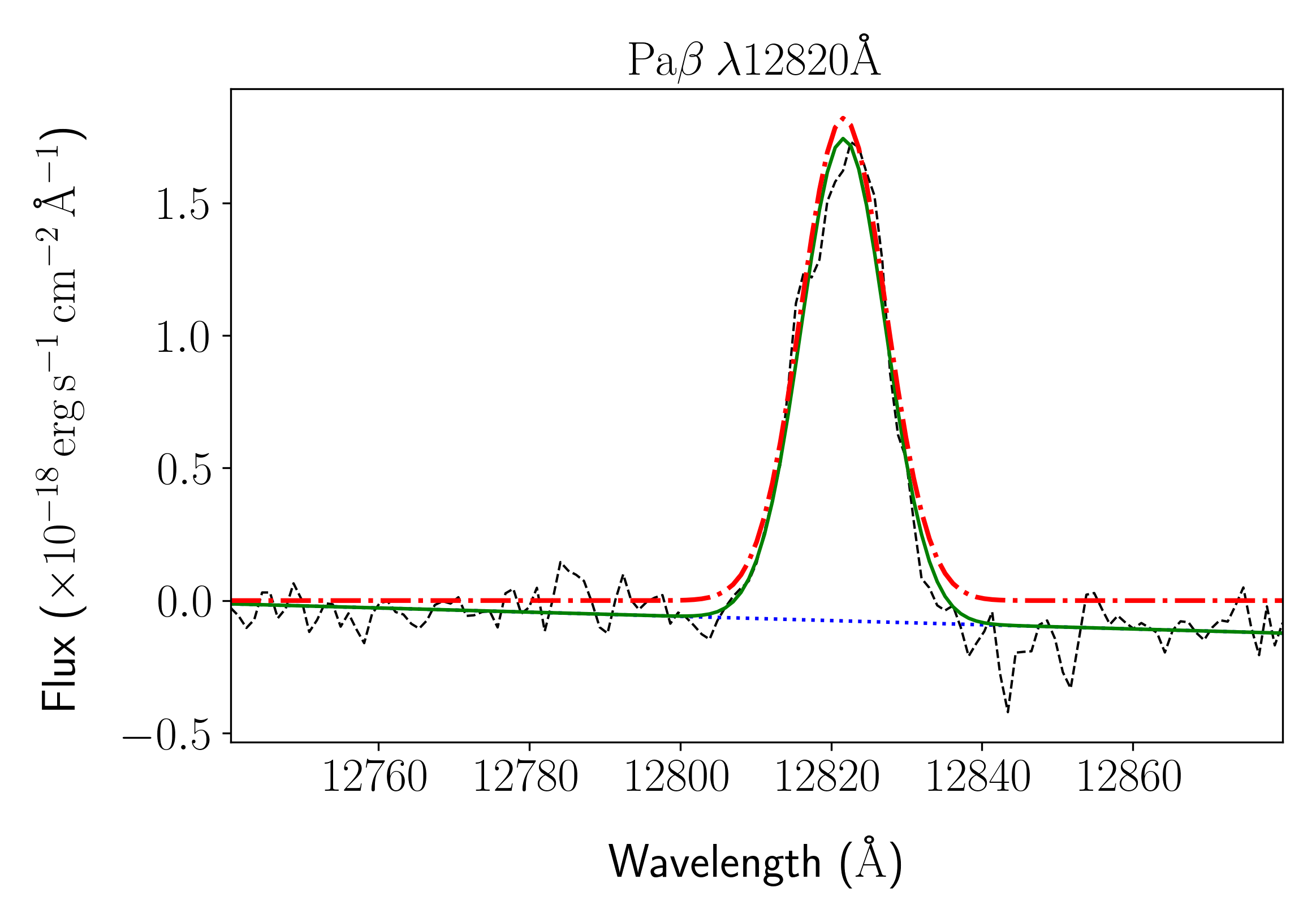}
	\end{minipage}
	\begin{minipage}{.33\textwidth}
		\includegraphics[width=\textwidth]{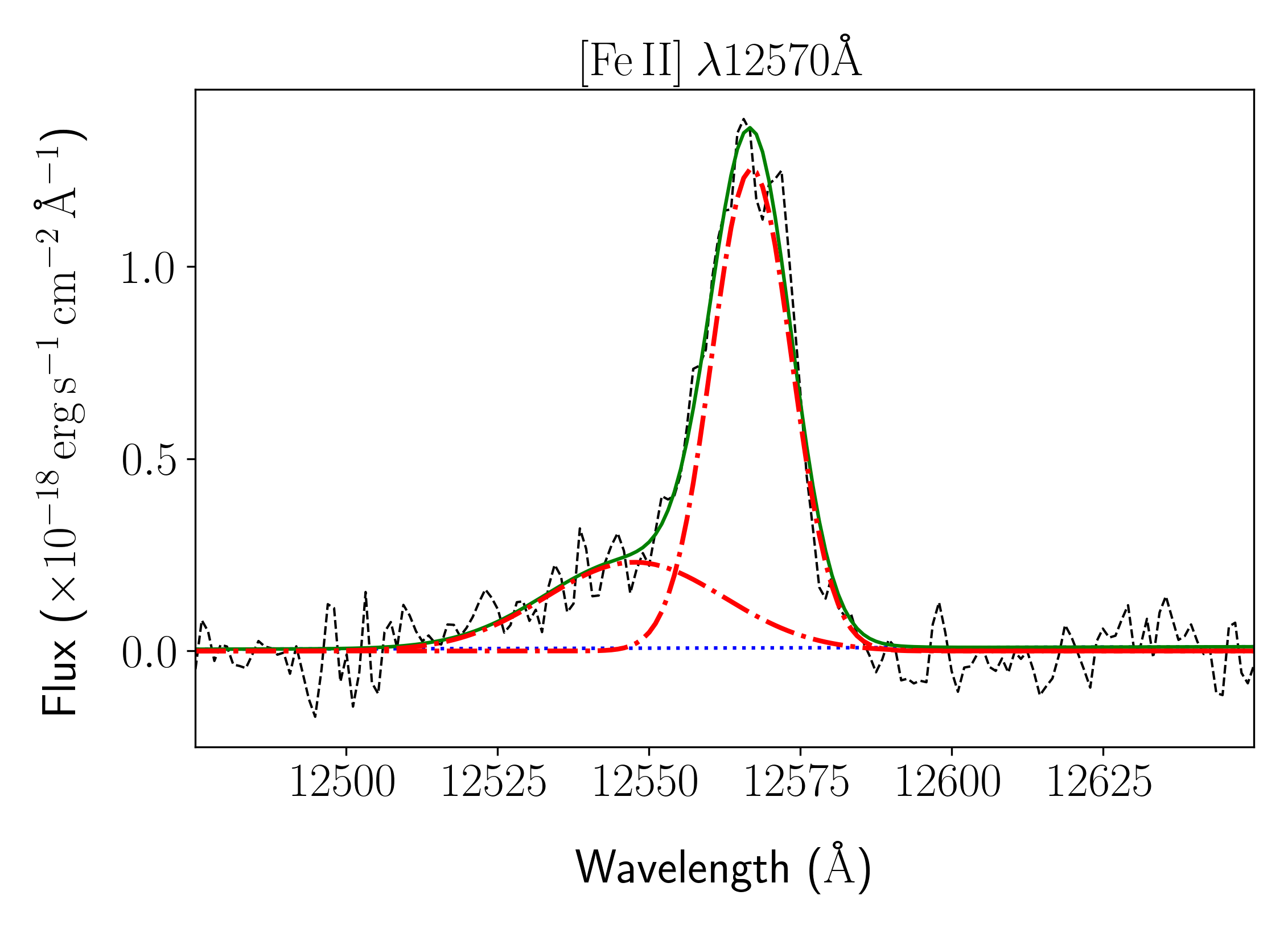}
	\end{minipage}
	\begin{minipage}{.33\textwidth}
		\includegraphics[width=\textwidth]{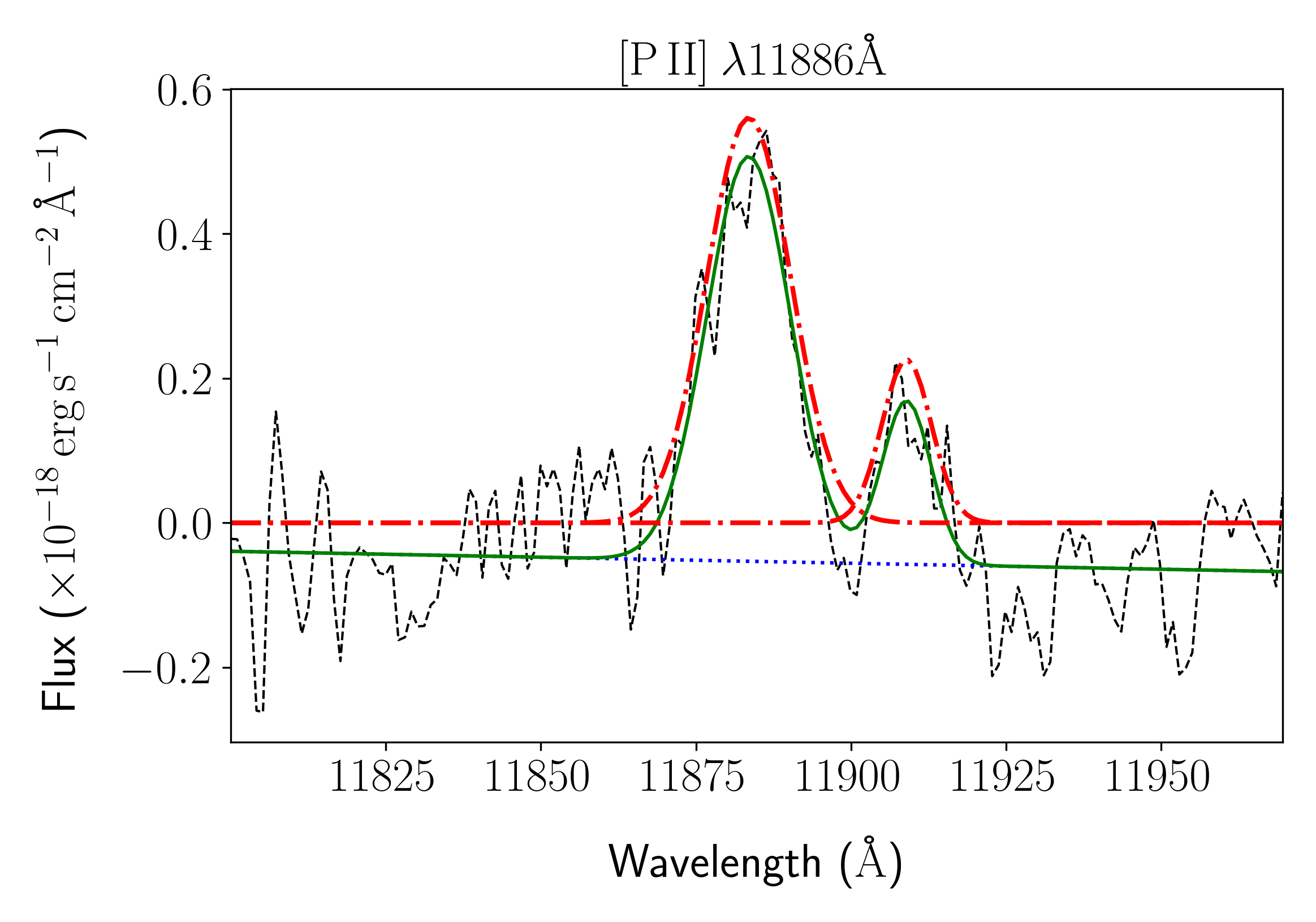}
	\end{minipage}
	\end{minipage}

	\begin{minipage}{\textwidth}
	\centering
	\begin{minipage}{.33\textwidth}
		\includegraphics[width=\textwidth]{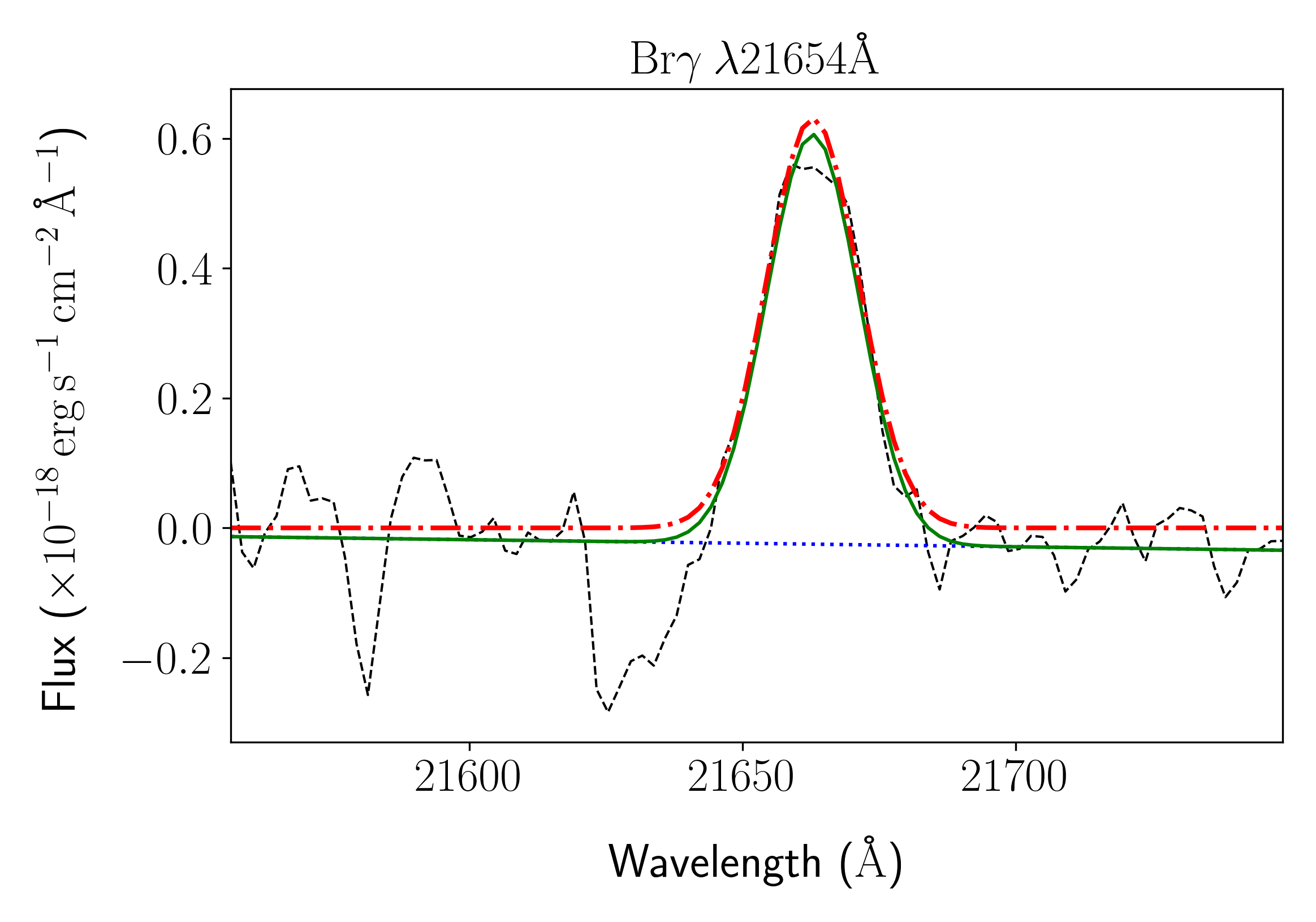}
	\end{minipage}
	\begin{minipage}{.33\textwidth}
		\includegraphics[width=\textwidth]{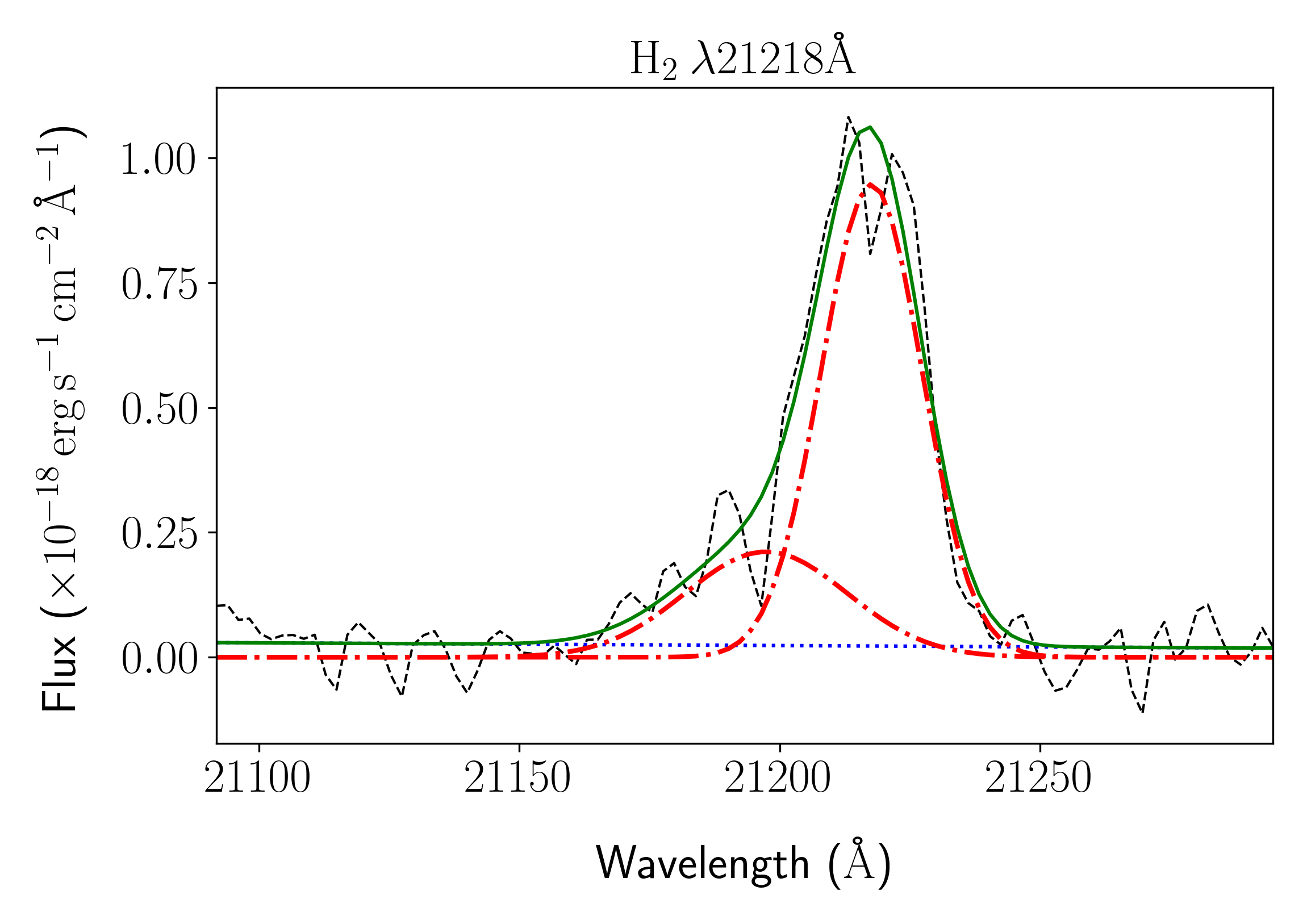}
	\end{minipage}
	\end{minipage}
	
\caption{Examples of fitting results using the package {\sc ifscube} for the spaxel corresponding to the nucleus of the galaxy. In each panel, the black-dashed and green-solid lines are the observed and modelled spectra, respectively. The fitted Gaussian components are shown by the red-dash-dotted lines and the blue-dotted line is the fitted residual continuum. The main fitted feature is indicated at the top of each panel. Note that the second component accompanying the $\mathrm{[\ion{P}{ii}]\,\lambda 11886}$\AA\ emission line corresponds to the $\mathrm{[\ion{Ni}{ii}]\,\lambda 11910}$\AA\ emission feature.
 }
 \label{fig:IfscubeFits}
\end{center}
\end{figure*}

We show in Fig.~\ref{fig:IfscubeFits} the resulting fits for the main emission lines for the spaxel corresponding to the nucleus of the galaxy. From these example fits, one can see that the broad, blue-shifted components accompanying the $\mathrm{[\ion{Fe}{ii}]\,\lambda 12570}$\AA\ and $\mathrm{H_2\,\lambda 21218}$\AA\ emission lines are clearly distinguished from their main narrow counterparts. Moreover, the panel displaying the resulting fit for the $\mathrm{[\ion{P}{ii}]\,\lambda 11886}$\AA\ emission line also shows a second, well separated feature corresponding to the $\mathrm{[\ion{Ni}{ii}]\,\lambda 11910}$\AA\ emission.  We did not attempt to model the $\mathrm{[\ion{P}{ii}]}$ emission at $\lambda 11470$\AA\ since it is barely seen in the integrated spectra show in Fig.~\ref{fig:HST_NIFS} as well as in the gas spectra shown in Fig.~\ref{fig:ppxf}.

\section{Results} \label{sec:Results}

\begin{figure*}
\begin{center}

	\begin{minipage}{\textwidth}
	\centering
	\begin{minipage}{.3\textwidth}
		\includegraphics[width=\textwidth]{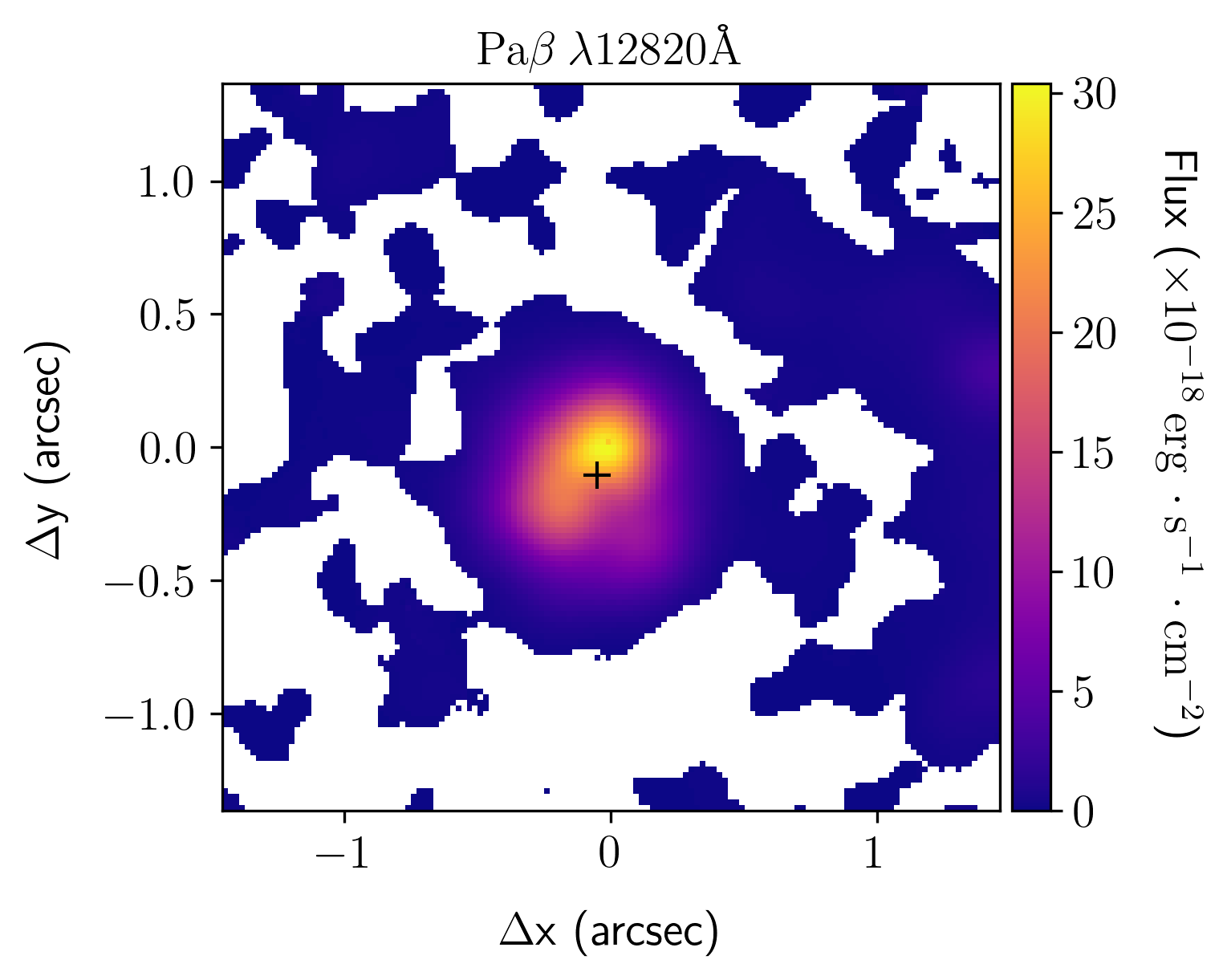}
	\end{minipage}
	\begin{minipage}{.3\textwidth}
		\includegraphics[width=\textwidth]{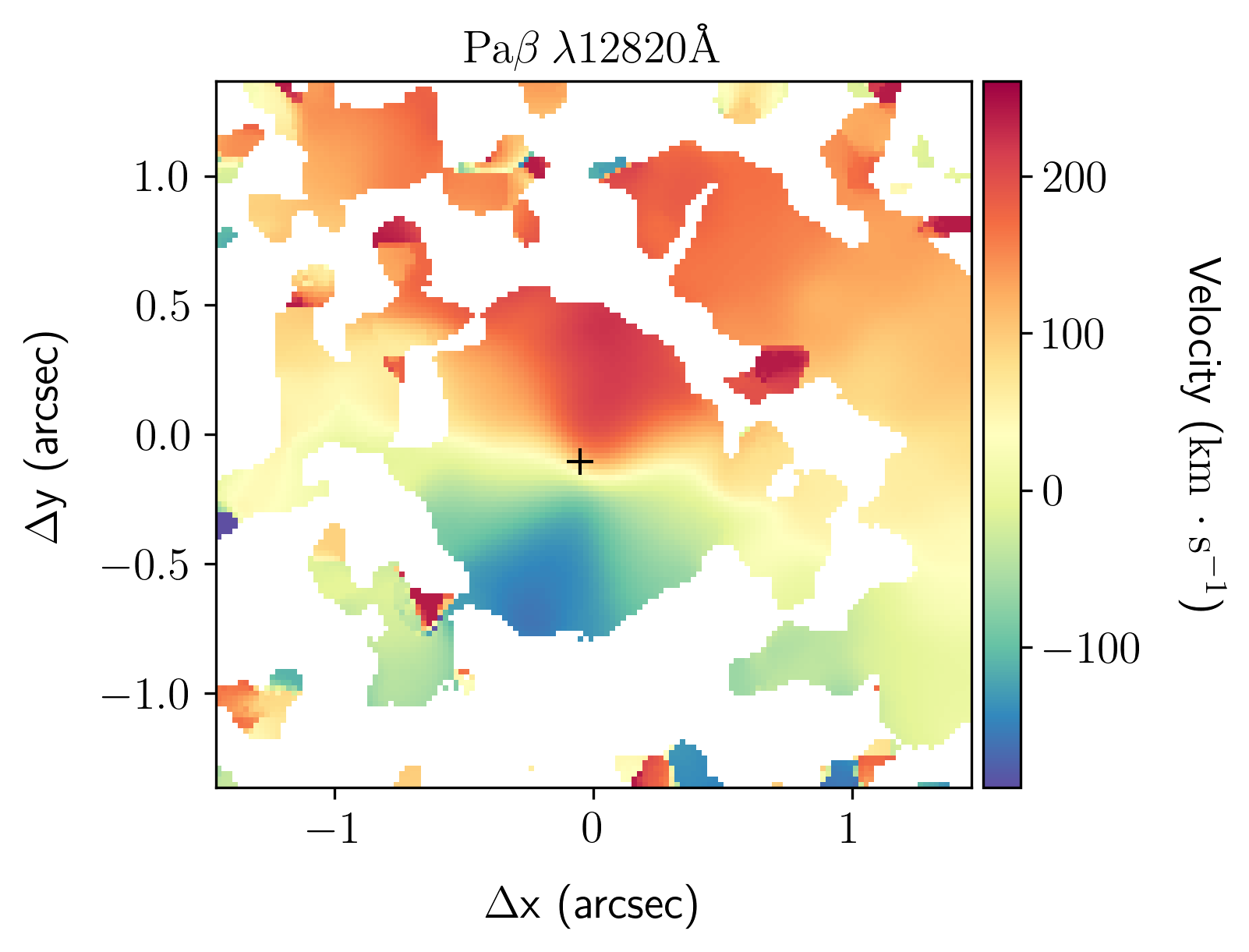}
	\end{minipage}
	\begin{minipage}{.3\textwidth}
		\includegraphics[width=\textwidth]{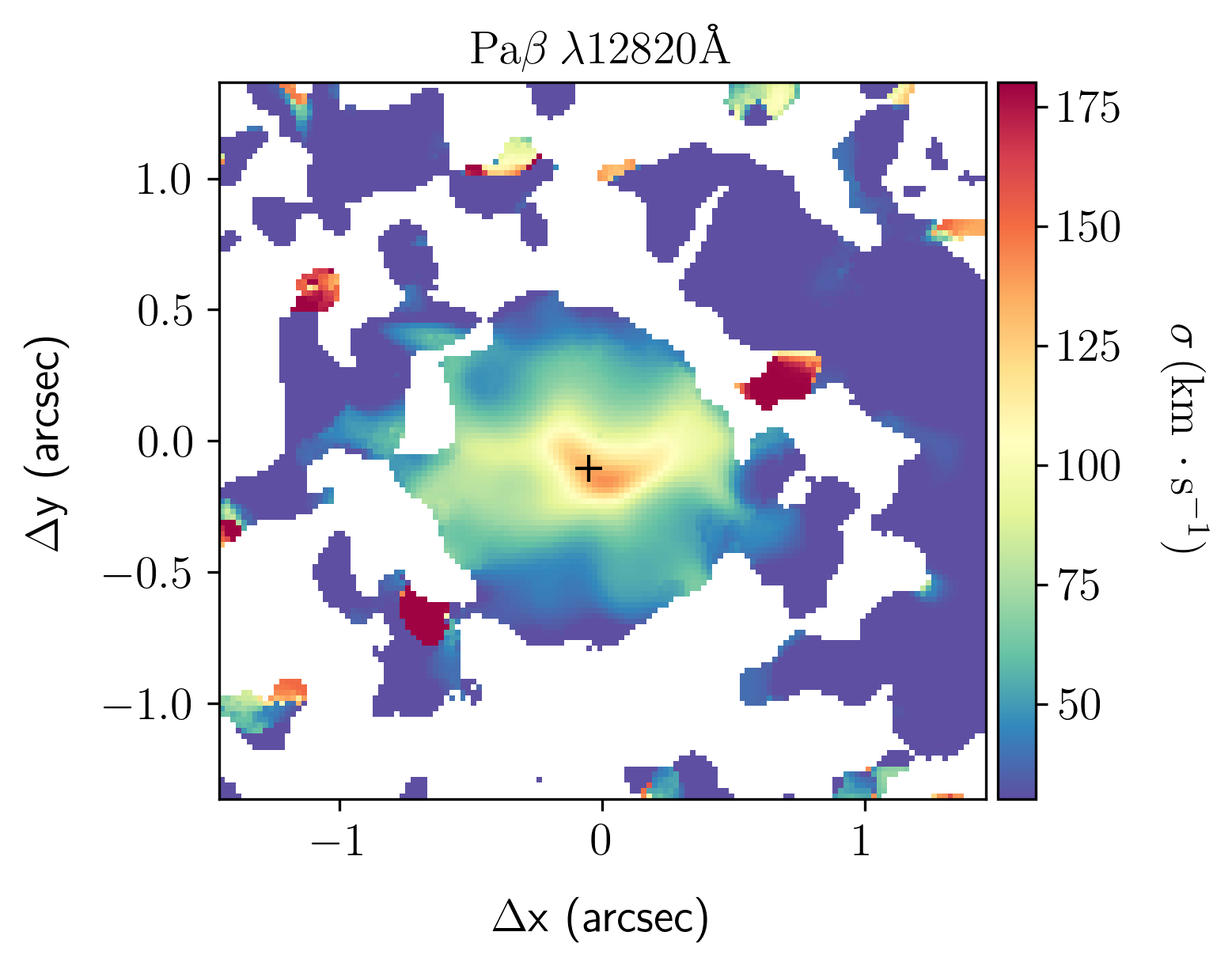}
	\end{minipage}
	\end{minipage}

	\begin{minipage}{\textwidth}
	\centering
	\begin{minipage}{.3\textwidth}
		\includegraphics[width=\textwidth]{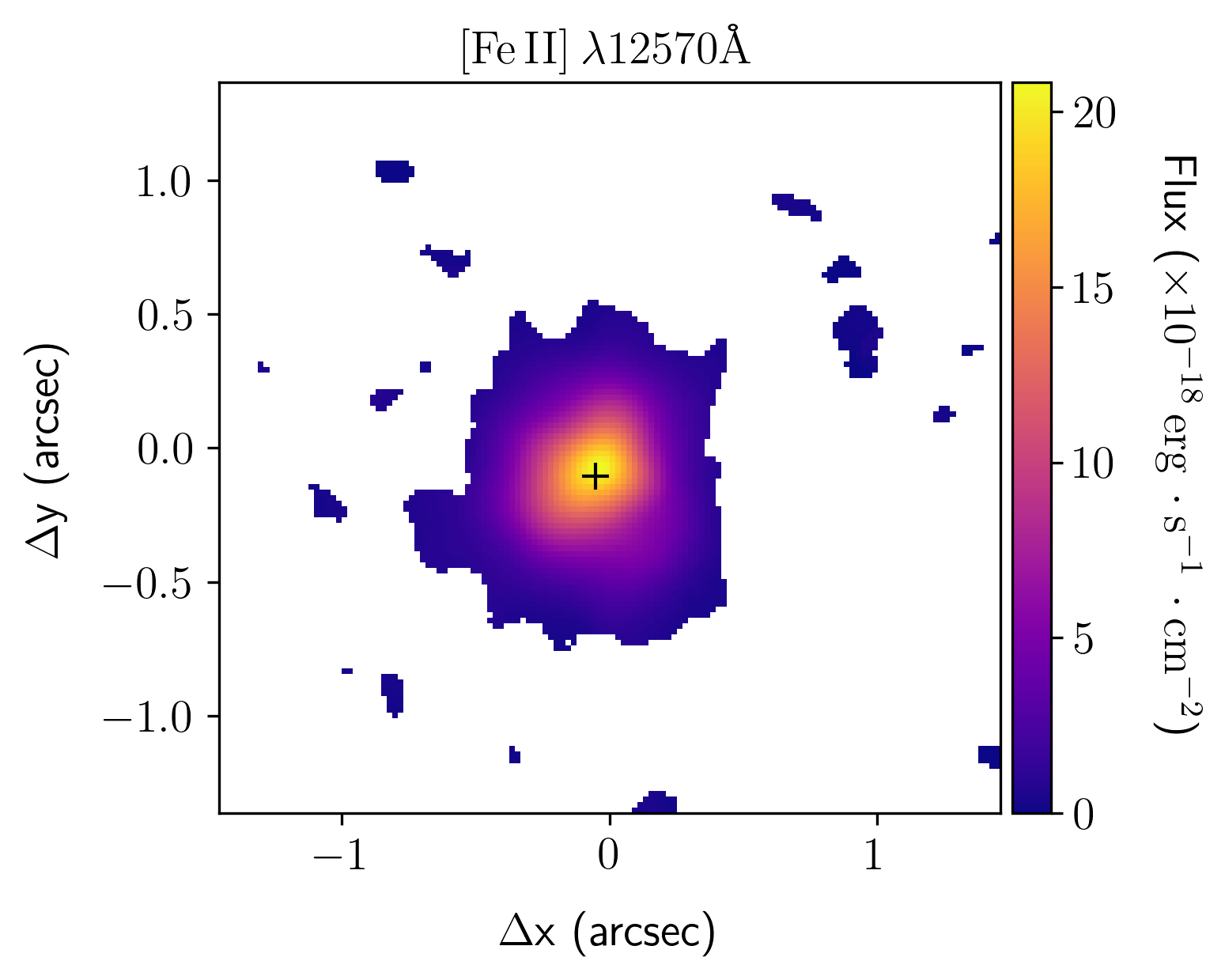}
	\end{minipage}
	\begin{minipage}{.3\textwidth}
		\includegraphics[width=\textwidth]{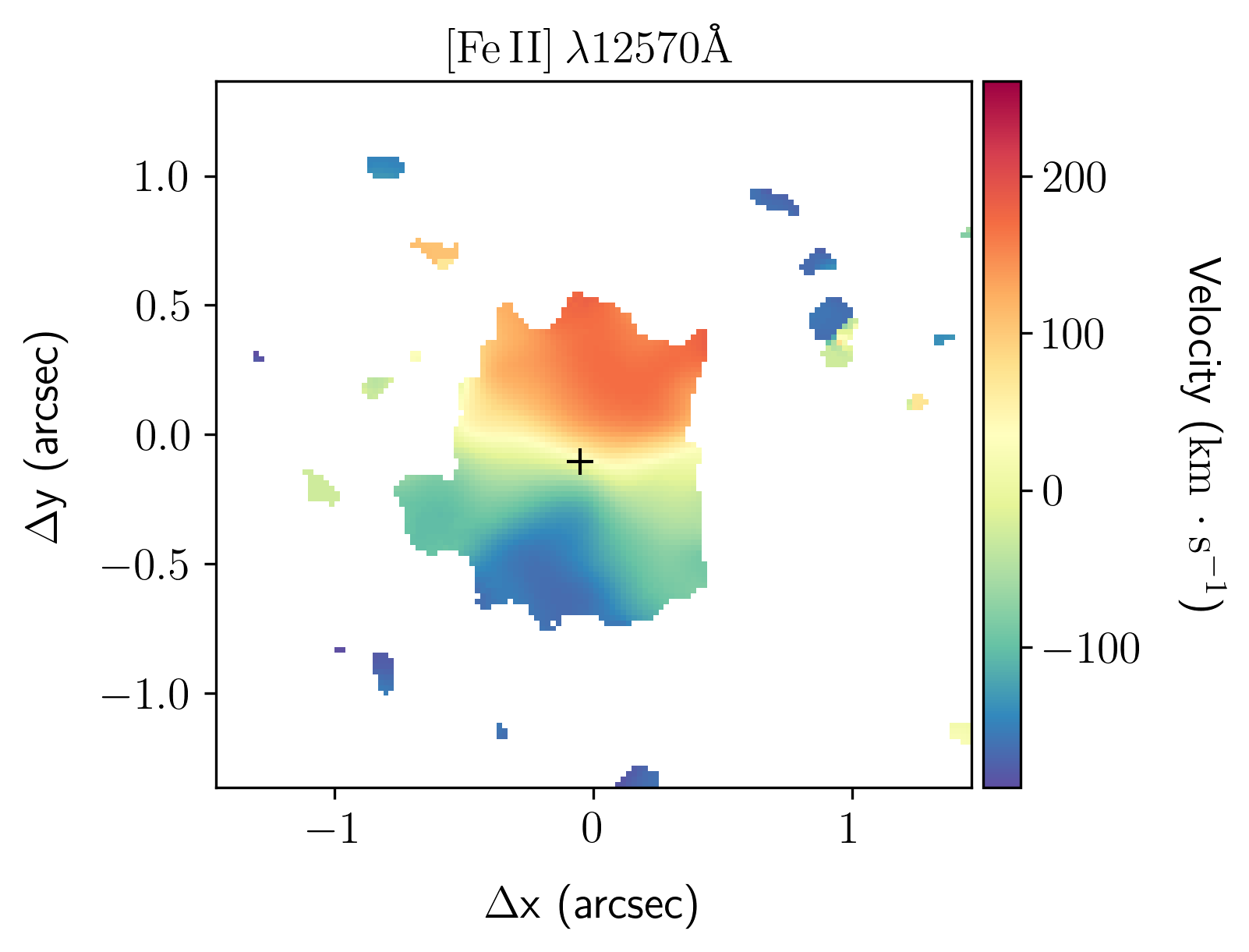}
	\end{minipage}
	\begin{minipage}{.3\textwidth}
		\includegraphics[width=\textwidth]{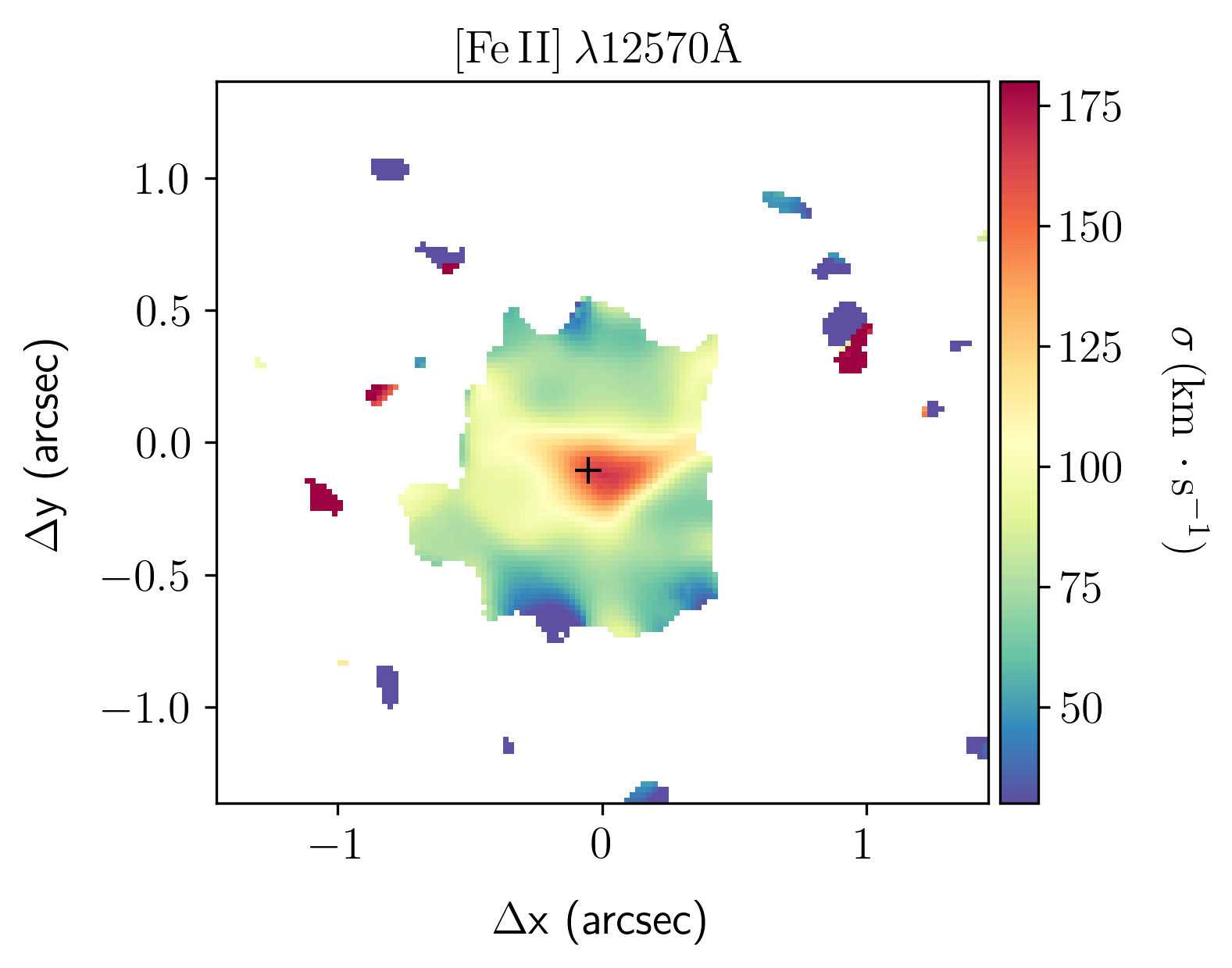}
	\end{minipage}
	\end{minipage}
	
	\begin{minipage}{\textwidth}
	\centering
	\begin{minipage}{.3\textwidth}
		\includegraphics[width=\textwidth]{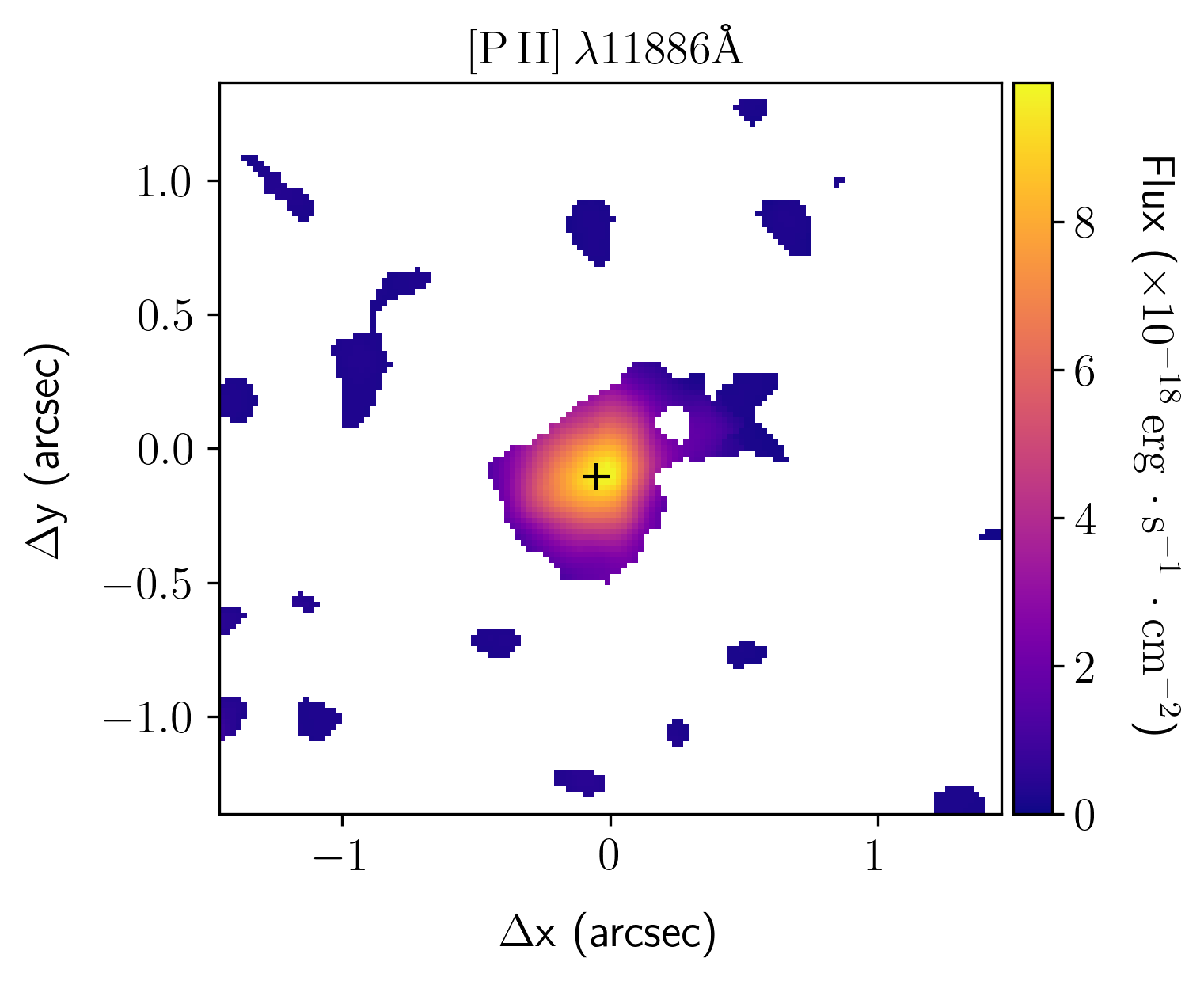}
	\end{minipage}
	\begin{minipage}{.3\textwidth}
		\includegraphics[width=\textwidth]{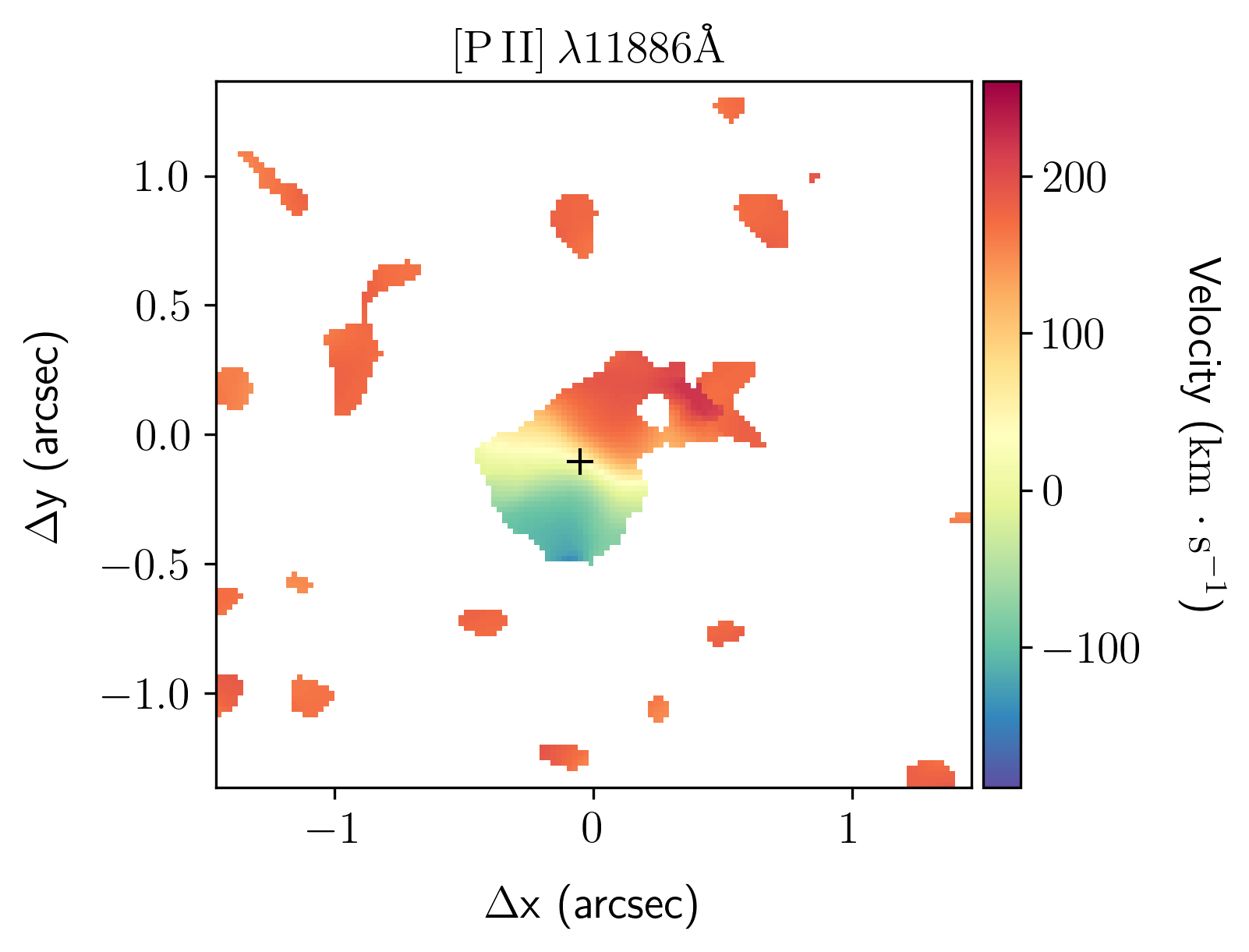}
	\end{minipage}
	\begin{minipage}{.3\textwidth}
		\includegraphics[width=\textwidth]{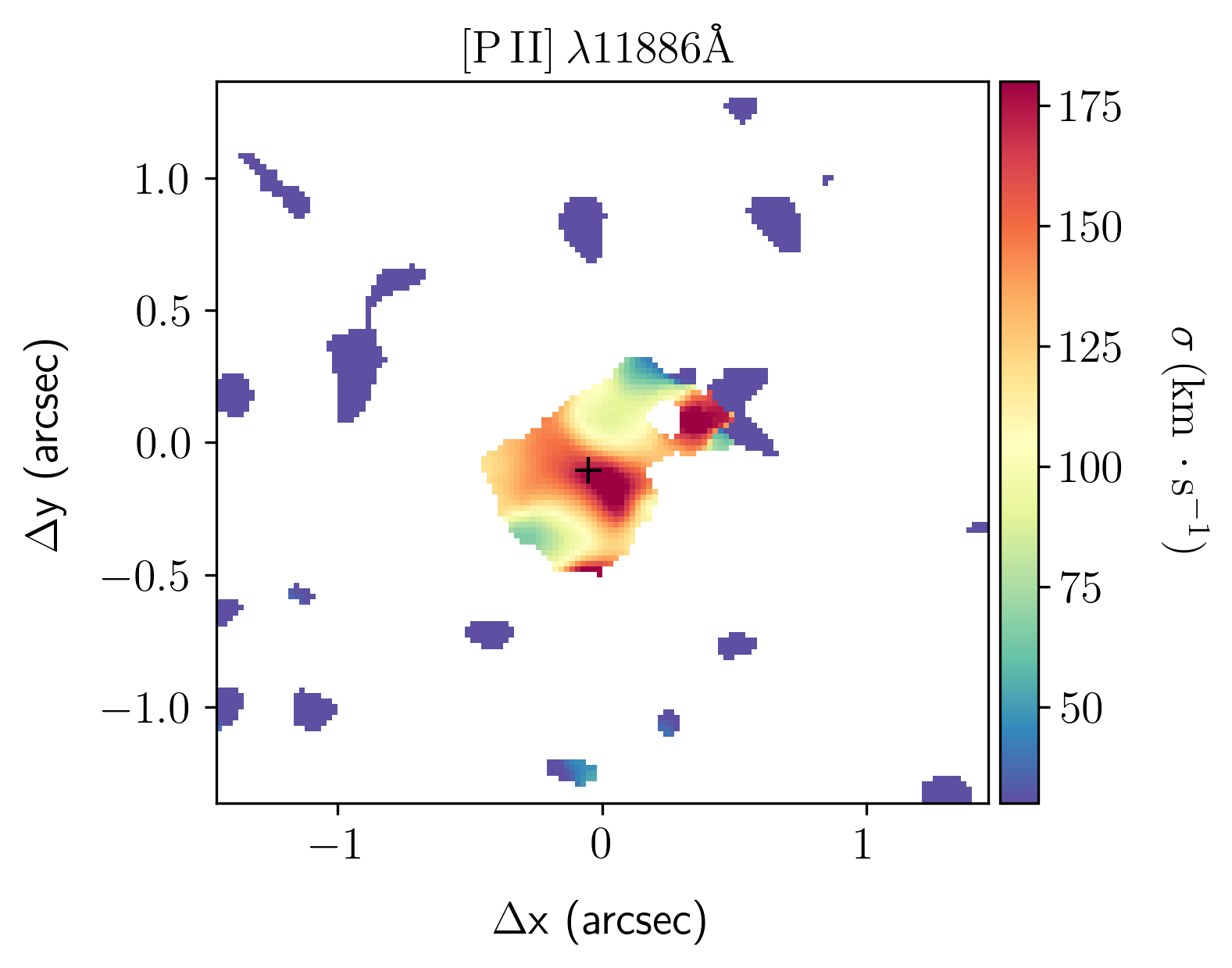}
	\end{minipage}
	\end{minipage}
	
	\begin{minipage}{\textwidth}
	\centering
	\begin{minipage}{.3\textwidth}
		\includegraphics[width=\textwidth]{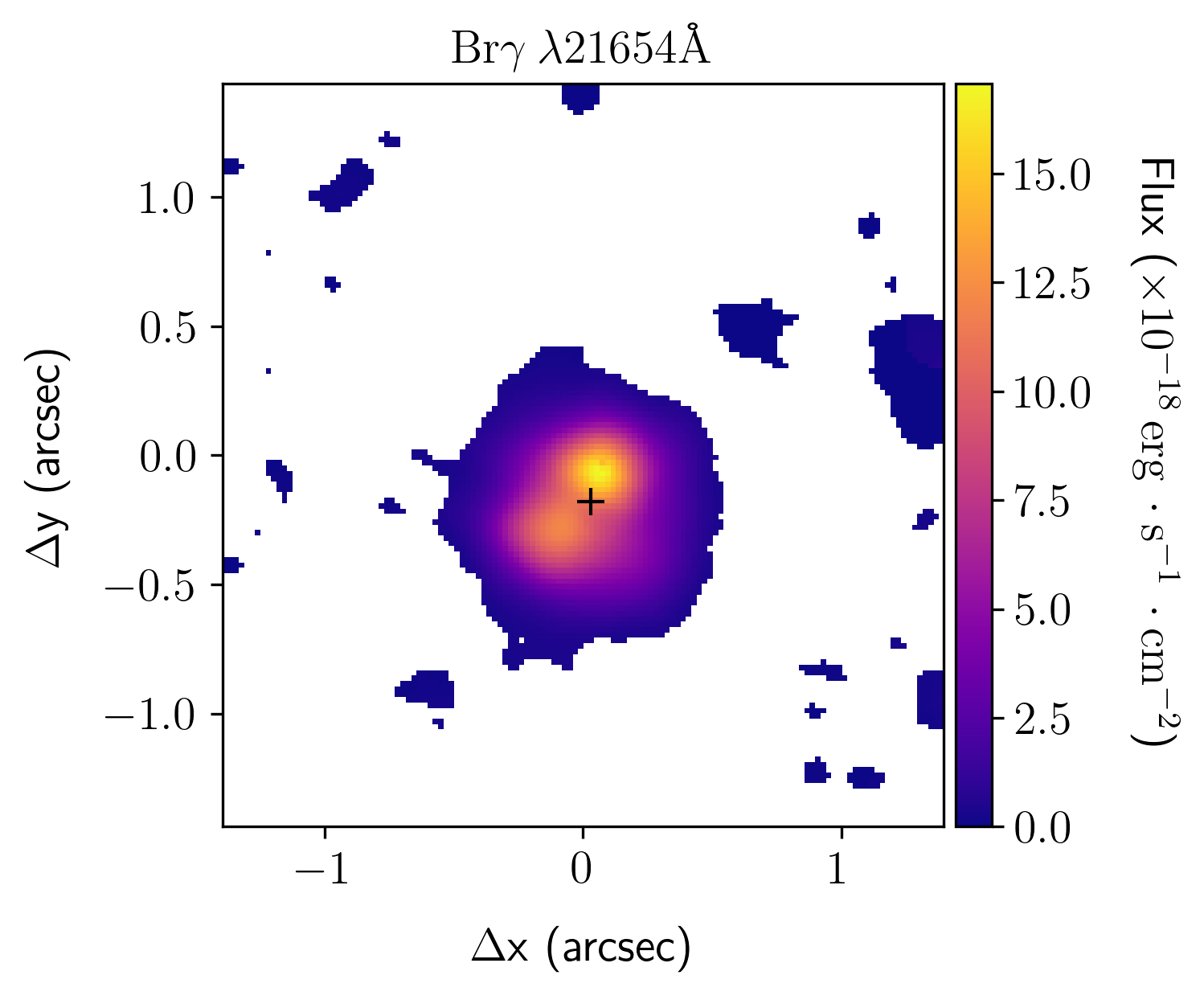}
	\end{minipage}
	\begin{minipage}{.3\textwidth}
		\includegraphics[width=\textwidth]{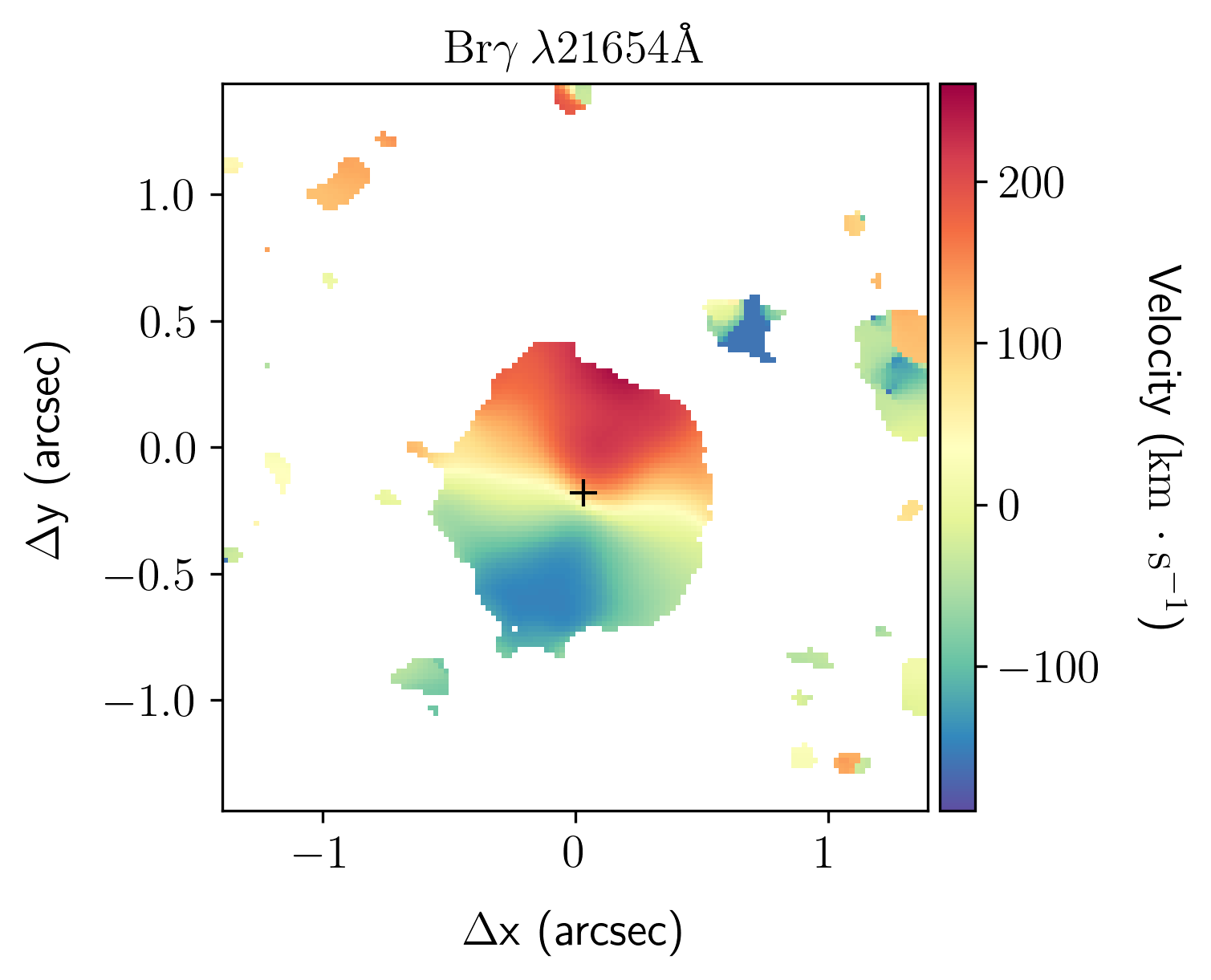}
	\end{minipage}
	\begin{minipage}{.3\textwidth}
		\includegraphics[width=\textwidth]{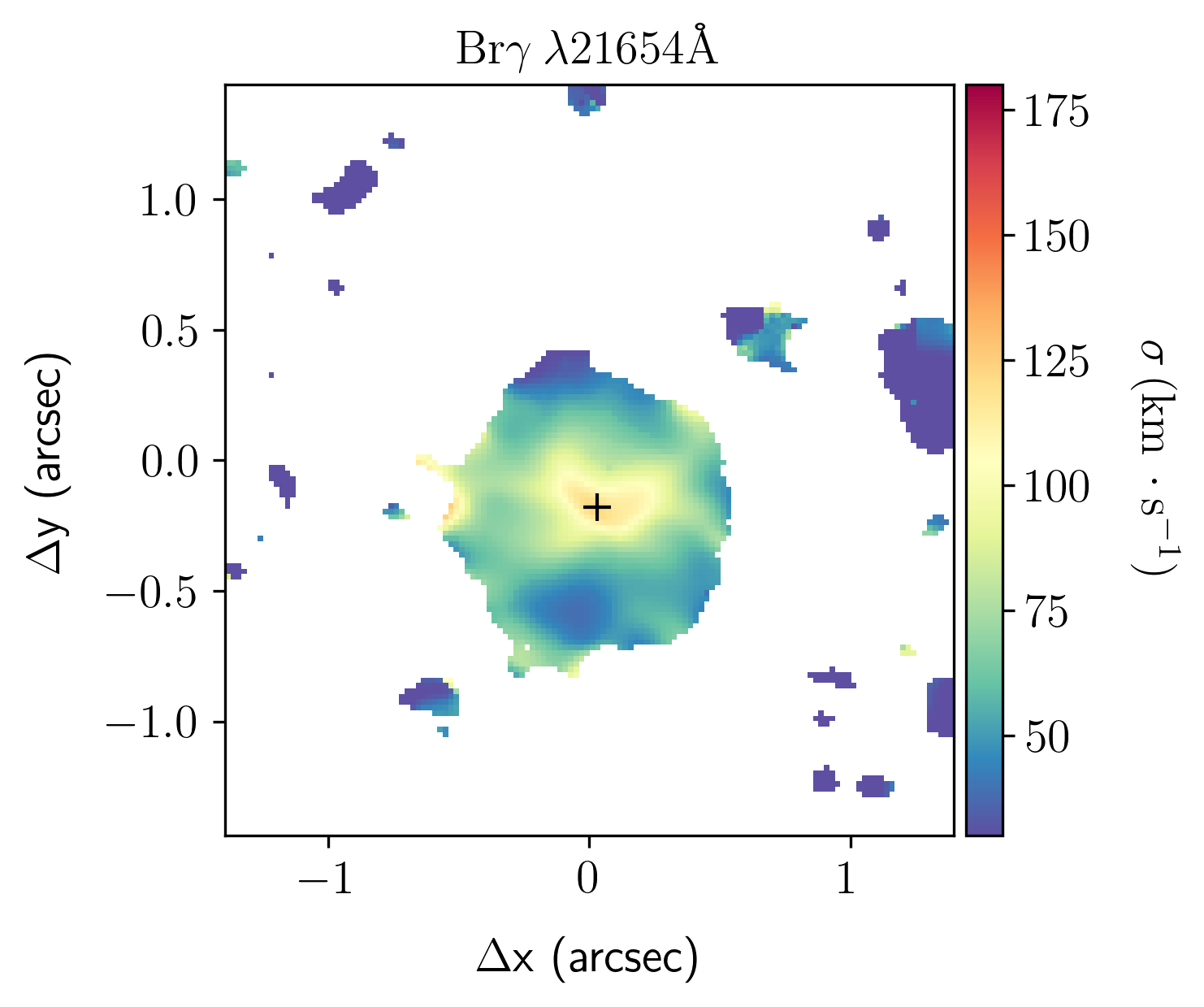}
	\end{minipage}
	\end{minipage}

	\begin{minipage}{\textwidth}
	\centering
	\begin{minipage}{.3\textwidth}
		\includegraphics[width=\textwidth]{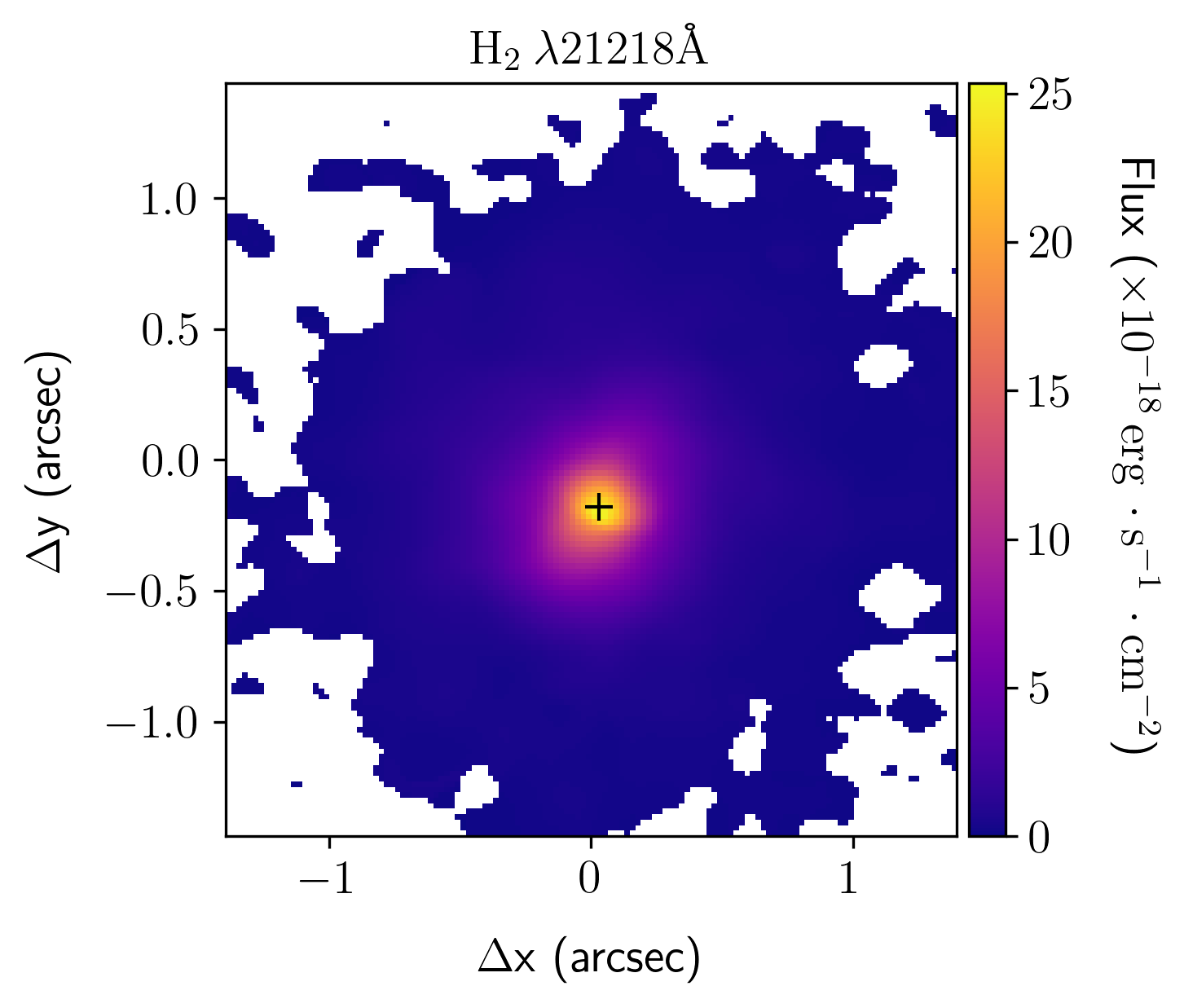}
	\end{minipage}
	\begin{minipage}{.3\textwidth}
		\includegraphics[width=\textwidth]{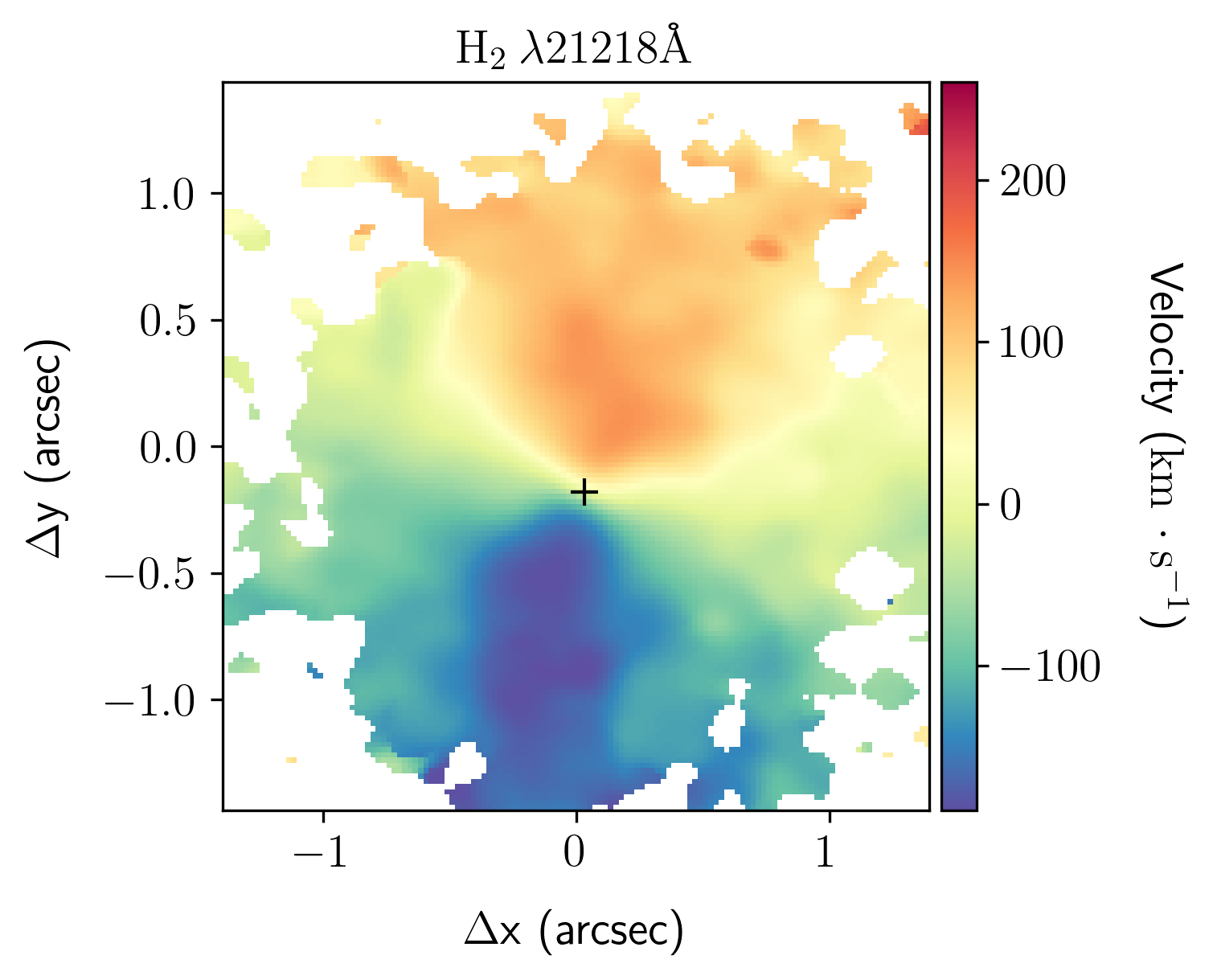}
	\end{minipage}
	\begin{minipage}{.3\textwidth}
		\includegraphics[width=\textwidth]{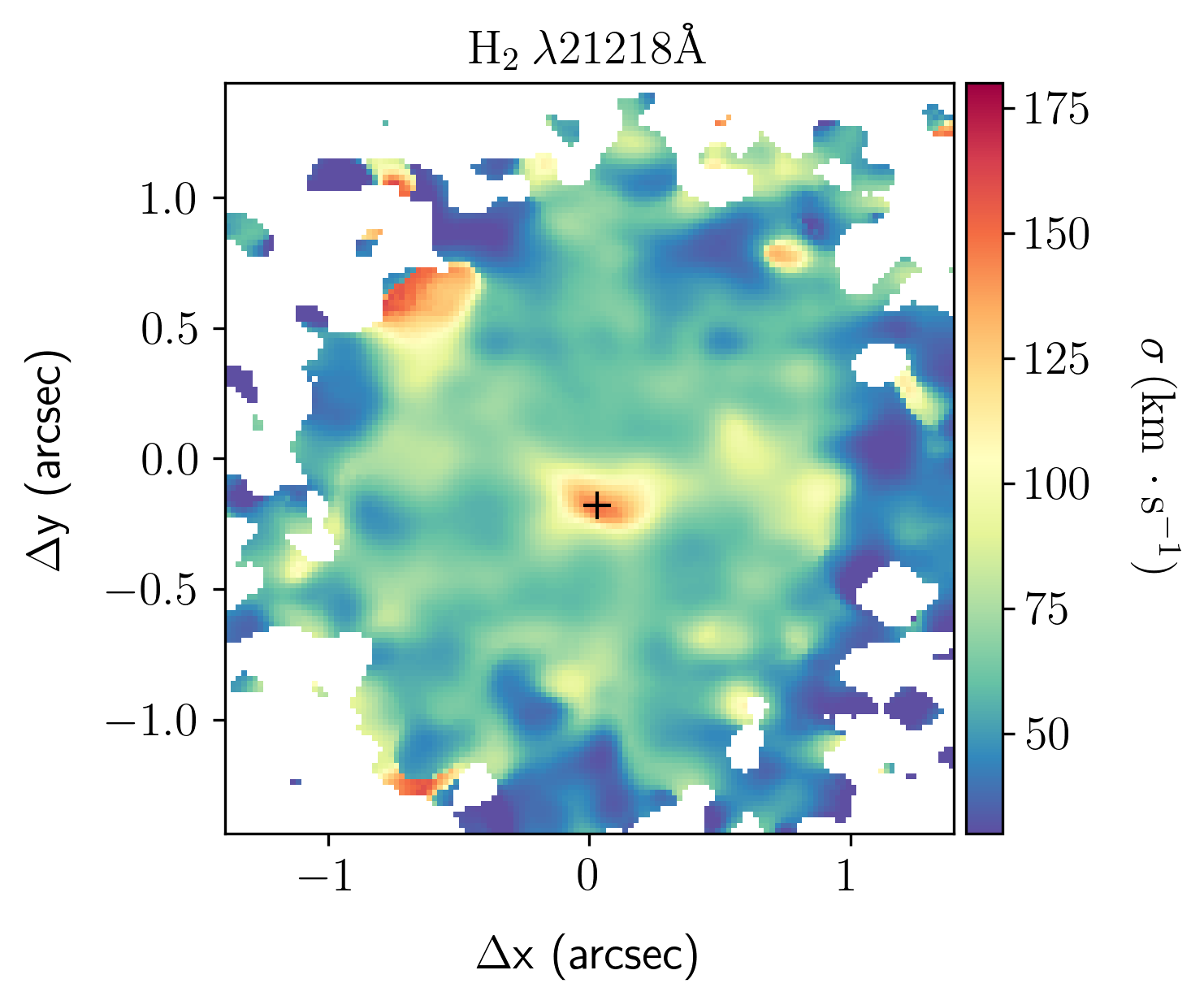}
	\end{minipage}
	\end{minipage}
	
\caption{Maps of flux, radial velocities and $\sigma$ of the $\mathrm{Pa\beta}$ (first row), narrow $\mathrm{[\ion{Fe}{ii}]\,\lambda 12570}$\AA\ (second row), $\mathrm{[\ion{P}{ii}]\,\lambda 11886}$\AA\ (third row), $\mathrm{Br\gamma}$ (fourth row) and narrow $\mathrm{H_2\,\lambda 21218}$\AA\ (fifth row) emission lines. The range of values are indicated by the accompanying color bars. The black cross in each panel indicates the nucleus of the galaxy. Pixels in white were not fitted. North is up and east is to the left in all maps.
 }
 \label{fig:Maps}
\end{center}
\end{figure*}

We show maps of flux distribution, radial velocities and $\sigma$ in the first, second and third columns, respectively, of Fig.~\ref{fig:Maps} for the  $\mathrm{Pa\beta}$, $\mathrm{[\ion{P}{ii}]\,\lambda 11886}$\AA\ and $\mathrm{Br\gamma}$ emission lines as well as for the $\mathrm{[\ion{Fe}{ii}]\,\lambda 12570}$\AA\ and $\mathrm{H_2\,\lambda 21218}$\AA\ narrow components. In all cases, velocities are relative to the corrected systemic velocity of the galaxy $v_{sys,cor} = 5841\,\mathrm{km\,s^{-1}}$ (see Sec.~\ref{sec:CND}), and maps of velocity dispersions were corrected for the instrumental broadening. In all panels north is up, east is to the left, the black cross indicates the nucleus of the galaxy, and pixels in white were not considered in the fit. In Fig.~\ref{fig:MapsOthers} we show the flux distribution of the $\mathrm{[\ion{Fe}{ii}]\,\lambda 12570}$\AA\ and $\mathrm{H_2\,\lambda 21218}$\AA\ broad, blue-shifted emission. The flux map of the $\mathrm{[\ion{Ni}{ii}]\,\lambda 11910}$\AA\ emission line is shown in Fig.~\ref{fig:flux_NiII}.

Hereafter, we refer to the $\mathrm{[\ion{Fe}{ii}]\,\lambda 12570}$\AA\ , $\mathrm{[\ion{P}{ii}]\,\lambda 11886}$\AA\ and $\mathrm{[\ion{Ni}{ii}]\,\lambda 11910}$\AA\ emission lines simply as $\mathrm{[\ion{Fe}{ii}]}$,  $\mathrm{[\ion{P}{ii}]}$ and  $\mathrm{[\ion{Ni}{ii}]}$, respectively.

\subsection{Flux distribution of the emission lines} \label{sec:FluxRes}

The $\mathrm{Pa\beta}$, $\mathrm{[\ion{Fe}{ii}]}$,  $\mathrm{[\ion{P}{ii}]}$ and $\mathrm{Br\gamma}$ emission features, as well as the $\mathrm{H_2\,\lambda 21218}$\AA\ molecular line show spatially resolved emission. The most compact flux distribution is found for $\mathrm{[\ion{P}{ii}]}$ which extends to roughly 0.4$\arcsec$ from the nucleus of the galaxy, while the $\mathrm{H_2\,\lambda 21218}$\AA\ emission line extends over almost the entire NIFS FoV. In the \textit{J}~band, both the $\mathrm{[\ion{Fe}{ii}]}$ and $\mathrm{[\ion{P}{ii}]}$ flux distributions peak close to the galaxy nucleus and display a slight elongation in the northwest-southeast direction. In the \textit{K}~band, H$_2$ emission peaks at the nucleus and smoothly decreases towards the edges of the FoV.

The $\mathrm{Pa\beta}$ and $\mathrm{Br\gamma}$ flux maps show that the flux distribution is asymmetric and that the peak is off-centred -- $\mathrm{Pa\beta}$ and $\mathrm{Br\gamma}$ fluxes peak northwest, $\mathrm{\approx 0.1\arcsec}$ of the nucleus, and decrease towards the south-east. This may indicate either the presence of a gradient in the amount of visual extinction at the galaxy or that these lines are tracing a circumnuclear structure. We note that extended $\mathrm{Pa\beta}$ emission is also seen along the right edge of the FoV.

\begin{figure*}
    \centering
	\begin{minipage}{\textwidth}
	\centering
	\begin{minipage}{.3\textwidth}
		\includegraphics[width=\textwidth]{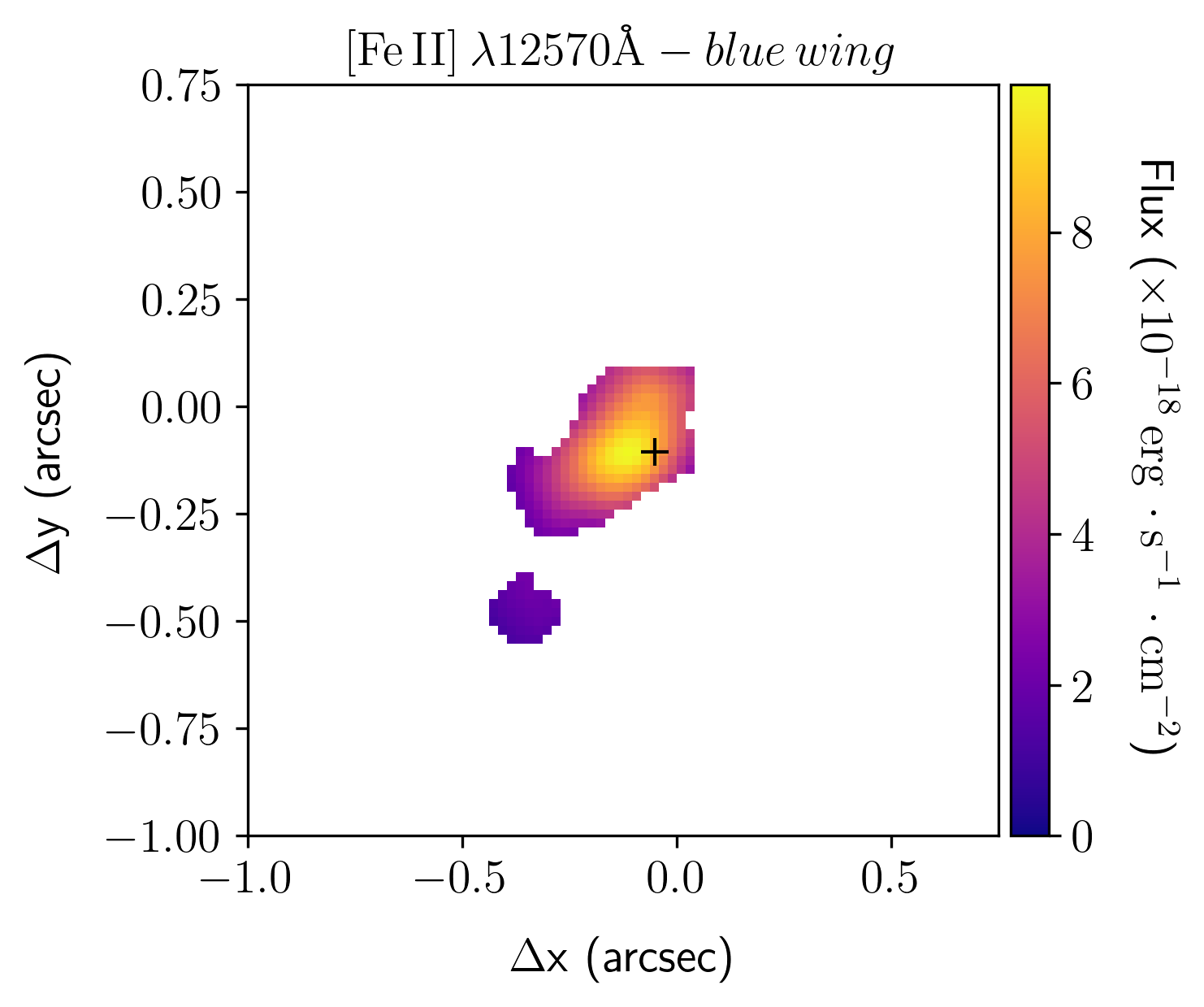}
	\end{minipage}
	\begin{minipage}{.3\textwidth}
		\includegraphics[width=\textwidth]{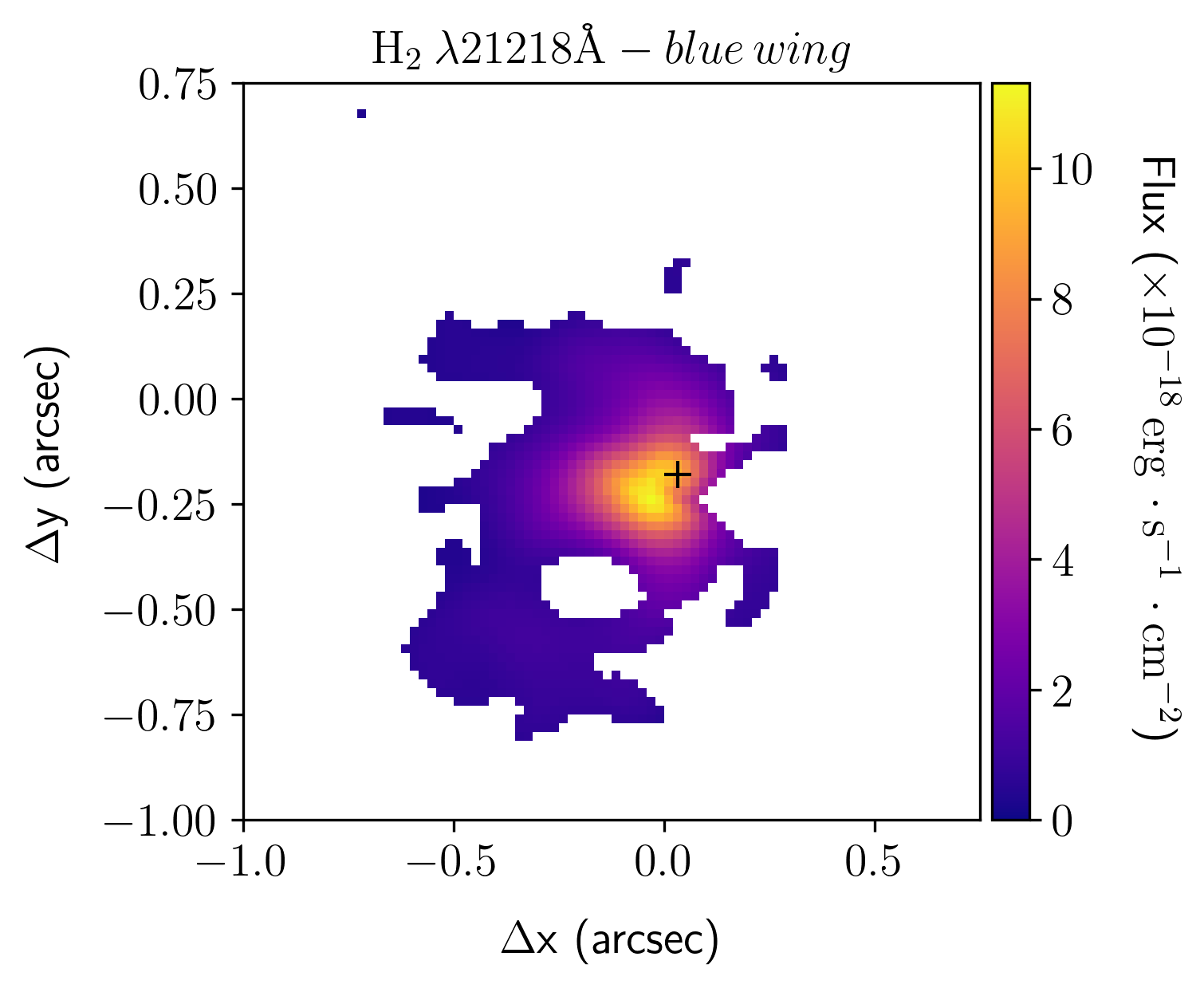}
	\end{minipage}
	\end{minipage}
    \caption{Flux maps of the $\mathrm{[\ion{Fe}{ii}]\,\lambda 12570}$\AA\ (left) and $\mathrm{H_2\,\lambda 21218}$\AA\ (right) broad, blue-shifted components.}
    \label{fig:MapsOthers}
\end{figure*}

The flux distributions of the broad, blue-shifted components associated with the $\mathrm{[\ion{Fe}{ii}]}$ and $\mathrm{H_2\,\lambda 21218}$\AA\ emission lines are shown in the left and right panels of Fig.~\ref{fig:MapsOthers}, respectively. We masked the spaxels where their peaks are smaller than three times the standard deviation of the adjacent continuum. These maps show that the main contribution of these broad, blue-shifted components to the observed emission is restricted to a compact region of radius $\mathrm{\approx 0.25\arcsec}$ close to the nucleus of the galaxy, and with slightly higher values towards the south-east.

\begin{figure}
\centering
	\includegraphics[width=0.65\linewidth]{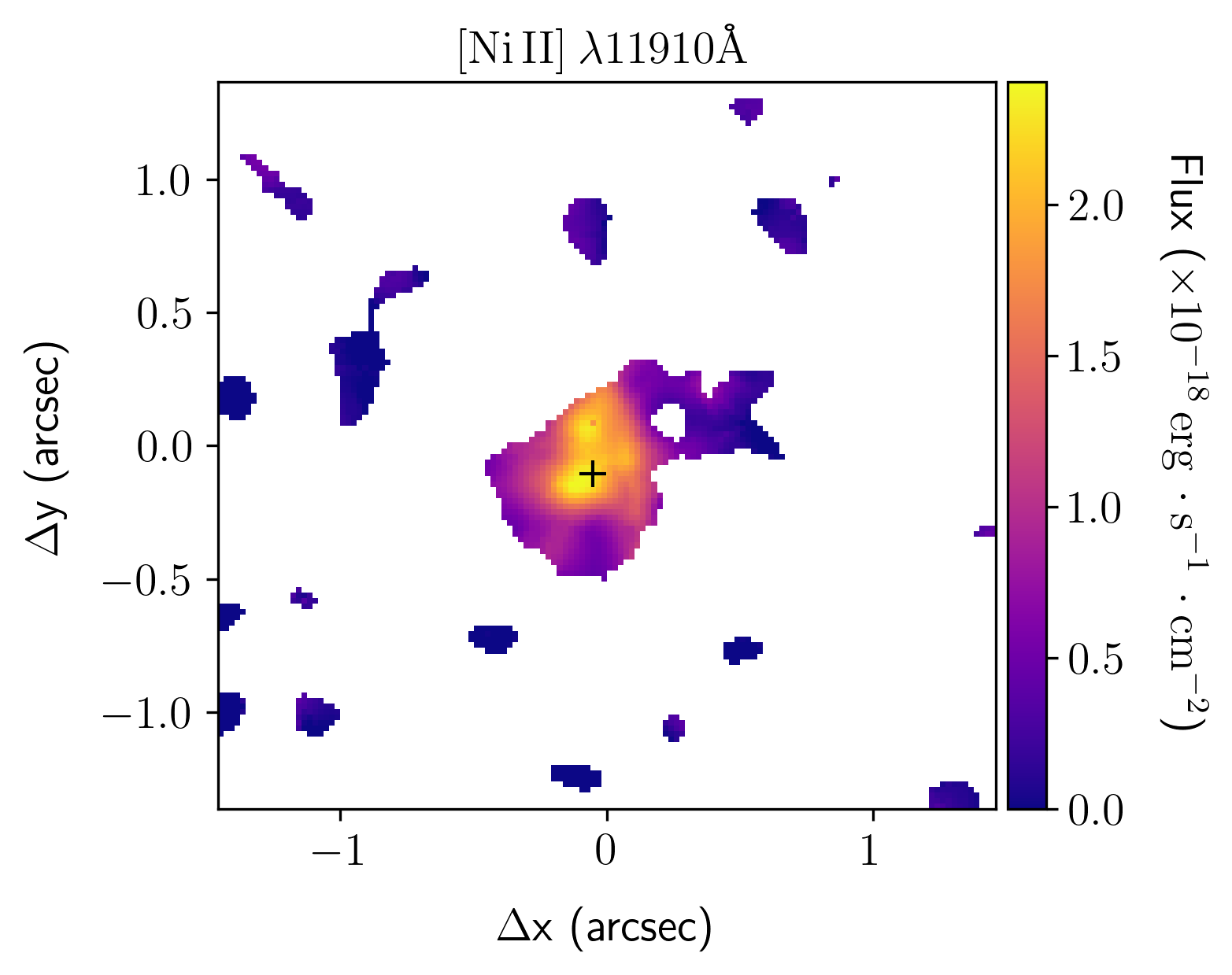}
	\caption{Flux distribution of the $\mathrm{[\ion{Ni}{ii}]\,\lambda 11910}$\AA\ emission line. }
   \label{fig:flux_NiII}
\end{figure}

Finally, the $\mathrm{[\ion{Ni}{ii}]}$ emission (Fig.~\ref{fig:flux_NiII}) is overall much weaker than that of $\mathrm{[\ion{P}{ii}]}$, although, same magnitude -- of the order of $\mathrm{10^{-18} erg\,s^{-1}\,cm^{-2}\,spaxel^{-1}}$ -- flux levels can be seen along a patchy distribution close to the galaxy nucleus. As a matter of fact, on average, the $\mathrm{[\ion{Ni}{ii}]/[\ion{P}{ii}]}$ ratio is $\approx 0.24$, reaching up to 0.3 in some locations. We have not found any report in the literature on the detection of $\mathrm{[\ion{Ni}{ii}]\,\lambda 11910}$\AA\ emission in NGC~34. However, $\mathrm{[\ion{Ni}{ii}]}$ emission at $\lambda 11910$\AA\ has been found in other AGN in a spectroscopic survey carried out by \citet{2017MNRAS.467..540L}. 

\subsection{Gas kinematics} \label{sec:KinRes}

The velocity maps for all the fitted emission lines consistently show a rotation signature with a kinematic major axis at the approximate PA of $-15^{\circ}$, with red-shifted velocities north of the galaxy nucleus, and blue-shifted velocities to the south. For the ionized hydrogen, velocities range from $-160$ to 240~$\mathrm{km\,s^{-1}}$. The ionized gas ($\mathrm{[\ion{Fe}{ii}]}$ and $\mathrm{[\ion{P}{ii}]}$) shows velocities between $-160$ and 170~$\mathrm{km\,s^{-1}}$, while the molecular hydrogen has velocities between $-190$ and 140~$\mathrm{km\,s^{-1}}$. We could not constrain the kinematics of the $\mathrm{[\ion{Ni}{ii}]}$ emission feature due to its patchy distribution.

Velocity dispersions peak close to the nucleus of the galaxy in all cases. The highest values, 160--185~$\mathrm{km\,s^{-1}}$, are found for the ionized gas. Both the ionized and molecular hydrogen show maximum velocity dispersions in the range 125--145~$\mathrm{km\,s^{-1}}$.

We recall that we included broad, blue-shifted components to the fitting of the $\mathrm{[\ion{Fe}{ii}]}$ and $\mathrm{H_2\,\lambda 21218}$\AA\ emission lines, since fitting them with a single Gaussian could not account for the presence of broad, blue wings in a compact region close to the nucleus of the galaxy. These components have fixed radial velocities at $-420$ and $-250$~$\mathrm{km\,s^{-1}}$, and fixed velocity dispersions of 350 and 215~$\mathrm{km\,s^{-1}}$ for the $\mathrm{[\ion{Fe}{ii}]}$ and $\mathrm{H_2\,\lambda 21218}$\AA\ emission lines, respectively.

\begin{figure*}
    \centering
    \includegraphics[width=\textwidth]{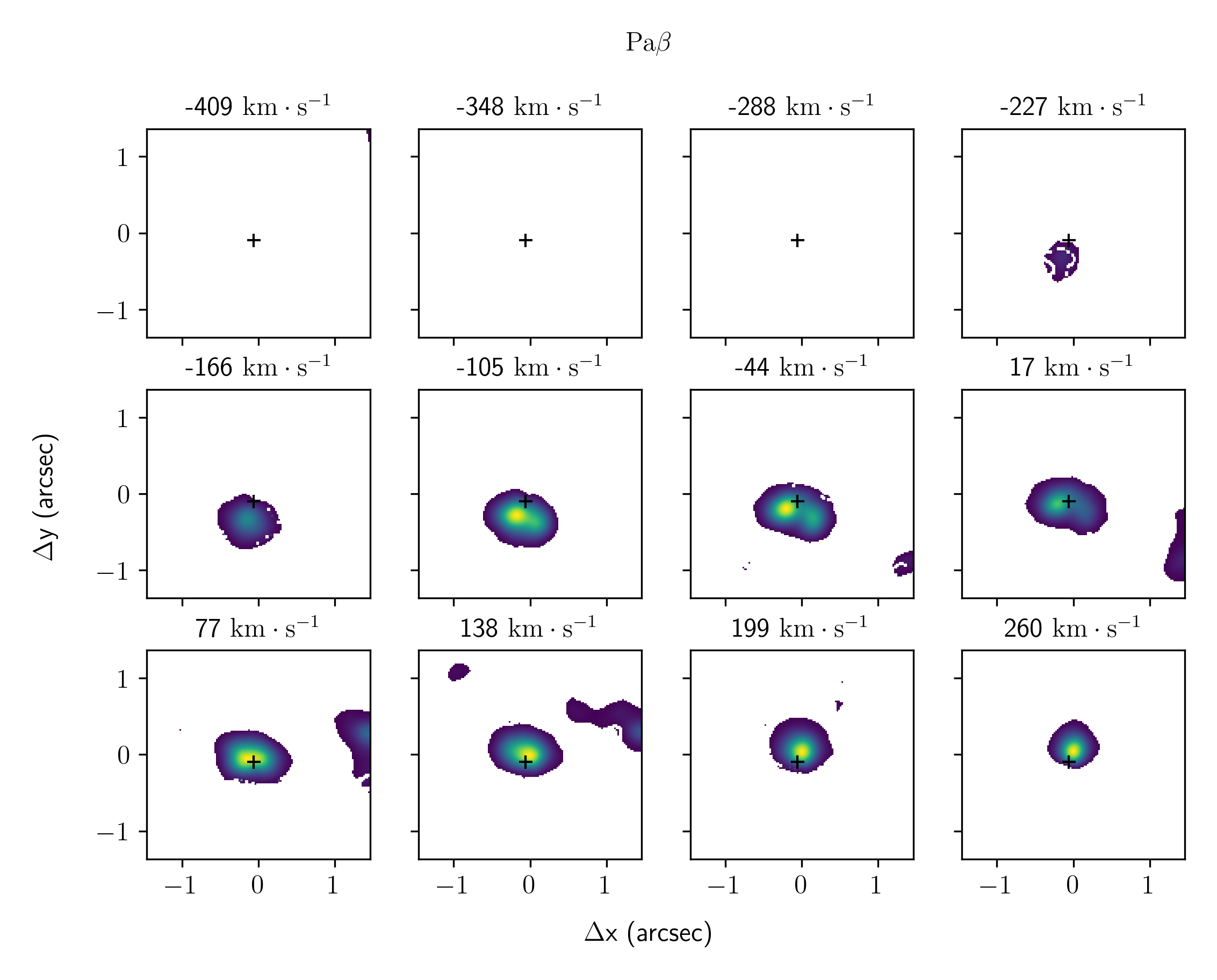}
    \caption{$\mathrm{Pa\beta}$ channel maps. The velocity is shown in top of each panel and the black cross marks the nucleus of the galaxy.}
    \label{fig:CM_PaB}
\end{figure*}

\begin{figure*}
    \centering
    \includegraphics[width=\textwidth]{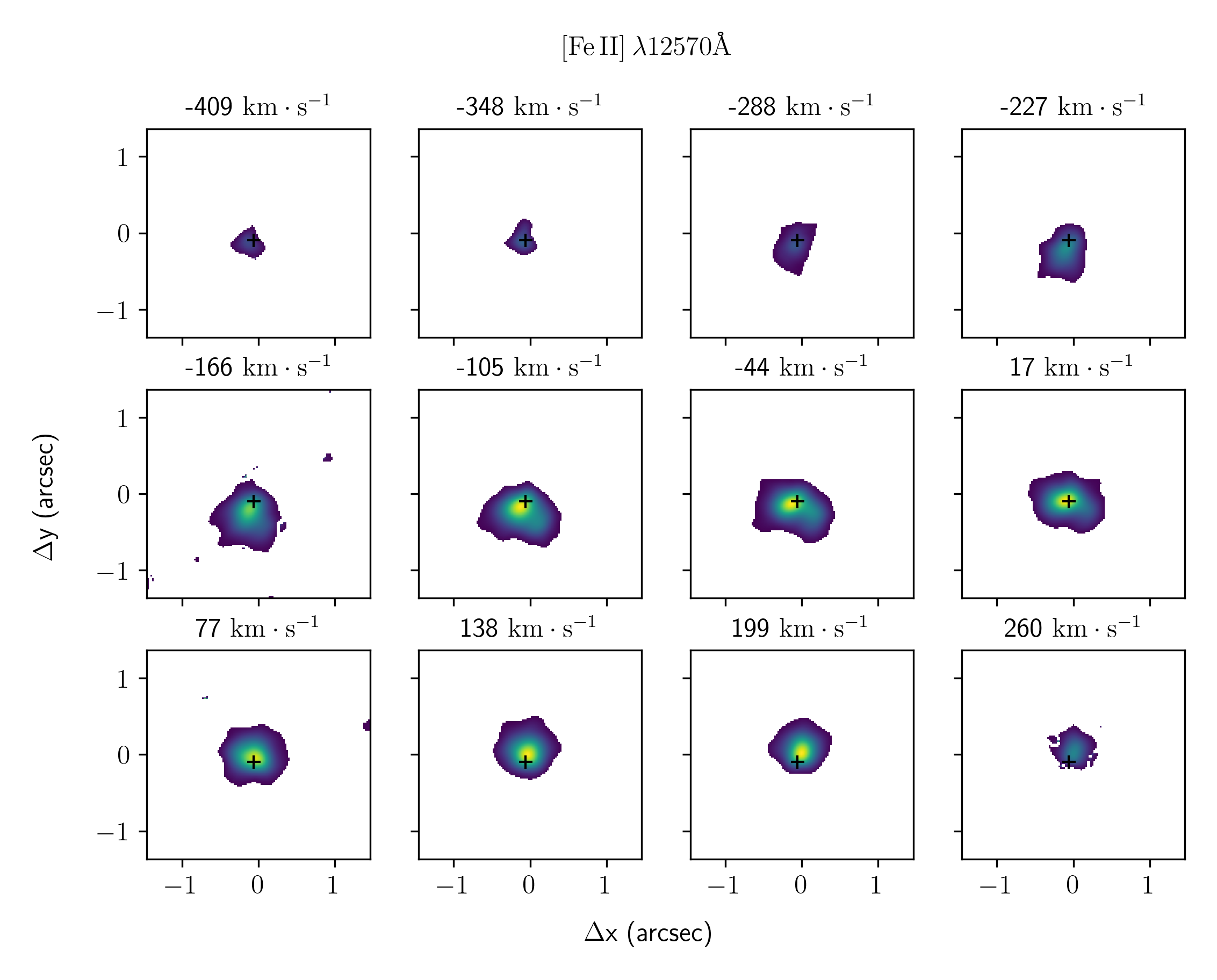}
    \caption{Same as in Fig.~\ref{fig:CM_PaB} for the $\mathrm{[\ion{Fe}{ii}]\,\lambda 12570}$\AA\ emission line.}
    \label{fig:CM_FeII}
\end{figure*}

\begin{figure*}
    \centering
    \includegraphics[width=\textwidth]{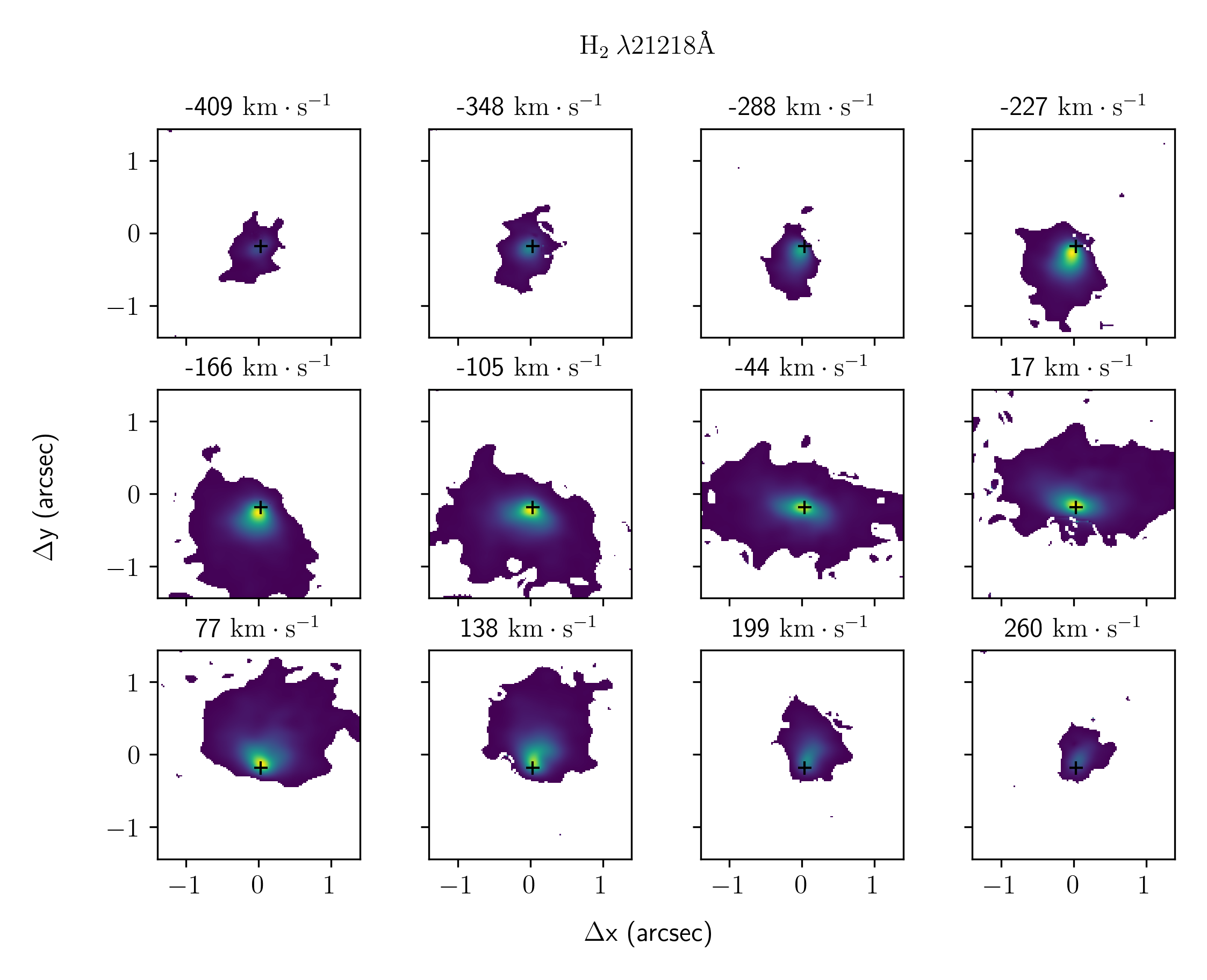}
    \caption{Same as in Fig.~\ref{fig:CM_PaB} for the $\mathrm{H_2\,\lambda 21218}$\AA\ emission line.}
    \label{fig:CM_H2}
\end{figure*}

We can further investigate the gas kinematics using channel maps, which is the mapping of the emission lines in constant velocity, or equivalently, wavelength bins. We built channel maps of the $\mathrm{Pa\beta}$, $\mathrm{[\ion{Fe}{ii}]}$ and $\mathrm{H_2\,\lambda 21218}$\AA\ emission lines using the \textit{channel\textunderscore maps module} of the {\sc ifscube} package. This module evaluates the adjacent continuum level and masks spaxels with flux values that after the continuum subtraction fall below a lower threshold defined by the user. Our channel maps are shown in Figs.~\ref{fig:CM_PaB}--\ref{fig:CM_H2}. Maps were built between $-440$ and 290~$\mathrm{km\,s^{-1}}$, with integrated fluxes in velocity bins of 61~$\mathrm{km\,s^{-1}}$ centred at the velocity shown at the top of each panel. We masked spaxels with fluxes lower than 1.0 and 0.1$\mathrm{\times 10^{-19} erg\,s^{-1}\,cm^{-2}\,\AA^{-1}}$ for the \textit{J} and \textit{K} bands, respectively.

All the channel maps confirm the transition from southern, blue-shifted to northern, red-shifted velocities. The rotation signature can be seen starting at the $-227$~$\mathrm{km\,s^{-1}}$ velocity channel. The channel maps of $\mathrm{Pa\beta}$ (Fig.~\ref{fig:CM_PaB}) confirm the asymmetric flux distribution. We can see that lower flux values can be found south of the nucleus, and that as one moves towards red-shifted velocities, higher fluxes do not coincide with the nucleus of the galaxy, but instead, the flux distribution seems to encircle it and reaches maximum values to the north. The maps between the 17 and 138~$\mathrm{km\,s^{-1}}$ velocity channels also show the extended emission seen at the right edge of the FoV in the $\mathrm{Pa\beta}$ flux map shown in Fig~\ref{fig:Maps} (top-left).

The $\mathrm{[\ion{Fe}{ii}]}$ and $\mathrm{H_2\,\lambda 21218}$\AA\ channel maps (Figs.~\ref{fig:CM_FeII} and \ref{fig:CM_H2}) show that for these emission lines, fluxes peak very close to the nucleus of the galaxy. More importantly, these maps confirm the presence of the broad, blue-shifted components starting at the $-409$~$\mathrm{km\,s^{-1}}$ velocity channel. For the $\mathrm{[\ion{Fe}{ii}]}$, this blue wing is very compact and located in the nucleus of the galaxy, while for the $\mathrm{H_2\,\lambda 21218}$\AA\, it is slightly elongated to the south.

\subsection{Reddening} \label{sec:Red}

To estimate the visual extinction $A_\mathrm{v}$ throughout the inner $\mathrm{1.2\,kpc \times 1.2\,kpc}$ of NGC~34, we followed \citet{2014MNRAS.443.1754D} who adopted the ratio of total to selective extinction as $\mathrm{R}_V = 4.05 \pm 0.80$ from \citet{2000ApJ...533..682C}, which has been shown to be the most suitable one when dealing with extinction in starbursts \citep{2000ApJ...533..682C,2003ApJ...599L..21F}. In the case B recombination for hydrogen (low-density limit, $T = 10^4 \mathrm{K}$, \citealt{2006agna.book.....O}), the intrinsic value of the $\mathrm{Pa\beta / Br\gamma}$ emission line ratio is 5.88, and the visual extinction $A_\mathrm{v}$ is:

\begin{equation}
    A_\mathrm{v}= -15.24 \log{\left( \frac{1}{5.88} \frac{\mathrm{Pa\beta}}{\mathrm{Br\gamma}} \right)}.
	\label{eq:Reddening}
\end{equation}

\begin{figure}
\centering
	\includegraphics[width=0.7\linewidth]{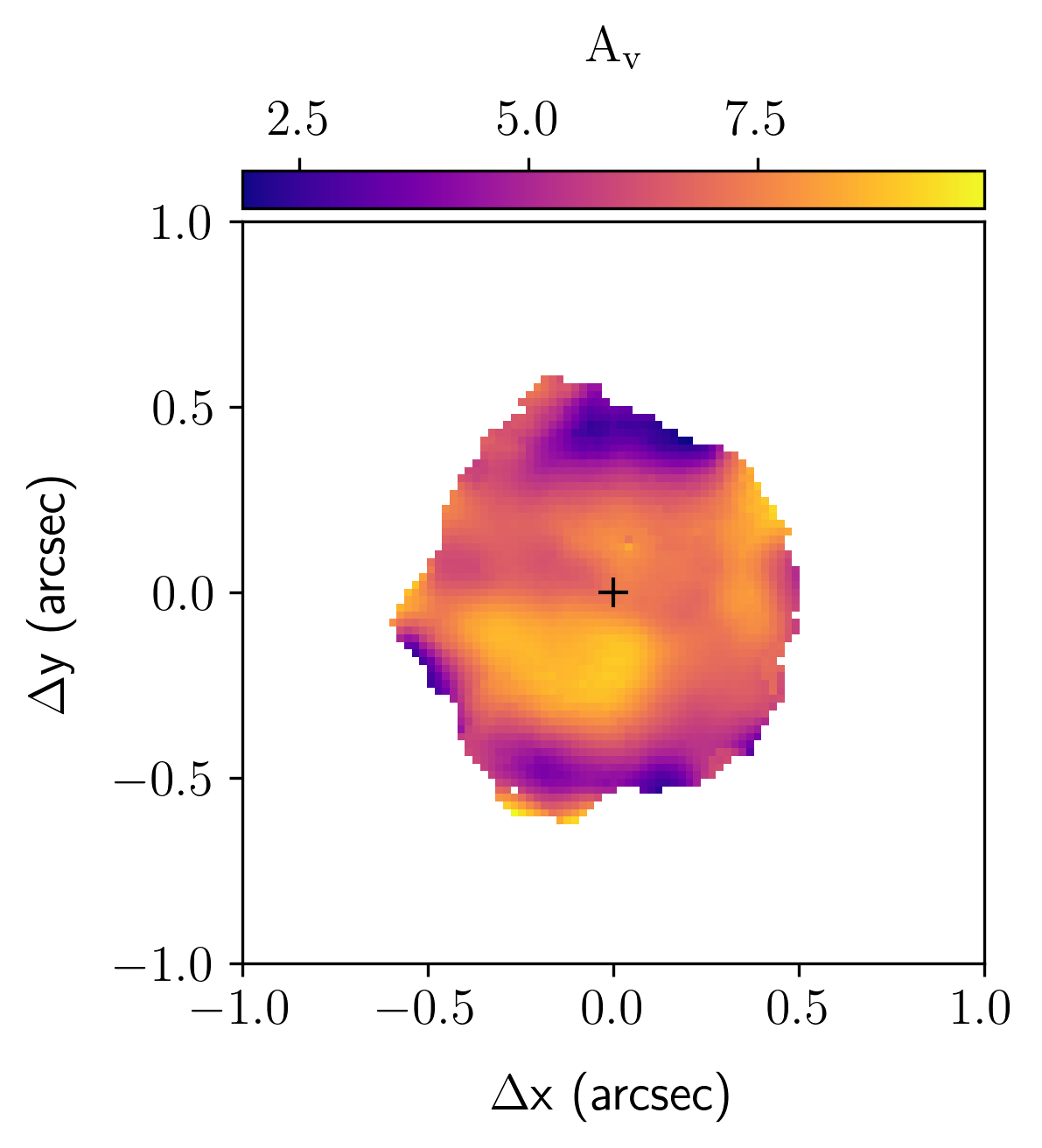}
	\caption{Map of visual extinction $A_\mathrm{v}$. }
   \label{fig:Reddening}
\end{figure}

The $A_\mathrm{v}$ distribution throughout NGC~34 is shown in Fig.~\ref{fig:Reddening}. The derived values range from $\sim$~1.9--9.0~mag. Higher extinction can be found towards the south of the galaxy. The mean $A_\mathrm{v}$ in the northern region is 6.78$\pm$0.86 and the median is 7.01, while in the south, the mean $A_\mathrm{v}$ is 7.74$\pm$0.95 and the median is 7.94. These results are in full agreement with long-slit studies \citep[see table~4 of][]{2014MNRAS.443.1754D}.

\citet{2006ApJ...650..835A} used \textit{HST} NICMOS (Near Infrared Camera and Multi-Object Spectrometer) observations to study the NIR and star-forming properties of a sample of local LIRGs. They derived extinctions in the range 2--6~mag to the stars (photometric extinctions), while extinctions to the gas, derived from the $\mathrm{H\alpha / Pa\alpha}$ and $\mathrm{Pa\alpha / Br\gamma}$ ratios, tend to be higher, ranging from 0.5 up to 15~mag. The values found for NGC~34 indicate that its inner regions are embedded in a heavily obscured environment, which could potentially affect optical studies of high spatial resolution aiming at investigating the nature of its nuclear activity.

\subsection{Emission-line ratios} \label{sec:LineRatios}

\begin{figure}
\begin{center}

		\includegraphics[width=0.7\linewidth]{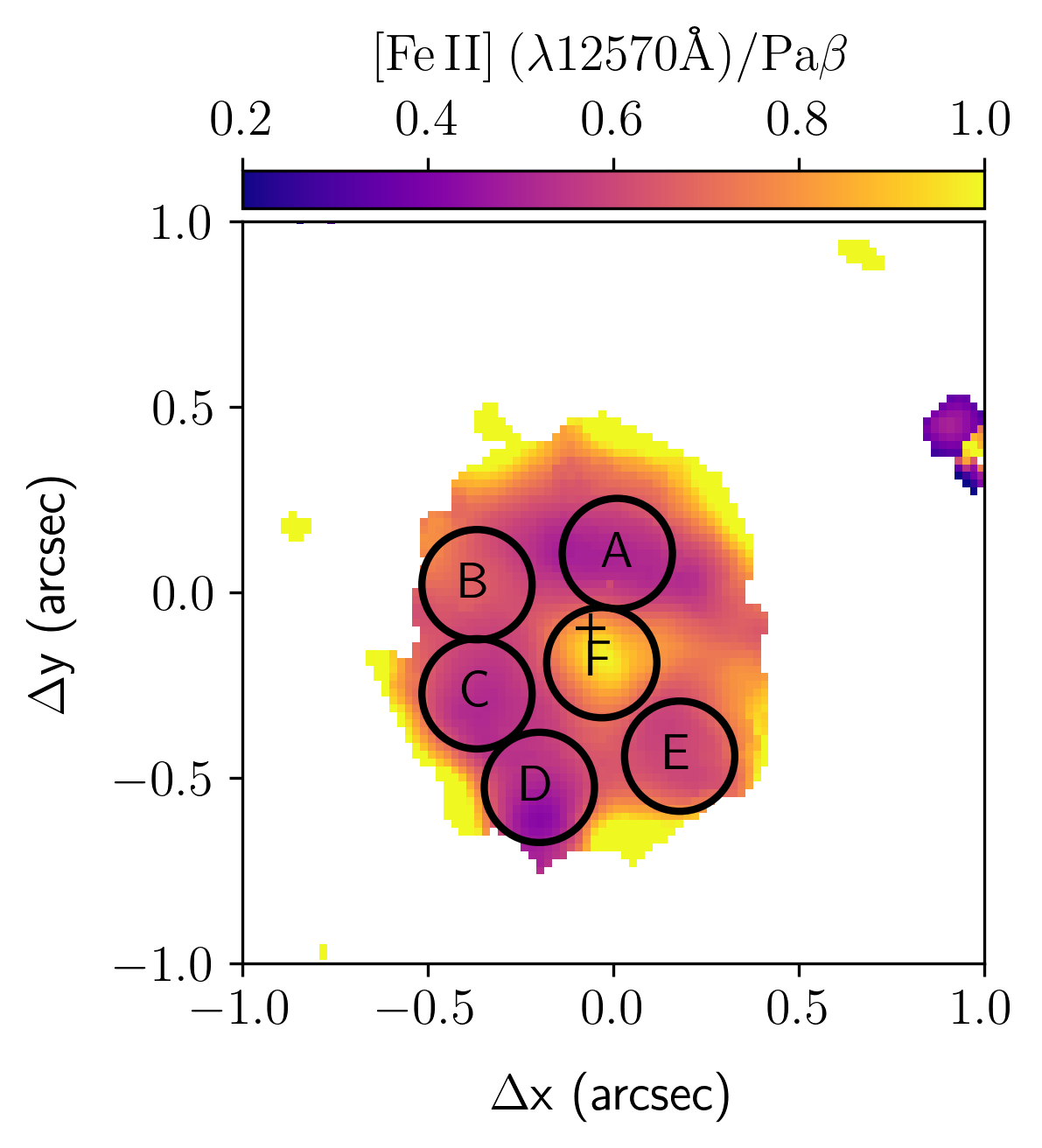}
		
		\includegraphics[width=0.7\linewidth]{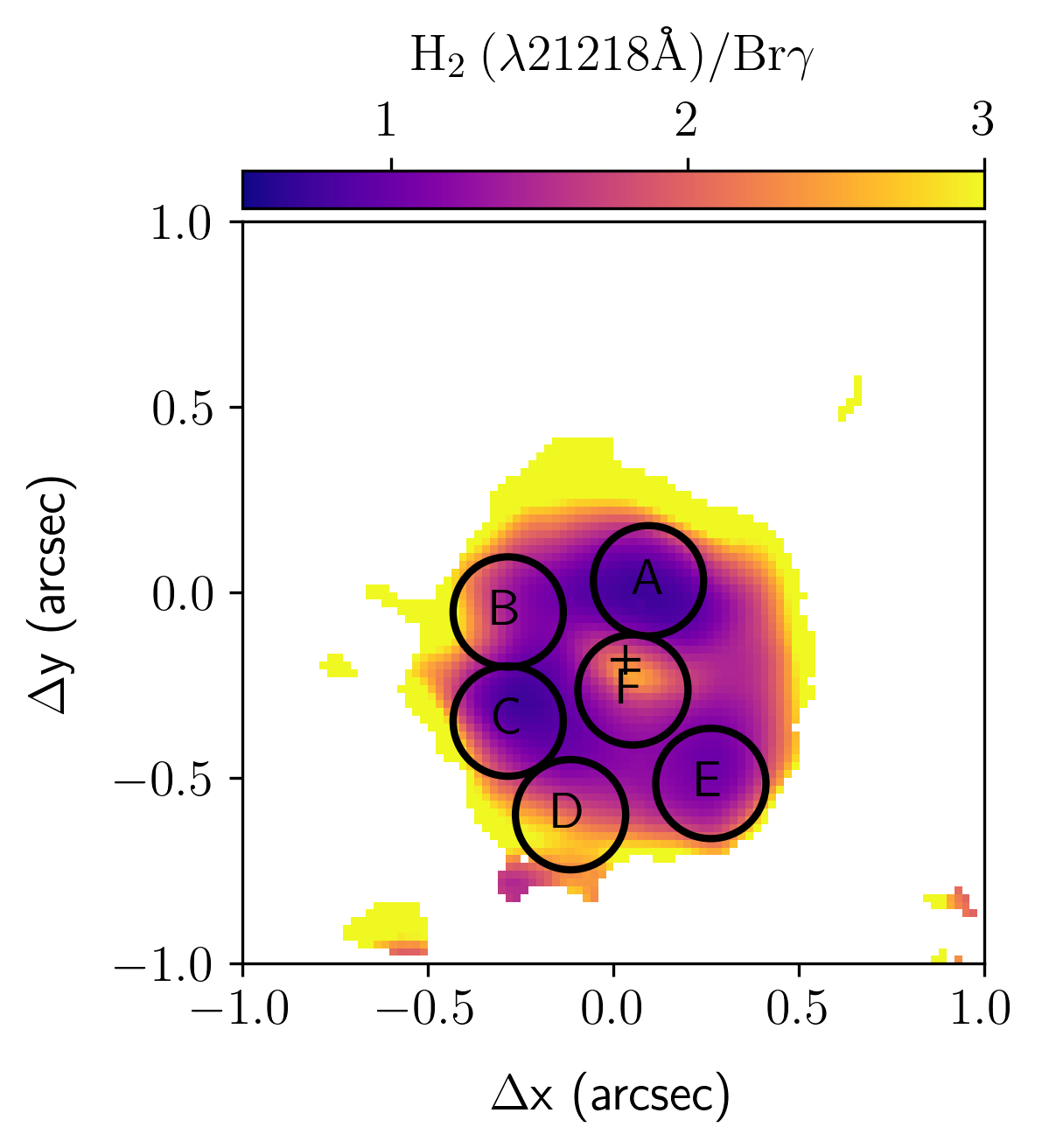}
	
\caption{Maps of the $\mathrm{[\ion{Fe}{ii}] / Pa\beta}$ (top) and $\mathrm{H_{2} / Br\gamma}$ (bottom) emission lines ratios. The black, open circles mark regions from A to F where integrated measurements were taken. The black cross marks the galaxy nucleus.
 }
 \label{fig:LineRatios}
\end{center}
\end{figure}

We show maps of the $\mathrm{[\ion{Fe}{ii}] / Pa\beta}$ (top panel) and $\mathrm{H_{2} / Br\gamma}$ (bottom panel) emission-line ratios in Fig.~\ref{fig:LineRatios} along with the locations of regions identified from A--F that correspond to apertures of 0.2$\arcsec \times$0.2$\arcsec$. We integrated the gas spectra in each aperture and fitted the emission lines using Gaussian functions and the package {\sc ifscube}, thus obtaining their amplitudes, $\sigma$ and respective uncertainties, which were then used to estimate the uncertainties of the integrated fluxes through error propagation. Fluxes of the emission lines in each region are presented in Table~\ref{tab:fluxes_regions} and the integrated spectra are shown in Appendix~\ref{appendix:RegionsSpectra}. 

The $\mathrm{[\ion{Fe}{ii}] / Pa\beta}$ and $\mathrm{H_{2} / Br\gamma}$ line-ratios are reddening insensitive and are used to build BPT-like diagrams in the NIR \citep{1998ApJS..114...59L,2004AA...425..457R,2005MNRAS.364.1041R,2013MNRAS.430.2002R,2015AA...578A..48C,2021MNRAS.tmp..778R}. The hydrogen recombination lines are tracers of UV-ionizing radiation from young OB-stars or from the AGN. The $\mathrm{[\ion{Fe}{ii}]}$ traces shocked, partially ionized regions and photo-dissociation regions, while the $\mathrm{H_2\,\lambda 21218}$\AA\ traces the hot molecular gas (T$\geq 10^3$K). Star-forming galaxies (SFGs) have $\mathrm{[\ion{Fe}{ii}] / Pa\beta < 0.6}$ and $\mathrm{H_{2} / Br\gamma < 0.4}$, AGN-dominated systems have $\mathrm{0.6 < [\ion{Fe}{ii}] / Pa\beta < 2.0}$ and $\mathrm{0.4 < H_{2} / Br\gamma < 6.0}$, and LINERs have $\mathrm{[\ion{Fe}{ii}] / Pa\beta > 2.0}$ and $\mathrm{H_{2} / Br\gamma > 6.0}$ (e.g.: \citealt{2013MNRAS.430.2002R}).

Our maps of $\mathrm{[\ion{Fe}{ii}] / Pa\beta}$ and $\mathrm{H_{2} / Br\gamma}$ for NGC~34 clearly show a ring-like structure with lower values of the ratios around the nucleus of the galaxy. Many locations in the $\mathrm{[\ion{Fe}{ii}] / Pa\beta}$ map are consistent with pure star-formation. Along the ring-shaped structure, $\mathrm{[\ion{Fe}{ii}] / Pa\beta}$ values range from $\sim$0.4--0.75 and reach values of $\sim$1.0 close to the nucleus of the galaxy in region F. In the case of the $\mathrm{H_{2} / Br\gamma}$, the obtained values for the emission-line ratios are all consistent with the presence of an AGN, including locations along the ring-shaped structure where values range from $\sim$0.7--1.4 and reach up to $\sim$2.5 in the nucleus. 

Overall, the fact that the structure in the shape of a ring around the nucleus of the galaxy can be seen in both maps of emission-line ratios, in addition to many locations displaying $\mathrm{[\ion{Fe}{ii}] / Pa\beta}$ values consistent with pure star-formation indicate that the NGC~34 nuclear starburst is distributed in a circumnuclear star-formation ring. However, larger values found for both line ratios, especially in the case of the $\mathrm{H_{2} / Br\gamma}$, indicate that additional mechanisms are required to explain the $\mathrm{[\ion{Fe}{ii}]}$ emission and the excitation of the H$_2$ molecule, as will be discussed in section~\ref{sec:FeIIH2Nature}.

\begin{table*}
\centering
	\caption{Observed emission lines fluxes for the regions indicated in Fig.~\ref{fig:LineRatios}. Fluxes are in units of $\mathrm{\times 10^{-15} erg\,cm^{-2}\,s^{-1} }$. Each region corresponds to an aperture of 0.2$\arcsec \times$0.2$\arcsec$.}
	\label{tab:fluxes_regions}
	\begin{tabular}{cccccccc} 
	    \hline
	    Region & $\mathrm{[\ion{P}{ii}]}$ & $\mathrm{[\ion{Fe}{ii}]}$ & $\mathrm{Pa\beta}$ & $\mathrm{H_2}$ & $\mathrm{H_2}$ & $\mathrm{H_2}$ & $\mathrm{Br\gamma}$  \\
		  & ( 11886\AA\ ) & ( 12570\AA\ ) & ( 12818\AA\ ) & ( 21218\AA\ ) & ( 22230\AA\ ) & ( 22470\AA\ ) & ( 21654\AA\ ) \\
		\hline
		A & 0.56$\pm$0.00 & 1.24$\pm$0.07 & 2.15$\pm$0.09 & 1.03$\pm$0.11 & 0.26$\pm$0.02 & 0.16$\pm$0.01 & 1.09$\pm$0.06 \\
       B & 0.27$\pm$0.02 & 0.56$\pm$0.03 & 0.86$\pm$0.03 & 0.55$\pm$0.04 & 0.18$\pm$0.01 & 0.08$\pm$0.00 & 0.40$\pm$0.02 \\
       C & 0.24$\pm$0.01 & 0.61$\pm$0.03 & 1.07$\pm$0.04 & 0.63$\pm$0.03 & 0.18$\pm$0.01 & 0.10$\pm$0.00 & 0.63$\pm$0.02 \\
       D & 0.11$\pm$0.01 & 0.27$\pm$0.01 & 0.50$\pm$0.01 & 0.39$\pm$0.03 & 0.13$\pm$0.01 & 0.05$\pm$0.00 & 0.23$\pm$0.01 \\
        E & 0.14$\pm$0.01 & 0.43$\pm$0.02 & 0.70$\pm$0.02 & 0.37$\pm$0.02 & 0.11$\pm$0.01 & 0.06$\pm$0.00 & 0.31$\pm$0.01 \\
        F & 0.82$\pm$0.02 & 1.81$\pm$0.13 & 1.98$\pm$0.08 & 2.25$\pm$0.43 & 0.55$\pm$0.09 & 0.22$\pm$0.03 & 1.08$\pm$0.05 \\
		\hline
	\end{tabular}
\end{table*}

\subsection{Principal Component Analysis Tomography} \label{sec:PCA}

We also analysed the data cubes of NGC~34 using Principal Component Analysis (PCA) Tomography \citep{2009MNRAS.395...64S}. This technique uses PCA applied to data cubes in order to extract useful information from a given object (see e.g. \citealt{2011ApJ...734L..10R,2014MNRAS.440.2419R,2015AA...576A..58R,2019MNRAS.489.5653D}). It searches for correlations between \textit{m} spectral pixels across \textit{n} spaxels. A set of \textit{m} eigenvectors is built as a combination of the \textit{m} spectral pixels and they are ordered by their contribution of the variance (obtained with the respective eigenvalues) of the data cube. The tomograms correspond to the projection of the eigenvectors on the data cubes. In practice, the eigenvectors, also called eigenspectra, show the correlations between the wavelengths, while the tomograms reveal where such correlations occur in the spatial dimension.  

\begin{figure}
    \includegraphics[scale=0.31]{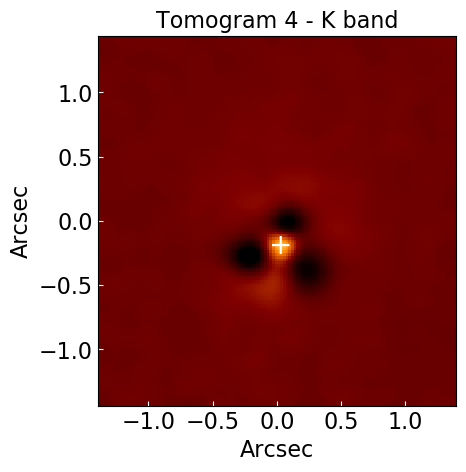}
    \includegraphics[scale=0.31]{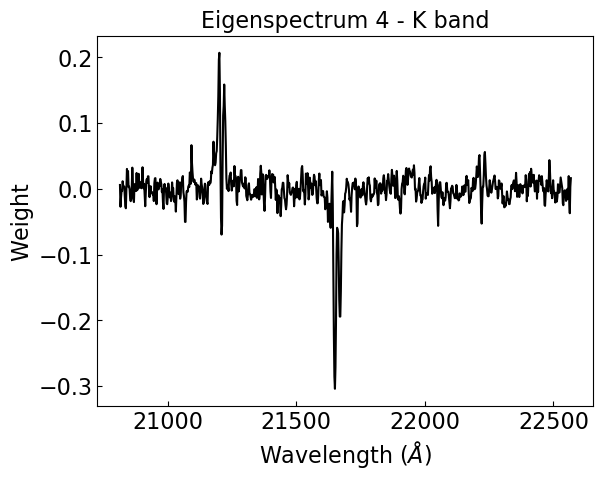}
    \includegraphics[scale=0.31]{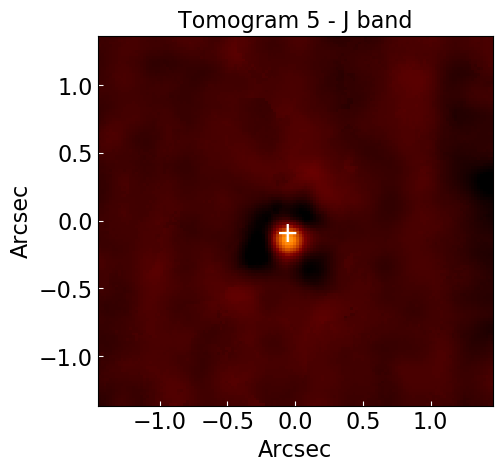}
    \includegraphics[scale=0.31]{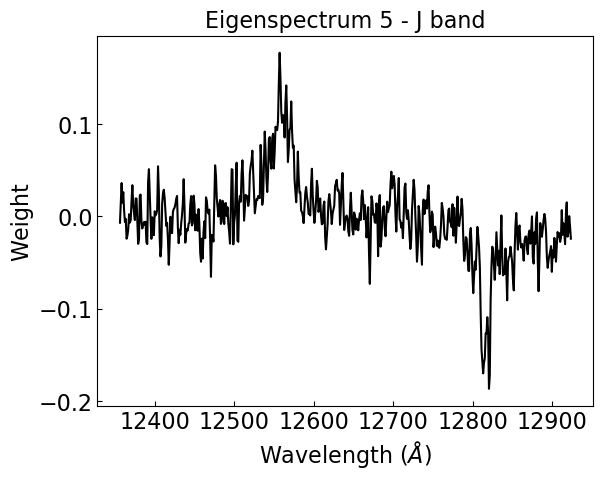}
    \caption{PCA Tomography results. In both tomograms (left panels), the white cross marks the position of the centre of the bulge of NGC 34. The tomograms identify two distinct structures in the central region of NGC~34: a nuclear component which is related to the $\mathrm{[\ion{Fe}{ii}]}$ and $\mathrm{H_2\,\lambda 21218}$\AA\ emission lines; and a circumnuclear structure associated with the $\mathrm{Pa\beta}$ and $\mathrm{Br\gamma}$ emission. The nuclear and circumnuclear structures appear anti-correlated in the eigenspectra (right panels).}
    \label{fig:PCA_results}
\end{figure}

We applied PCA Tomography in both \textit{J} and \textit{K} bands data cubes after the subtraction of the stellar component of NGC~34. In the case of the \textit{K} band data cube, we used the spectral range 20814--22568~\AA, which contains all H$_2$ and the Br$\gamma$ lines. For the \textit{J} band data cube, a spectral range of 12356--12923~\AA\ was used, containing both $\mathrm{[\ion{Fe}{ii}]}$ and Pa$\beta$ lines. The most relevant results revealed by the PCA Tomography correspond to the fourth eigenspectrum of the \textit{K} band data cube and the fifth eigenspectrum of the \textit{J} band data cube, shown in Fig.~\ref{fig:PCA_results}. Eigenspectrum 4 of the \textit{K} band data cube explains 1.76~per cent of the variance and is characterised by an anti-correlation between the Br$\gamma$ and the H$_2\lambda$21218\AA\ lines. Its tomogram reveals a nuclear object, which is related to the H$_2\lambda$21218\AA\ line, and a circumnuclear structure, which corresponds to the Br$\gamma$ line. A similar scheme is seen in the fifth tomogram of the \textit{J} band data cube. In this case, the eigenspectrum 5 explains 1.21~per cent of the variance and is characterised by a broad $\mathrm{[\ion{Fe}{ii}]}$ component anti-correlated with the Pa$\beta$ line. The nuclear object seen in tomogram 5 is related to the broad $\mathrm{[\ion{Fe}{ii}]}$ component, while the circumnuclear structure is associated with the Pa$\beta$ emission. 

\section{Discussion} \label{sec:Disc}

\subsection{Nuclear disc of ionized and molecular gas} \label{sec:CND}

The maps of radial velocities of the fitted emission lines presented in Sec.~\ref{sec:KinRes} clearly show signature of rotation in the ionized and molecular gas  phases. In order to compare the kinematics of the gas in NGC~34 with that of the stars, we carried out a new spectral synthesis in the \textit{K} band data cube using the code {\sc ppxf} and the Gemini library of late spectral templates \citep{2009ApJS..185..186W}, which was especially designed to support stellar kinematics studies in external galaxies. We also followed the recommendations of \citet{Cappellari2017} to include only additive polynomials when using {\sc ppxf} to fit the stellar kinematics. We then used the code {\sc pafit}\footnote{Available at https://pypi.org/project/pafit/}, which implements the method presented in Appendix C of \citet{2006MNRAS.366..787K}, to determine: (1) a correction ($v_{offset}$) to be applied to the adopted systemic velocity of the galaxy determined from the stellar velocity field ($v_{sys,cor} = v_{sys} + v_{offset}$) and (2) the global kinematic PA of the stellar and gas discs.
 
\begin{figure*}
    \centering
	\begin{minipage}{\textwidth}
	\centering
	\begin{minipage}{.33\textwidth}
		\includegraphics[width=\textwidth]{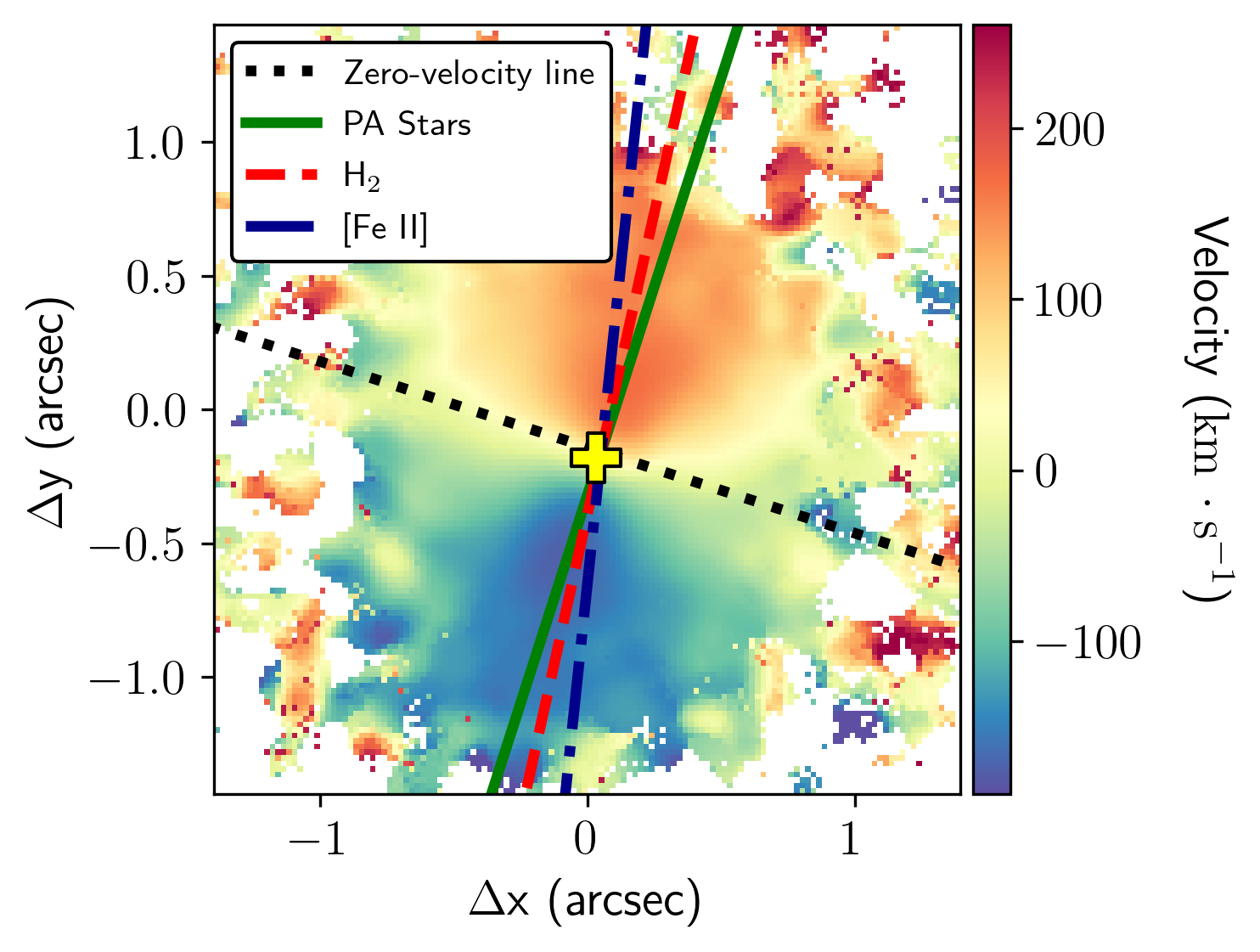}
	\end{minipage}
	\begin{minipage}{.33\textwidth}
	\includegraphics[width=\textwidth]{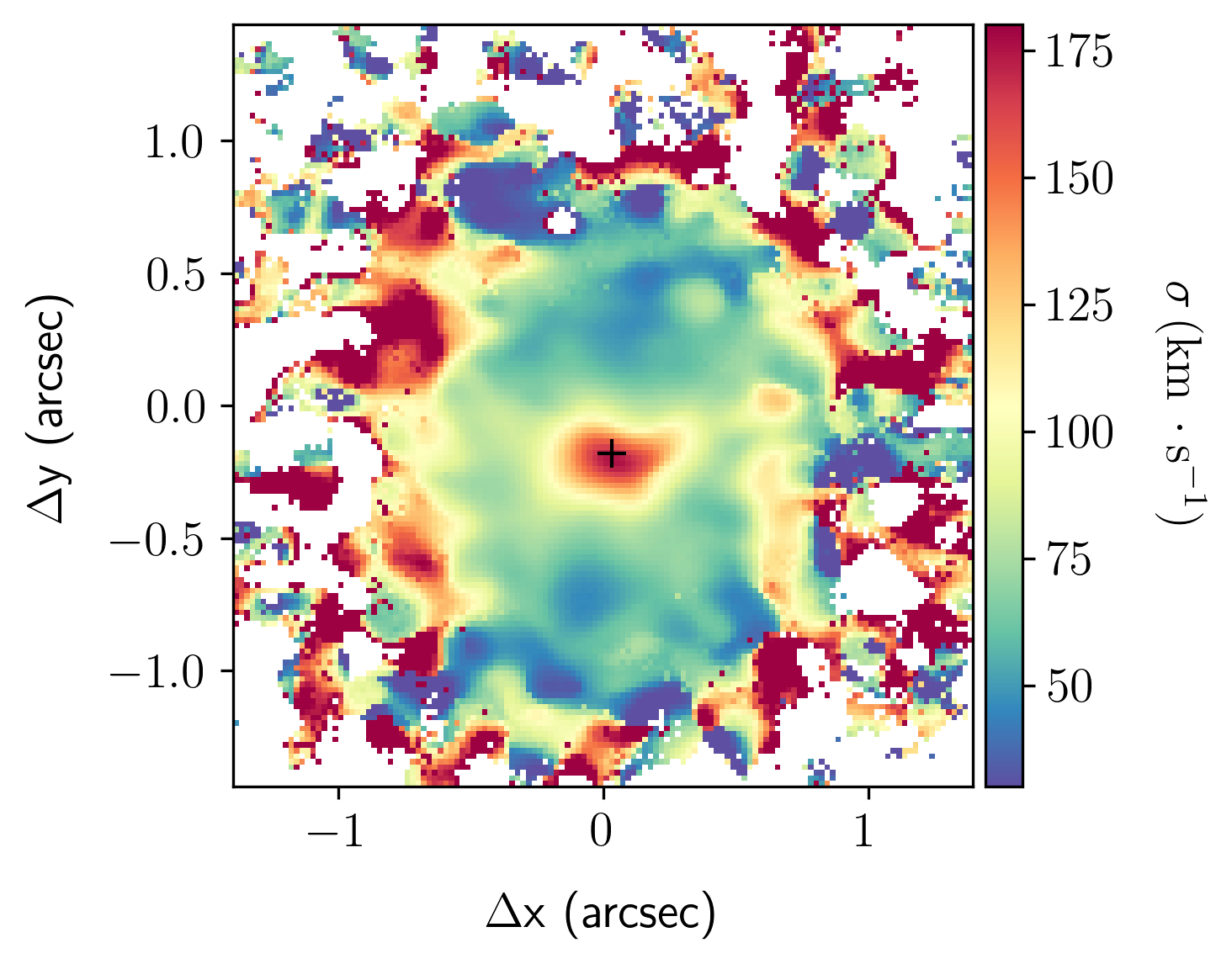}
	\end{minipage}
	\begin{minipage}{.33\textwidth}
		\includegraphics[width=\textwidth]{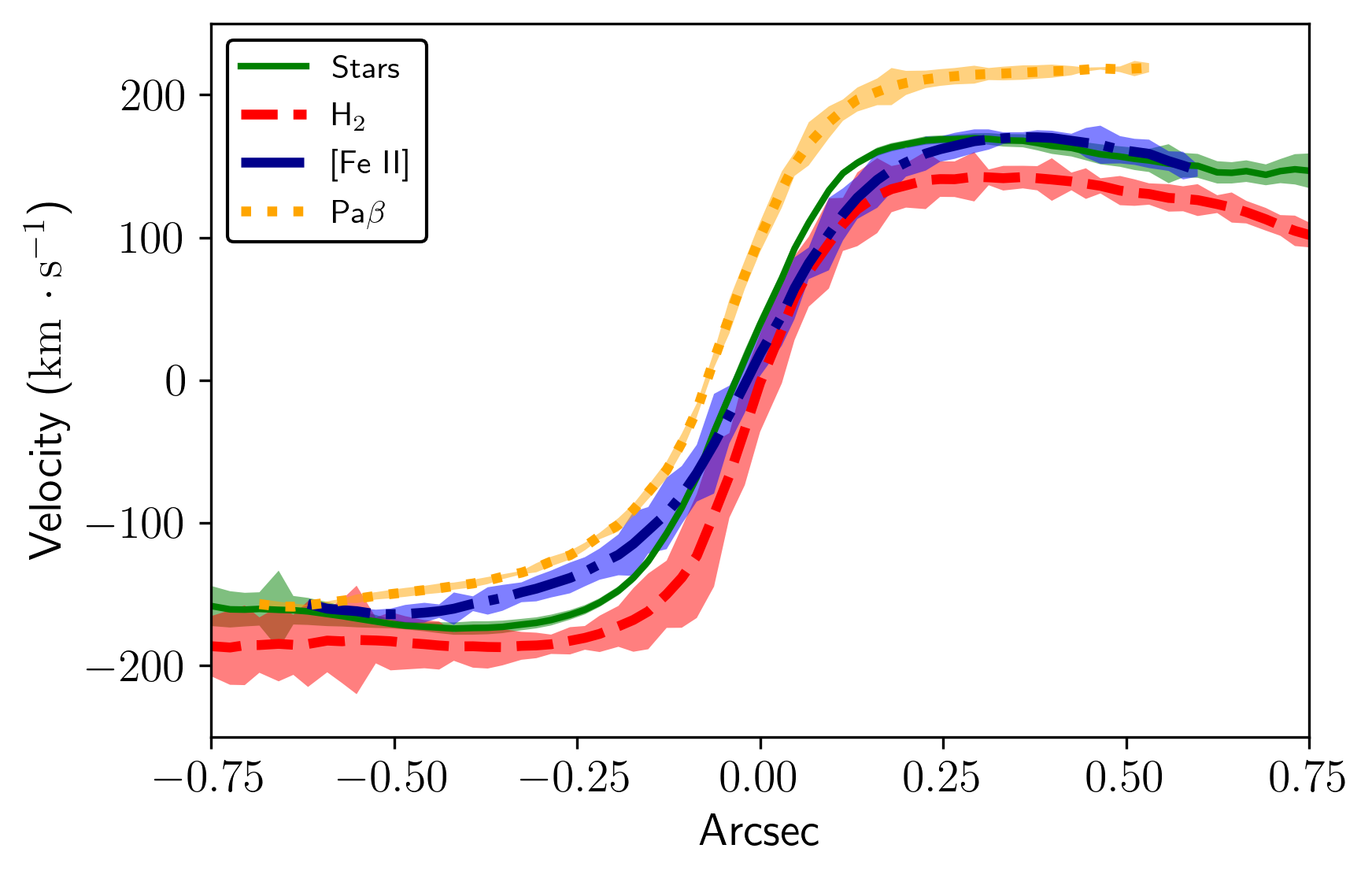}
	\end{minipage}
	\end{minipage}
    \caption{Left panel: Map of the stellar velocity field. The solid green, blue dashdot and dashed red lines show the global kinematic PA of the stellar, ionized and molecular gas discs, respectively. The black dotted line indicates the zero-velocity line and the yellow cross marks the galaxy nucleus. Middle panel: field of stellar velocity dispersion. In both fields, white pixels correspond to regions where uncertainties in the velocities are $> 50\,\mathrm{km\,s^{-1}}$. Right panel: velocities profiles of the gas and stars along the kinematic PA of the stellar disc. The shaded areas are 3$\sigma$ uncertainties.}
    \label{fig:StellarDisc}
\end{figure*}

The NGC~34 stellar velocity field is shown in the left panel of Fig.~\ref{fig:StellarDisc} and the correction to the systemic velocity is -40~$\mathrm{km\,s^{-1}}$, resulting in $v_{sys,cor} = 5841\,\mathrm{km\,s^{-1}}$. In general, one can see that the range of velocities as well as the PA of the rotation signature of the stars are very similar to the ones found for the ionized and molecular gas, indicating that they are under the influence of the same gravitational potential. This can be straightforwardly associated with the presence of a nuclear disc (ND) of stars and gas. For the stars, the kinematic PA of the disc is $\mathrm{-18^{\circ}.0 \pm 1^{\circ}.2}$. The global kinematic PAs of the ionized and molecular gas discs are $\mathrm{-6^{\circ}.0 \pm 0^{\circ}.5}$ and $\mathrm{-12^{\circ}.5 \pm 1^{\circ}.8}$, respectively, resulting in a mean PA of $\mathrm{-9^{\circ}.2 \pm 0^{\circ}.9}$. The PA of the stellar, ionized and molecular gas discs are indicated in Fig.~\ref{fig:StellarDisc} by the solid green, blue dashdot and dashed red lines, respectively. The field of stellar velocity dispersion (middle panel) show higher values in the galaxy nucleus. In the right panel of Fig.~\ref{fig:StellarDisc}, we show stellar and gas velocity profiles taken along the kinematic PA of the stellar disc. Velocities rise steeply and reach maximum values in a radius smaller than 0.5$\arcsec$ ($\approx$200~pc) from the nucleus of the galaxy, indicating that the disc is compact.

Our results are in agreement with previous works presented by \citet{2014AJ....147...74F} and \citet{2014ApJ...787...48X}, thus confirming the presence of a ND of ionized and molecular gas in NGC~34 with a northern receding and a southern approaching side. However, \citet{2014ApJ...787...48X} point out that the $\sim$2~kpc~CO(1-0) disc found by \citet{2014AJ....147...74F} is associated with diffuse gas emission, while the much more compact $\sim$200~pc~CO(6-5) disc detected by them (see Fig~1 in \citealt{2014ApJ...787...48X}) has the same distribution of the nuclear starburst traced by radio continuum emission. The kinematic PA of the CO(6-5) disc is $\mathrm{-15^{\circ}}$ and the rotation velocity also rises up to a radius of 0.5$\arcsec$ and then flattens, indicating that our NIR IFU data and the ALMA observations of \citet{2014ApJ...787...48X} share the same kinematics. 

\subsection{Resolved NIR diagnostic diagram and the NGC~34 power source} \label{sec:DiagDiagram}

As previously mentioned, the nature of the NGC~34 nuclear emission line spectrum is still under debate, with different works classifying it as due to a starburst \citep{1996ApJS..102..309M,2006AA...457...61R} and/or AGN \citep{2010ApJ...709..884Y,2011MNRAS.414.3084B}. In this work, we characterise the nuclear activity in NGC~34 by means of high spatial resolution observations of NIR emission lines, namely the $\mathrm{Pa\beta}$, $\mathrm{[\ion{Fe}{ii}]}$, $\mathrm{Br\gamma}$ and $\mathrm{H_2\,\lambda 21218}$\AA\ emission features, which are more suited for probing the dust-embedded environments typically found in (U)LIRGs. While the $\mathrm{Pa\beta}$ and $\mathrm{Br\gamma}$ emissions are mainly due to the UV-ionizing flux produced by young OB-stars (ages $\sim$10~Myr) or by the AGN, emissions by the $\mathrm{[\ion{Fe}{ii}]}$ and H$_2$ species have different origin.

Excitation of the H$_2$ molecule can be due to: (1) UV pumping (\textit{fluorescence}), also regarded as \textit{non-thermal} emission, where H$_2$ molecules are electronically excited by the absorption of UV photons in the Lyman-Werner band (918--1108\AA\ ) in photo-dissociation regions (PDRs), followed by rapid transitions to ro-vibrational excited levels of the ground state (e.g. \citealt{1987ApJ...322..412B}); and the \textit{thermal} processes either through (2) direct heating of the molecular gas by shocks (e.g. \citealt{1989ApJ...342..306H}) or by (3) X-rays, for example, produced by an AGN (e.g. \citealt{1996ApJ...466..561M}). Early works have already shown that for a single object, the observed H$_2$ spectrum is often the result of multiple excitation mechanisms that are at play (e.g \citealt{1994ApJ...427..777M}). Similarly, $\mathrm{[\ion{Fe}{ii}]}$ emission can also be traced to shock dominated regions, associated either with radio jets, nuclear outflows and/or supernova remnants (SNRs), or to photo-ionization by a central X-ray source (AGN, \citealt{2000ApJ...528..186M}).

In this sense, NIR diagnostic diagrams based on the $\mathrm{[\ion{Fe}{ii}] / Pa\beta}$ and $\mathrm{H_{2} / Br\gamma}$ line-ratios provide a progression in the observed values from pure photo-ionized (such as in $\mathrm{\ion{H}{ii}}$ regions) to pure shock excited regions (for example, shocks in SNRs), with star-forming (SF) regions displaying low values for both ratios, while regions dominated by the AGN have intermediate values \citep{1998ApJS..114...59L,2004AA...425..457R,2005MNRAS.364.1041R,2013MNRAS.430.2002R,2015AA...578A..48C,2021MNRAS.tmp..778R}.

\begin{figure*}
\begin{center}

	\begin{minipage}{\textwidth}
	\centering
	\begin{minipage}{.48\textwidth}
		\includegraphics[width=\textwidth]{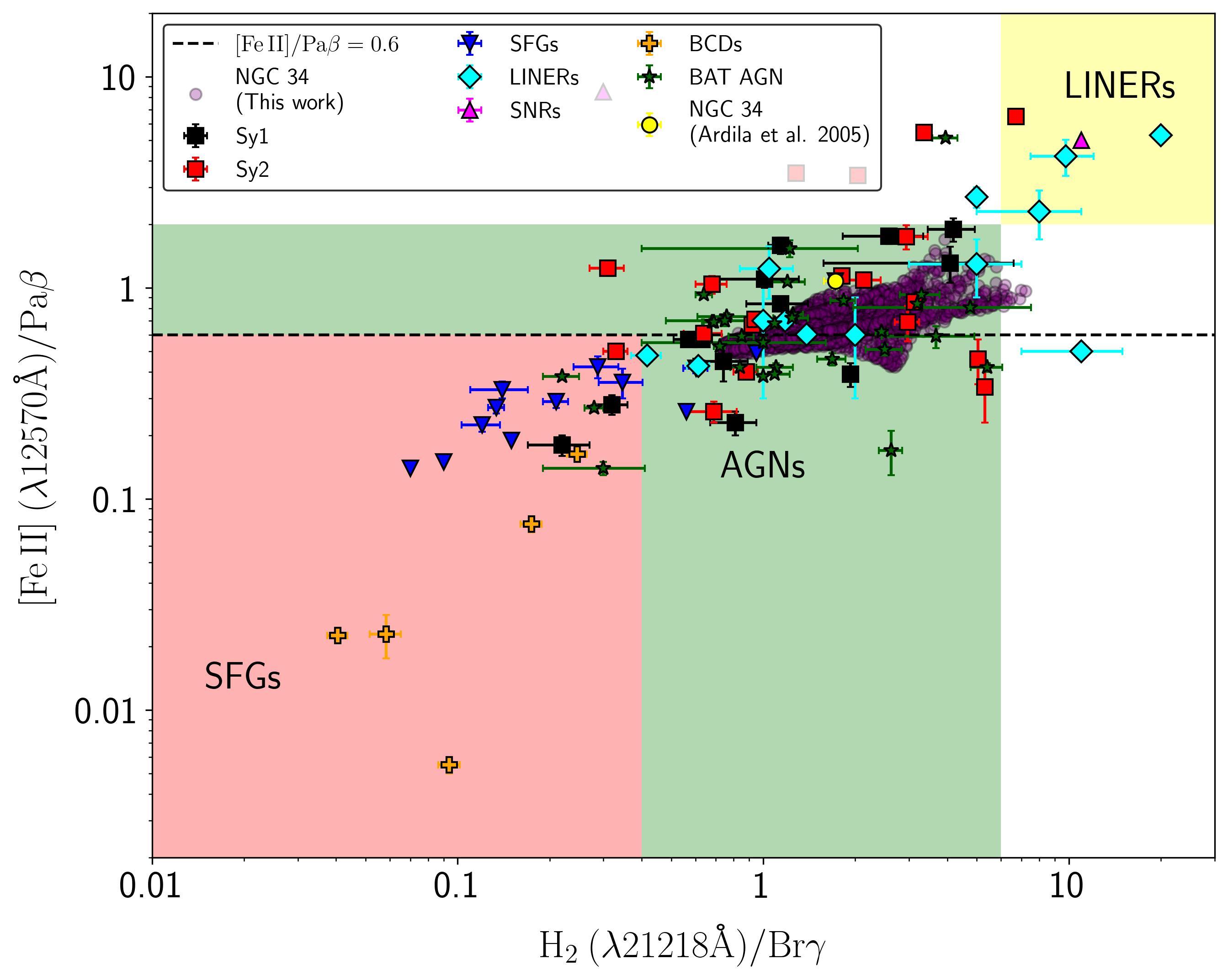}
	\end{minipage}
	\quad
	\begin{minipage}{.48\textwidth}
		\includegraphics[width=\textwidth]{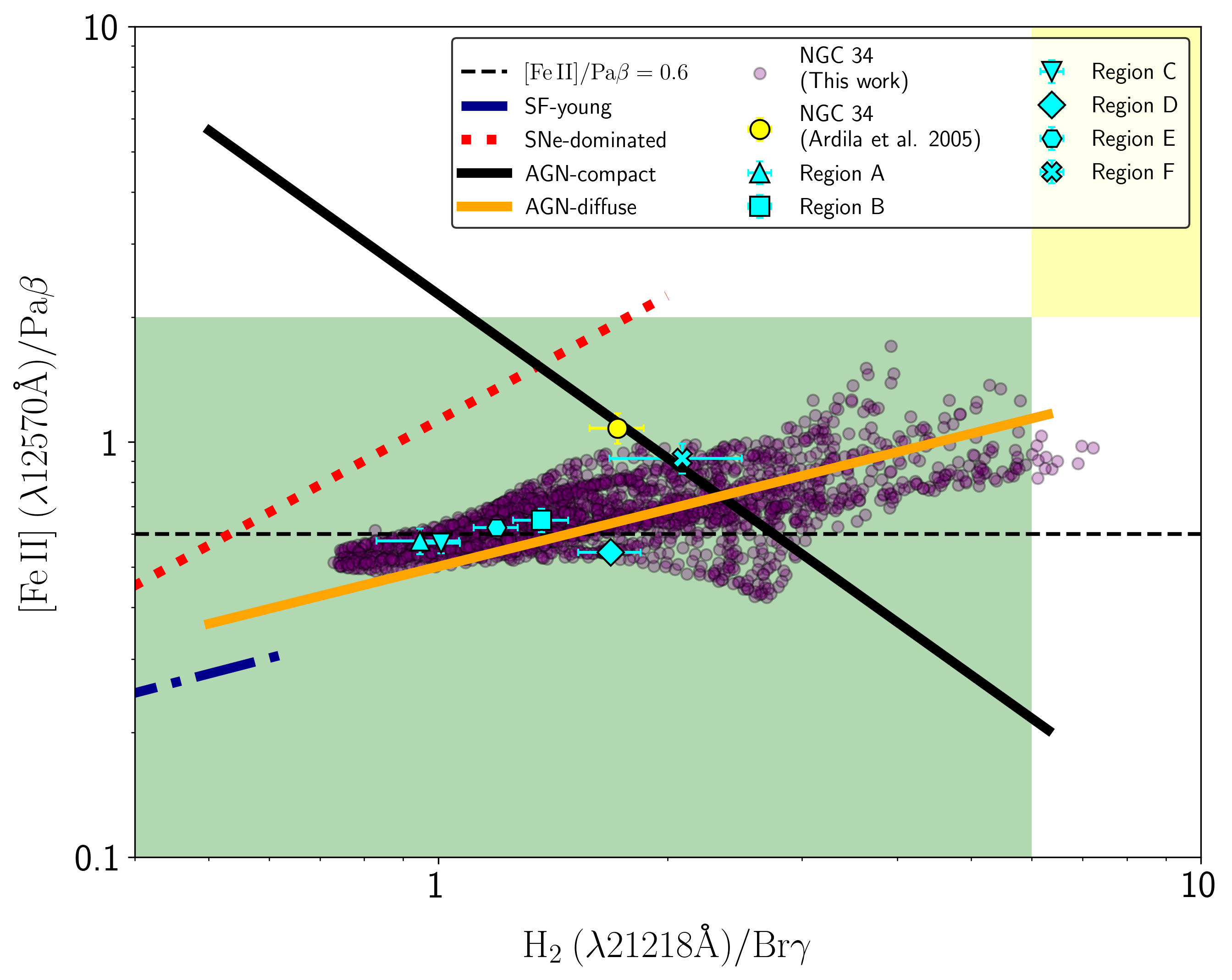}
	\end{minipage}
	\end{minipage}
	
\caption{NIR diagnostic diagram based on the $\mathrm{[\ion{Fe}{ii}] / Pa\beta}$ and $\mathrm{H_{2} / Br\gamma}$ emission lines ratios. In both panels, the horizontal, dashed line corresponds to $\mathrm{[\ion{Fe}{ii}] / Pa\beta = 0.6}$ from \citet{2013MNRAS.430.2002R}. The shaded areas divide the diagram into the regions dominated by Star-Forming Galaxies (SFGs), AGNs and Low Ionization Nuclear Emission Regions (LINERs) following \citet{2013MNRAS.430.2002R}. Left: The light-purple circles are the values obtained from this work for the spaxels where the four emission lines were fitted. Data from the literature for the Seyfert galaxies, types 1 and 2 (Sy1 and Sy2, respectively), are from \citet{2004AA...425..457R}, \citet{2005MNRAS.364.1041R}, \citet{2006AA...457...61R}, \citet{2013MNRAS.430.2002R}, \citet{2002MNRAS.331..154R}, \citet{2001AJ....122..764K} and \citet{2021MNRAS.tmp..778R}. Data for the BAT AGN Spectroscopic Survey are from \citet{2017MNRAS.467..540L}. Data for SFGs are from \citet{2005MNRAS.364.1041R}, \citet{2013MNRAS.430.2002R}, \citet{2004ApJ...601..813D} and \citet{1998ApJS..114...59L}. For LINERs, data are from \citet{1998ApJS..114...59L} and \citet{2013MNRAS.430.2002R}. Supernova remnants (SNRs) are from  \citet{1998ApJS..114...59L} and Blue Compact Dwarfs (BCDs) data points are from \citet{2011ApJ...734...82I}. Right: zoomed version of the left panel showing our resolved NIR diagnostic diagram for NGC~34 and the results obtained from the integrated regions of Fig.\ref{fig:LineRatios} along with the linear relations defined by \citet{2015AA...578A..48C} for young star-forming regions (SF-young), SNe-dominated regions, AGN-dominated and AGN-diffuse regions.
 }
 \label{fig:DiagnosticDiagrams}
\end{center}
\end{figure*}

In Fig.~\ref{fig:DiagnosticDiagrams}, the left panel shows our spatially resolved NIR diagnostic diagram of NGC~34 along with data collected from the literature for other sources (see Appendix~\ref{appendix:DataLiter} for the complete list of sources, values of the data points and references). The limiting ratio $\mathrm{[\ion{Fe}{ii}] / Pa\beta < 0.6}$ from \citet{2013MNRAS.430.2002R} for SFGs is indicated by the horizontal, black, dashed line. A zoomed version of the diagram is presented in the right panel along with the values of line-ratios obtained in the integrated regions shown in Fig.~\ref{fig:LineRatios} and listed in Table~\ref{tab:fluxes_regions}. We also included the linear relations between the $\mathrm{[\ion{Fe}{ii}] / Pa\beta}$ and $\mathrm{H_{2} / Br\gamma}$ line ratios defined by \citet{2015AA...578A..48C}, who investigated the two-dimensional excitation structure of the interstellar medium in a sample of low-z LIRGs and Seyferts, using NIR integral field spectroscopy\footnote{Note that the linear relations presented by \citet{2015AA...578A..48C} are defined on the basis of the $\mathrm{[\ion{Fe}{ii}]1.64\mu m / Br\gamma}$ ratio. Assuming the theoretical ratio $\mathrm{[\ion{Fe}{ii}]1.64\mu m / [\ion{Fe}{ii}]1.26\mu m} = 0.7646$ and case B recombination, we can calculate the $\mathrm{[\ion{Fe}{ii}]1.26\mu m / Pa\beta}$ line ratio using $\mathrm{[\ion{Fe}{ii}]1.64\mu m / Br\gamma} = 4.4974 \times \mathrm{[\ion{Fe}{ii}]1.26\mu m / Pa\beta}$ (see Sec.~2.2 of \citealt{2015AA...578A..48C}).}. In their work (see Sec.~4.1 of \citealt{2015AA...578A..48C}), distinct, linear relations in the $\mathrm{[\ion{Fe}{ii}] / Pa\beta}$--$\mathrm{H_{2} / Br\gamma}$ plane were defined for regions identified as (1) young star-forming regions (SF-young, ages $\leq$6~Myr); (2) aged, supernovae dominated regions (SNe-dominated, ages $\sim$8--40~Myr); (3) compact regions associated with nuclear AGNs (AGN-compact) and (4) diffuse, extended emitting regions where the AGN radiation field is still detected (AGN-diffuse).

Following the diagnostics presented by \citet{2013MNRAS.430.2002R}, our results indicate that all the NGC~34 spaxels fall in the AGN dominated region. However, one should note that considering only the $\mathrm{[\ion{Fe}{ii}] / Pa\beta}$ line ratio, some spaxels fall bellow the 0.6 limiting value for SFGs. This is better illustrated by the integrated measurements taken in the regions A--F (right panel of Fig.~\ref{fig:DiagnosticDiagrams}). Recall that a clear ring-shaped structure around the nucleus of the galaxy can be seen in the maps of the $\mathrm{[\ion{Fe}{ii}] / Pa\beta}$ and $\mathrm{H_{2} / Br\gamma}$ line ratios shown in Fig.~\ref{fig:LineRatios}. Regions A--E are all located on this circumnuclear ring. In Fig.~\ref{fig:DiagnosticDiagrams}, regions A, C and D are below the $\mathrm{[\ion{Fe}{ii}] / Pa\beta = 0.6}$ curve, while regions B and E are slightly above it, although considering the measured uncertainties, they could also be below this line. Region F, which is surrounded by the ring structure, is well separated from the other points in the NIR diagnostic diagram and occupies a portion of the diagram where both the $\mathrm{[\ion{Fe}{ii}]}$ and $\mathrm{H_2\,\lambda 21218}$\AA\ emission lines are enhanced. Our interpretation of these findings is that the low $\mathrm{[\ion{Fe}{ii}] / Pa\beta}$ values found in regions A--E are due to a combination of excitation mainly caused by the stars with some contribution of the AGN, and that the nuclear starburst in NGC~34 is currently distributed in the circumnuclear star-forming ring (CNSFR) traced by these regions. The higher ratio found in region F reveals the presence of the AGN. Results presented by \citet{2005MNRAS.364.1041R} had already placed NGC~34 in the AGN dominated region of the diagram (yellow circle in Fig.~\ref{fig:DiagnosticDiagrams}). However, since these results were based on one-dimensional spectra, they could not isolate the CNSFR. In addition, when considering the results found by \citet{2015AA...578A..48C}, we can see that region F, which we associate with the AGN, does follow the linear relation defined for AGN-dominated regions. The other regions in the galaxy follow the linear relation found for regions where the radiation field of the AGN acts as a contributor to the observed emission.

Our interpretation is corroborated by the results of the PCA Tomography presented in Sec.~\ref{sec:PCA}, which clearly identify two distinct structures in the central region of NGC~34: a nuclear component which is related to the $\mathrm{[\ion{Fe}{ii}]}$ and $\mathrm{H_2\,\lambda 21218}$\AA\ emission lines; and a circumnuclear structure associated with the $\mathrm{Pa\beta}$ and $\mathrm{Br\gamma}$ emission. 

\subsection{The nature of the $\mathrm{[\ion{Fe}{ii}]}$ and $\mathrm{H_2\,\lambda 21218}$\AA\ emission lines} \label{sec:FeIIH2Nature}

As can be seen by the NIR diagnostic diagram in Fig.~\ref{fig:DiagnosticDiagrams}, emission by $\mathrm{[\ion{Fe}{ii}]}$ and H$_2$ is detected in objects displaying several degrees of nuclear activity. In starburst galaxies, $\mathrm{[\ion{Fe}{ii}]}$ emission is enhanced by supernova-driven shocks. For this reason, as the starburst ages, the level of $\mathrm{[\ion{Fe}{ii}]}$ is expected to rise due to the increasing SN activity, while the levels of the hydrogen recombination lines decrease following the aging of the stellar population. In Seyfert objects, the X-ray emission produced by the central engine is able to penetrate deeper in the molecular clouds, thus creating extensive partially ionized zones where both $\mathrm{[\ion{Fe}{ii}]}$ and H$_2$ emissions are enhanced. Moreover, shocks associated with nuclear outflows also contribute as an additional excitation mechanism \citep{2000ApJ...528..186M}.

\citet{1998ApJS..114...59L} point out that while H$_2$ can be easily destroyed, 98~per cent of the iron is tied up in dust grains. Therefore, in order to invoke a single mechanism to power both lines, it must not destroy the H$_2$ molecules and must free up iron through dust grain destruction, both of which can be achieved in partially ionized zones created by the nuclear X-ray emission.

The centre of NGC~34 hosts a hard X-ray source associated with an obscured Seyfert 2 nucleus \citep{2012MNRAS.423..185E}. The fact that region F not only shows higher $\mathrm{[\ion{Fe}{ii}] / Pa\beta}$ and $\mathrm{H_{2} / Br\gamma}$ values but also clearly lies in a different position in the NIR diagnostic diagram in Fig.\ref{fig:DiagnosticDiagrams} can be explained by an enhancement of $\mathrm{[\ion{Fe}{ii}]}$ and H$_2$ due to the AGN X-ray emission. Moreover, although contribution from the ionizing flux related to the circumnuclear stellar activity likely plays a role in powering both the $\mathrm{[\ion{Fe}{ii}]}$ and H$_2$ emissions in the locations where the $\mathrm{[\ion{Fe}{ii}] / Pa\beta}$ values are low, an additional mechanism is necessary mainly to explain the enhanced H$_2$ emission in these same locations. Such an additional excitation component could also be due to the X-rays emitted by the central engine. 

Following \citet{2008MNRAS.385.1129R} and \citet{2007A&A...466..451Z}, we can use the models of \citet{1996ApJ...466..561M} to verify if X-ray heating by a source with intrinsic luminosity $L_X$ is a viable H$_2$ excitation mechanism. Considering a cloud at a distance $d$ from a hard X-ray source and with gas density $n$, the emergent $\mathrm{H_2\,\lambda 21218}$\AA\ flux can be obtained from figure 6 in \citet{1996ApJ...466..561M} through the determination of the effective ionization parameter $\xi_{eff}$:

\begin{equation}
\xi_{eff} = 1.26\times 10^{-4} \frac{f_x}{n_5 N^{0.9}_{22}} \, ,
\label{eq:effionparam}
\end{equation} 
where $f_x$ is the incident X-ray flux at the distance $d$, $n_5$ is total hydrogen gas density in units of 10$^5$~cm$^{-3}$ and $N_{22}$ is the attenuating column density between the AGN and the gas cloud in units of 10$^{22}$~cm$^{-2}$. The value of $f_x$ can be determined using $L_x / 4\pi d^2$. We calculated $\xi_{eff}$ for three different distances from the AGN (40, 80 and 120~pc) adopting the intrinsic hard X-ray luminosity $L_X = 1.4 \times 10^{42} \mathrm{erg\,s^{-1}}$ \citep{2012MNRAS.423..185E}, $N_H = 4.2 \times 10^{20} \mathrm{cm^{-2}}$ \citep{2014AJ....147...74F} and for a gas density\footnote{The models of \citet{1996ApJ...466..561M} of X-ray irradiated molecular gas were calculated for $n = 10^{3}$ and $n = 10^{5}$~cm$^{-3}$. However, the former does not provide predictions on the observed H$_2$ fluxes in every region of interest of the source. Therefore, for completeness, we chose $n = 10^{5}$ to perform our calculations.} of $n = 10^{5}$~cm$^{-3}$. With the aid of figure 6(a) in \citet{1996ApJ...466..561M}, we determined the emergent H$_2$ flux in $\mathrm{erg \cdot cm^{-2} \cdot s^{-1}}$ for an aperture of $\mathrm{0.021\arcsec \times 0.021\arcsec}$ (the dimensions of a spaxel in our datacubes), which corresponds to a solid angle of 1.03~$\times$~10$^{-14}$~sr. Our results are shown in Table~\ref{tab:modelsMaloney}. The observed $\mathrm{H_2\,\lambda 21218}$\AA\ fluxes were determined from three different apertures centred in region F as follows: (1) at $d = 40$~pc, log(F$_{H_2}$) corresponds to the mean flux value in a spaxel located in a circular aperture of radius 40~pc, (2) at $d = 80$~pc, log(F$_{H_2}$) corresponds to the mean flux value in a spaxel located in a ring with inner and outer radii of 60 and 100~pc, respectively, and (3) at $d = 120$~pc, log(F$_{H_2}$) corresponds to the mean flux value in a spaxel located in a ring with inner and outer radii of 100 and 140~pc, respectively.

Our estimates indicate that X-ray heating can fully account for the H$_2$ emission seen in region F, which is probably associated with the AGN. For larger distances, although X-rays contribute with some amount to the observed $\mathrm{H_2\,\lambda 21218}$\AA\ fluxes, they are not the dominant excitation mechanism. Therefore we conclude that, in NGC~34, the observed $[\ion{Fe}{ii}]$ and H$_2$ emissions are due to a combination of photo-ionization by young stars, excitation by X-rays produced by the AGN and shocks. \citet{2018MNRAS.474.3640M}, when modelling the CO spectral line energy distribution, did not find that shocks contribute significantly to the heating of the molecular gas in NGC~34. However, our results provide kinematic signatures of shocks, as evidenced by the broad $[\ion{Fe}{ii}]$ and H$_2$ components, indicating that shocks do contribute to the observed emission.

Note that our calculations using equation~\ref{eq:effionparam} were carried out assuming a column density derived from radio observations of the atomic hydrogen (\ion{H}{i} at 21~cm, \citealt{2014AJ....147...74F}). The column densities obtained by \citet{2012MNRAS.423..185E} from X-rays and by \citet{2014ApJ...787...48X} using ALMA CO(6-5) observations are of the order $10^{23} \mathrm{cm^{-2}}$. In this case, the predicted H$_2$ intensities at $d = 40$~pc would be two orders of magnitude lower than the ones shown in Table~\ref{tab:modelsMaloney}, indicating that even at the nucleus of NGC~34, the AGN would have a minor role in the observed H$_2$ emission. This is unlikely considering the discussion presented in Sec.~\ref{sec:DiagDiagram}. \citet{2004AA...425..457R} also estimated the H$_2$ fluxes for a list of sources using the models of \citet{1996ApJ...466..561M} and found that the calculations using $N_H$ derived from X-rays result in a poorer fit compared to the calculations using $N_H$ derived from radio observations. They concluded that this is because $N_H$ derived from X-rays probes obscuring material in the innermost regions of active galaxies, while H$_2$ and also $[\ion{Fe}{ii}]$ emissions must arise farther out in the narrow line region, thus justifying the adoption of a column density obtained from 21~cm observations.

We further investigate the nature of the H$_2$ emission using the ratio between the $\mathrm{H_2\,\lambda 22470}$\AA\ and $\mathrm{H_2\,\lambda 21218}$\AA\ emission lines to distinguish thermal ($\sim 0.1-0.2$) from non-thermal ($\sim 0.55$) excitation mechanisms of the H$_2$ molecule \citep{1994ApJ...427..777M}. In NGC~34, $F_{2.247 \mu \mathrm{m}}/F_{2.122 \mu \mathrm{m}} < 0.2$ in all regions, thus confirming that the \textit{K}-band H$_2$ lines are predominantly excited by thermal processes (shocks and/or X-ray heating). We can not, however, rule out the contribution of UV-fluorescence to the observed H$_2$ spectrum since the H$_2$ emission at 21218\AA\ is itself much more sensitive to \textit{thermal} processes than the $\mathrm{H_2\,\lambda 22470}$\AA\ line, which is a tracer of \textit{fluorescence} \citep{1987ApJ...322..412B}.  

The predominance of \textit{thermal} processes in the excitation of the H$_2$  molecule is further illustrated by the fact that a broad, blue-shifted component is only detected in the H$_2$ emission at 21218\AA\ . The velocity and line width of this \textit{blue wing}, as well as of that seen in $\mathrm{[\ion{Fe}{ii}]}$, indicate that the central regions of NGC~34 also drive a nuclear outflow of molecular and ionized gas. Although the outflow could, in principle, be driven either by the AGN, by winds associated with the circumnuclear starburst or by SN-driven shocks, we favour the first one since (1) the highest fluxes of these broad components coincide, or almost coincide, with the nucleus of the galaxy and (2) we do not observe the high $\mathrm{[\ion{Fe}{ii}] / Pa\beta}$ values expected according to the linear relation defined by \citet{2015AA...578A..48C} for SNe-dominated regions, where SN-driven shocks are responsible for enhancing the $\mathrm{[\ion{Fe}{ii}]}$ emission. In the analysis carried out by \citet{2015AA...578A..48C} for the composite LIRG NGC~5135 for example, they found a value larger than 1.5 for the $\mathrm{[\ion{Fe}{ii}] / Pa\beta}$ ratio at the location of $\mathrm{[\ion{Fe}{ii}]}$ peak associated with a SNe-dominated circumnuclear star-forming clump. Therefore, we conclude that in NGC~34, an AGN-driven nuclear outflow is the most likely primary source of shock excitation of the H$_2$ and $\mathrm{[\ion{Fe}{ii}]}$ species. 

Finally, we can also estimate the H$_2$ vibrational excitation temperature $T_{\mathrm{vib}}$ using $T_{\mathrm{vib}} \simeq 5600 / \mathrm{ln}(1.355 \times F_{2.122 \mu \mathrm{m}}/F_{2.247 \mu \mathrm{m}})$ \citep{2002MNRAS.331..154R}. We find a mean value of 2504$\pm$174~K, which is in the range of values found for the H$_2$ thermal components in other sources (1800--2700~K, \citealt{2002MNRAS.331..154R}). \citet{2005MNRAS.364.1041R} derived an upper limit of $T_{\mathrm{vib}} <$1800~K for NGC~34. However, this result was based on one-dimensional spectra and includes contribution from the broad H$_2$ component. If we include the contribution of the broad, blue-shifted component to the observed $\mathrm{H_2\,\lambda 21218}$\AA\ flux, we find $T_{\mathrm{vib}}$~=~2196$\pm$154~K. The values found for $T_{\mathrm{vib}}$ support the assumption $T_{\mathrm{vib}} = 2000\,\mathrm{K}$ used in eq.~\ref{eq:MassH2} to estimate the mass of hot H$_2$. 

\begin{table}
    \centering
        \caption{Observed and expected $\mathrm{H_2\,\lambda 21218}$\AA\ fluxes according to the X-ray heating models of \citet{1996ApJ...466..561M} for an aperture of $\mathrm{0.021\arcsec \times 0.021\arcsec}$. Fluxes are in units of $\mathrm{erg \cdot cm^{-2} \cdot s^{-1}}$.}
    \label{tab:modelsMaloney}
    \begin{tabular}{ccccc}
\hline
\multicolumn{1}{c}{} & \multicolumn{1}{c}{$d$} & \multicolumn{1}{c}{Observed} & \multicolumn{2}{c}{$n = 10^{5} \, \mathrm{cm^{-3}}$} \\
 & (pc) & log(F$_{H_2}$) & log~$\xi_{eff}$ & log(F$_{H_2}$) \\
\hline
(1) & 40 & -16.7 & -1.8 & -16.2 \\
(2) & 80 & -17.0 & -2.4 & -18.7 \\
(3) & 120 & -17.2 & -2.8 & -18.5 \\
\hline
    \end{tabular}
\end{table}

\subsection{Circumnuclear star formation ring} \label{sec:CNSFR}

Our maps of emission line ratios and the \textit{J} and \textit{K}-bands PCA Tomography show that the nuclear starburst in NGC~34 is distributed in a CNSFR. We can use the $\mathrm{Br\gamma}$ luminosity ($L_{\mathrm{Br\gamma}}$) to estimate the SFR in the ring adopting the following relation \citep{1998ARA&A..36..189K}:  
 
 \begin{equation}
     SFR\, (\mathrm{M_{\sun}\, yr^{-1}}) = 8.2 \times 10^{-40} L_{\mathrm{Br\gamma}}(\mathrm{erg\, s^{-1}})\,.
     \label{eq:SFR}
 \end{equation}
 
The integrated, extinction corrected $\mathrm{Br\gamma}$ flux in the continuous ring A--E--A, with inner and outer radii of 60 and 180~pc, respectively, is $F_\mathrm{Br\gamma} \approx 20 \times 10^{-15} \mathrm{erg\, cm^{-2}\, s^{-1}}$, and $L_{\mathrm{Br\gamma}} \approx 1.6 \times 10^{40} \mathrm{erg\, s^{-1}}$. From equation~\ref{eq:SFR}, we derive $\mathrm{SFR} \approx 13\, \mathrm{M_{\sun}\, yr^{-1}}$. Also based on the $\mathrm{Br\gamma}$ luminosity, \citet{2005AA...434..149V} estimated a SFR~=~26.7~$\mathrm{M_{\sun}\, yr^{-1}}$, and \citet{2014MNRAS.443.1754D} found SFR~=~9.63~$\mathrm{M_{\sun}\, yr^{-1}}$.

\citet{2005AA...434..149V} carried out NIR medium-resolution spectroscopy of a sample of (U)LIRGs and found that the $\mathrm{Br\gamma}$ luminosity provides SFRs which are on average 60~per~cent of those derived from the far infrared luminosity. In the case of NGC~34, they found SFR(L$_{FIR}$)~=~49.9~$\mathrm{M_{\sun}\, yr^{-1}}$. They stated that although the AGN might contribute to the infrared emission, this contribution would have to exceed 80~per~cent to explain this large discrepancy. \citet{2006ApJ...650..835A} studied the NIR and star-forming properties of a sample of local LIRGs and also reported that the SFRs derived from the number of ionizing photons are on average 0.2--0.3~dex lower than those inferred from the total IR luminosity. This is because the luminosities of hydrogen recombination lines trace the most recent SFR, and there is a tendency for the measured IR luminosity to include some contribution from older stars. In addition, \citet{2012ApJ...744....2A} performed the spectral decomposition of a complete-volume-limited sample of local LIRGs and found that for the majority of them, the total AGN bolometric contribution to the observed IR luminosities has an upper limit of 5~per~cent. Therefore, most of their IR luminosities are related to star formation processes.

We point out that the SFR of NGC~34 derived by us is a lower limit since our observations provide a smaller aperture compared to previous works, and also due to the fact that the circumnuclear ring may extend to a larger radius. CNSFRs have been detected in a variety of LIRGs with diameters ranging from 0.7 to 2~kpc (e.g \citealt{2006ApJ...650..835A}). 

\subsection{Mass of ionized and molecular gas} \label{sec:MassCalc}

We can estimate the mass of ionized and molecular gas in NGC~34 using integrated, extinction corrected fluxes of $\mathrm{Br\gamma}$ and $\mathrm{H_2\,\lambda 21218}$\AA\ . Following \citet{2009MNRAS.394.1148S} and assuming an electron temperature of $T = 10^4\, \mathrm{K}$, the mass of ionized hydrogen, in solar masses, is given by:

\begin{equation}
    M_{\ion{H}{ii}} \approx 3\times 10^{19} \left(\frac{F_{\mathrm{Br\gamma}}}{\mathrm{erg\, cm^{-2}\, s^{-1}}} \right) \left(\frac{D}{\mathrm{Mpc}}\right)^{2} \left(\frac{N_e}{\mathrm{cm^{-3}}}\right)^{-1} ,
    \label{eq:MassHii}
\end{equation}
where $F_{\mathrm{Br\gamma}}$ is the $\mathrm{Br\gamma}$ integrated flux, $D$ is the distance to NGC~34 and we have assumed an electron density $N_e = 160\, \mathrm{cm^{-3}}$, which is the mean value calculated by \citet{2018A&A...618A...6K} using the $\mathrm{[\ion{S}{ii}]\lambda\lambda6716,6731}$ lines for a sample of nearby ($z < 0.02$) Seyfert~1s, Seyfert~2s and AGN-Starburst composite systems.

The mass of hot H$_2$ in solar masses can be estimated as follows \citep{1982ApJ...253..136S}:

\begin{align}\label{eq:MassH2}
    M_{\mathrm{H_2}}&= \frac{2\,m_p\, F_{\mathrm{H_{2}\lambda 21218}}\, 4\,\pi\, d^2}{f_{(\nu=1,J=3)}\, A_{S(1)}\, h\, \nu} \nonumber \\
     &= 5.0776 \times 10^{13} \left(\frac{F_{\mathrm{H_{2}\lambda 21218}}}{\mathrm{erg\, cm^{-2}\, s^{-1}}} \right) \left(\frac{D}{\mathrm{Mpc}}\right)^{2} ,
\end{align}
where, $m_p$ is the proton mass, $F_{\mathrm{H_{2}\lambda 21218}}$ is the integrated flux of the $\mathrm{H_2\,\lambda 21218}$\AA\ emission line, $D$ is the distance to the galaxy, $h$ is the Planck constant and $\nu$ is the frequency of the H$_2$ line. Adopting a typical vibrational excitation temperature of $T_{\mathrm{vib}} = 2000\,\mathrm{K}$, which is consistent with the values we have obtained in Sec.~\ref{sec:FeIIH2Nature}, the population fraction is $f_{(\nu=1,J=3)} = 1.22 \times 10^{-2}$ and the transition probability is $A_{S(1)} = 3.47\times 10^{-7} \mathrm{s}^{-1}$ \citep{1977ApJS...35..281T,1982ApJ...253..136S}.

We carried out the mass estimates for the inner $\mathrm{1.0\arcsec \times 1.0\arcsec}$ of the galaxy, since this corresponds to the region where we have derived visual extinction values to be applied to the observed fluxes. The integrated, extinction-corrected $\mathrm{Br\gamma}$ and $\mathrm{H_2\,\lambda 21218}$\AA\ fluxes are $F_\mathrm{Br\gamma} \approx 26 \times 10^{-15} \mathrm{erg\, cm^{-2}\, s^{-1}}$ and $F_\mathrm{H_2\,\lambda 21218} \approx 34 \times 10^{-15} \mathrm{erg\, cm^{-2}\, s^{-1}}$, respectively, resulting in $M_{\ion{H}{ii}} \approx 3 \times 10^7 \mathrm{M_{\sun}}$ and $M_{\mathrm{H_2}} \approx 11570 \mathrm{M_{\sun}}$. Although M$_{\ion{H}{ii}}$ is nearly 10$^3$ times larger than $M_{\mathrm{H_2}}$, we note that the later refers only to the hot molecular gas that gives rise to the observed NIR emission lines. The value derived by us for the hot molecular gas mass is similar to the one presented by \citet{2005MNRAS.364.1041R} who found $M_{\mathrm{H_2}} = 12600 \mathrm{M_{\sun}}$.    

Following \citet{2013MNRAS.428.2389M}, we can also estimate the mass of cold molecular gas using:

\begin{equation}
    \frac{M_{\mathrm{H_2 cold}}}{\mathrm{M_{\sun}}} \approx 1174 \times \left( \frac{L_{\mathrm{H_{2}\lambda 21218}}}{\mathrm{L_{\sun}}} \right) ,
    \label{eq:Mcold}
\end{equation}
where $L_{\mathrm{H_{2}\lambda 21218}}$ is the luminosity of the H$_2$ line. We find $M_{\mathrm{H_2 cold}} \approx 8 \times 10^9 \mathrm{M_{\sun}}$ for NGC~34, which is roughly six orders of magnitude higher than that derived for the hot H$_2$. This is in agreement with previous works that estimated the mass of cold molecular gas in NGC~34 from observations of the CO molecule ($M_{\mathrm{H_2 cold}} \approx 7 \times 10^9 \mathrm{M_{\sun}}$, \citealt{1992AA...255...87C,1990AA...229...17K,2003AA...398..493K}).

\section{Conclusions} \label{sec:Conclusions}

The galaxy NGC~34 is a local LIRG whose nature of its emission line features has been explained either as due to a pure starburst or due to a starburst--AGN composite source. In this work, we used AO-assisted IFU observations carried out with the NIFS instrument in the \textit{J} and \textit{K} bands to map the inner $\mathrm{1.2\,kpc \times 1.2\,kpc}$ of the galaxy in order to investigate the excitation mechanisms of its NIR spectrum. We summarise our main findings as follows:

\begin{itemize}
\item The NGC~34 NIR spectra are characterised by the  $[\ion{P}{ii}]\,\lambda 11470$\AA\ , $[\ion{P}{ii}]\,\lambda 11886$\AA\ , $[\ion{Fe}{ii}]\,\lambda 12570$\AA\ and $\mathrm{Pa\beta}$ emission features in the \textit{J}-band, and by the $\mathrm{H_2\,\lambda 21218}$\AA\ , $\mathrm{Br\gamma}$, $\mathrm{H_2\,\lambda 22230}$\AA\ and $\mathrm{H_2\,\lambda 22470}$\AA\ emission lines in the \textit{K}-band;
\item We report the detection of $[\ion{Ni}{ii}]$ emission at $\lambda 11910$\AA\ ;
\item The fluxes of all emission features but $\mathrm{Pa\beta}$ and $\mathrm{Br\gamma}$ peak at the nucleus of the galaxy. The $\mathrm{Pa\beta}$ and $\mathrm{Br\gamma}$ flux distributions are asymmetric and peak north-west. This asymmetry is confirmed by the channel maps and indicates that the hydrogen recombination lines are tracing a circumnuclear structure.
\item Through the analysis of the gas kinematics, we confirm the presence of a ND of ionized and molecular gas in NGC~34, with a northern-receding and a southern-approaching side.  We determined that the mean kinematic PA of the gas disc is $\mathrm{-9^{\circ}.2 \pm 0^{\circ}.9}$;
\item We find that the NGC~34 nuclear starburst is distributed in a CNSFR with approximate inner and outer radii of 60 and 180~pc, respectively, as revealed by maps of the $\mathrm{[\ion{Fe}{ii}] / Pa\beta}$ and $\mathrm{H_{2} / Br\gamma}$ emission-line ratios. These maps clearly show a ring-like structure with lower values around the nucleus of the galaxy, especially for the $\mathrm{[\ion{Fe}{ii}] / Pa\beta}$ ratios, where many locations display values that are consistent with pure star formation ($\mathrm{[\ion{Fe}{ii}] / Pa\beta \leq 0.6}$, \citealt{2013MNRAS.430.2002R}). The higher values found for both ratios indicate that multiple excitation mechanisms are at play in NGC~34. 
\item The presence of a CNSFR is corroborated by the PCA Tomography, which shows a nuclear object associated with the $\mathrm{H_2\,\lambda 21218}$\AA\ and $[\ion{Fe}{ii}]$ emission lines, and a circumnuclear structure associated with the $\mathrm{Pa\beta}$ and $\mathrm{Br\gamma}$ lines.
\item The resolved NIR diagnostic diagram based on the $\mathrm{[\ion{Fe}{ii}] / Pa\beta}$ and $\mathrm{H_{2} / Br\gamma}$ emission-line ratios show that all the NGC~34 spaxels fall in the AGN dominated region. Integrated measurements taken in different regions A--F also confirm that the inner regions of NGC~34 are characterised by a nuclear and circumnuclear structures. Regions A--E trace the CNSFR since the $\mathrm{[\ion{Fe}{ii}] / Pa\beta}$ ratios lie bellow or slightly above the limiting value of 0.6 for pure star formation defined by \citet{2013MNRAS.430.2002R}. Region F, in the nucleus of the galaxy, clearly occupies a portion of the diagram where both the $[\ion{Fe}{ii}]$ and $\mathrm{H_2\,\lambda 21218}$\AA\ lines are enhanced relative to the hydrogen recombination lines. Therefore, we associate region F with the AGN. NGC~34 hosts a hard X-ray source associated with an obscured Seyfert~2 nucleus \citep{2012MNRAS.423..185E}, and whose incident flux on the gas clouds could contribute to their enhancement. Moreover, the line ratios observed in region F follow the linear relation found by \citet{2015AA...578A..48C} for AGN-dominated regions;
\item Using the models of \citet{1996ApJ...466..561M}, we concluded that X-ray heating can fully account for the $\mathrm{H_2\,\lambda 21218}$\AA\ emission in the nucleus of the galaxy (region F), but it is not the predominant excitation mechanism at larger distances. Therefore we state that, in NGC~34, emission by $[\ion{Fe}{ii}]$ and H$_2$ is due to a combination of photo-ionization by young stars, excitation by X-rays produced by the AGN and shocks as evidenced by the kinematic signatures of outflows in the ionized and molecular gas phases;
\item We detected broad, blue-shifted components associated with the $[\ion{Fe}{ii}]$ and $\mathrm{H_2\,\lambda 21218}$\AA\ emission lines at $-420$ and $-250$~$\mathrm{km \cdot s^{-1}}$, respectively, that can be interpreted as a nuclear outflow of ionized and molecular gas. Our results indicate that the outflow is AGN-driven and is the most likely primary source of shock excitation of the H$_2$ and $\mathrm{[\ion{Fe}{ii}]}$ species;
\item From the $\mathrm{Br\gamma}$ luminosity of the CNSFR we estimated a lower limit of $\mathrm{SFR} \approx 13\, \mathrm{M_{\sun}\, yr^{-1}}$;
\item The mass of ionized hydrogen in NGC~34 is $M_{\ion{H}{ii}} \approx 3 \times 10^7 \mathrm{M_{\sun}}$, and the mass of cold molecular gas is $M_{\mathrm{H_2 cold}} \approx 8 \times 10^9 \mathrm{M_{\sun}}$, in agreement with previous works. 
\end{itemize}

Our results indicate that the merger remnant NGC~34 is a gas-rich system that hosts an AGN surrounded by a ring of star-formation. The AGN and the CNSFR are embedded in a compact, highly obscured environment making it difficult for optical studies to probe such inner, dusty regions. A common, evolutionary picture for (U)LIRGs is that mergers of gas-rich galaxies are responsible for funneling large amounts of molecular gas towards the centre of the merger, thus providing fuel to the starburst and AGN, both of which act on heating the surrounding dust. As the starburst declines, and the combined effects of SN explosions and AGN feedback clear out the nuclear dust, these objects become optically selected quasars (e.g.  \citealt{1988ApJ...325...74S}). In the case of NGC~34, we, therefore, conclude that we may be witnessing the early, evolutionary stage of a dust-enshrouded AGN.

\section*{Acknowledgements}

JCM thanks Coordena\c{c}\~{a}o de Aperfei\c{c}oamento de Pessoal de N\'{i}vel Superior (CAPES) for the financial support under the grant 88882.316156/2019-01. RR and RAR thanks Conselho Nacional de Desenvolvimento Cient\'{i}fico e Tecnol\'{o}gico (CNPq), CAPES and Funda\c{c}\~{a}o de Amparo \`{a} Pesquisa do Estado do Rio Grande do Sul  (FAPERGS) for the financial support. TVR also thanks CNPq for the financial support under the grant 306790/2019-0. NZD acknowledges partial support from FONDECYT through project 3190769. We thank Dr. Jo\~{a}o Evangelista Steiner\textdagger (1950--2020) for enlightening us with comments and suggestions before embarking on his journey to the stars. The authors also thank the anonymous referee for his/her careful revision that helped us to improve the quality of this manuscript. Based on observations obtained at the international Gemini Observatory, which is managed by the Association of Universities for Research in Astronomy (AURA) under a cooperative agreement with the National Science Foundation, on behalf of the Gemini Observatory partnership: the National Science Foundation (United States), National Research Council (Canada), Agencia Nacional de Investigaci\'{o}n y Desarrollo (Chile), Ministerio de Ciencia, Tecnolog\'{i}a e Innovaci\'{o}n (Argentina), Minist\'{e}rio da Ci\^{e}ncia, Tecnologia, Inova\c{c}\~{o}es e Comunica\c{c}\~{o}es (Brazil), and Korea Astronomy and Space Science Institute (Republic of Korea). This research made use of observations made with the NASA/ESA Hubble Space Telescope, and obtained from the Hubble Legacy Archive, which is a collaboration between the Space Telescope Science Institute (STScI/NASA), the Space Telescope European Coordinating Facility (ST-ECF/ESAC/ESA) and the Canadian Astronomy Data Centre (CADC/NRC/CSA). This research made use of: \textit{Astropy},\footnote{http://www.astropy.org} a community-developed core Python package for Astronomy \citep{astropy:2013, astropy:2018}; \textit{Photutils}, an \textit{Astropy} package for detection and photometry of astronomical sources \citep{larry_bradley_2021_4453725}; the \textit{NumPy} \citep{harris2020array} and \textit{Matplotlib} \citep{Hunter:2007} Python libraries.

\section*{Data availability}

The data underlying this article are available in the Gemini Observatory Archive at \url{https://archive.gemini.edu/searchform}, and can be accessed with the Program ID GN-2011B-Q-71.




\bibliographystyle{mnras}
\bibliography{Bib} 




\appendix

\section{Spectra of regions A--F}
\label{appendix:RegionsSpectra}

\begin{figure*}
\begin{center}

	\begin{minipage}{\textwidth}
	\centering
	\begin{minipage}{.33\textwidth}
		\includegraphics[width=\textwidth]{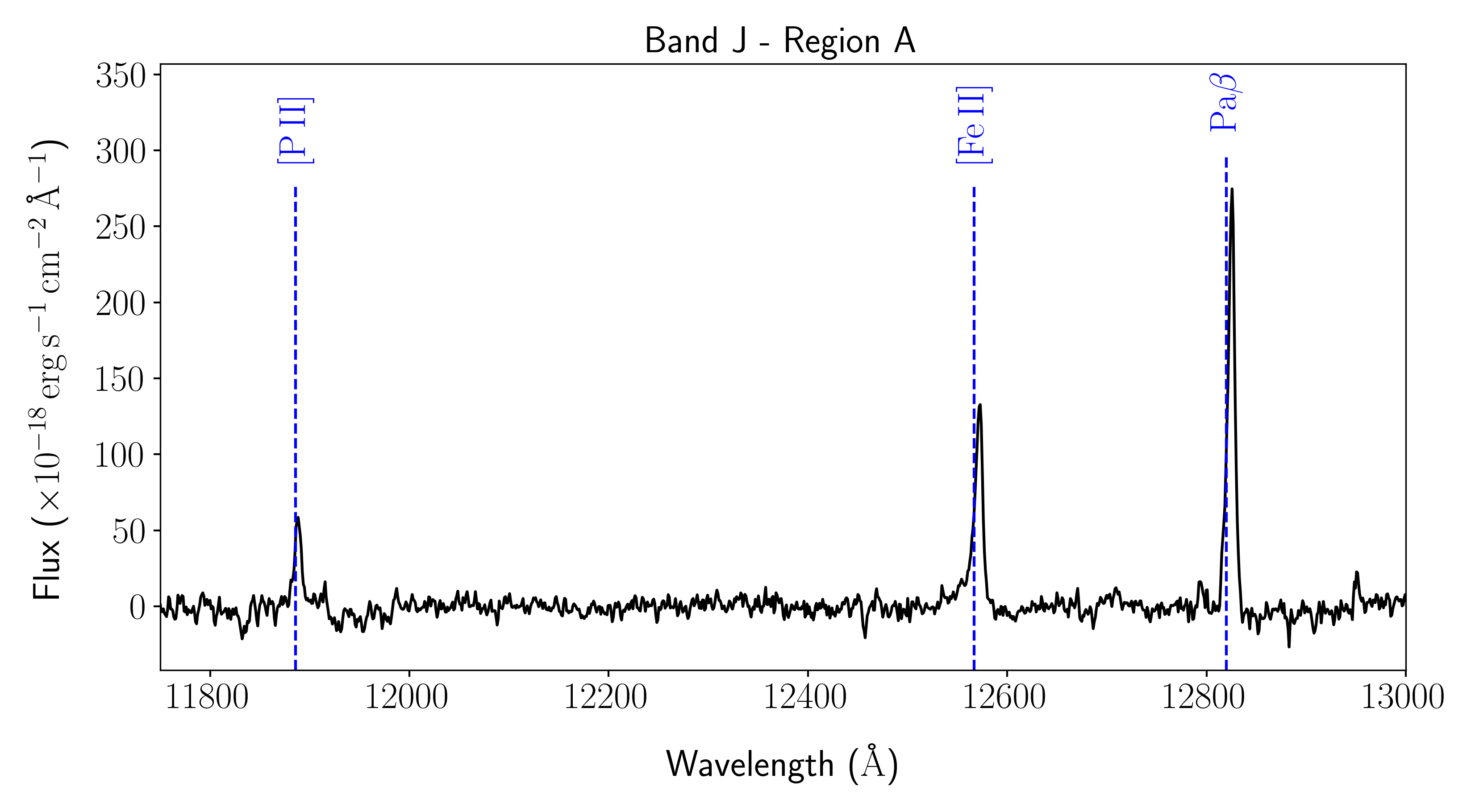}
	\end{minipage}
	\begin{minipage}{.33\textwidth}
		\includegraphics[width=\textwidth]{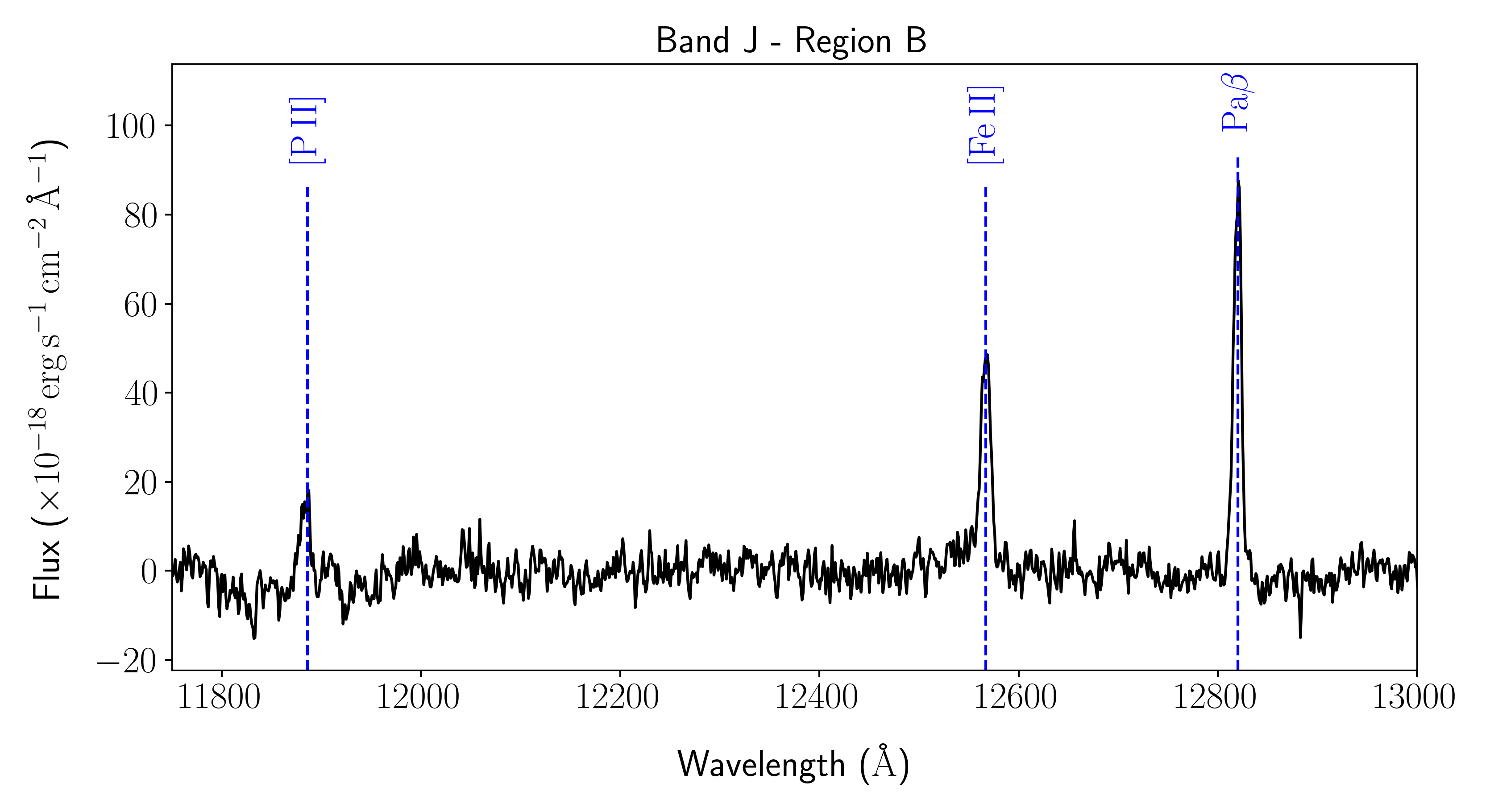}
	\end{minipage}
	\begin{minipage}{.33\textwidth}
		\includegraphics[width=\textwidth]{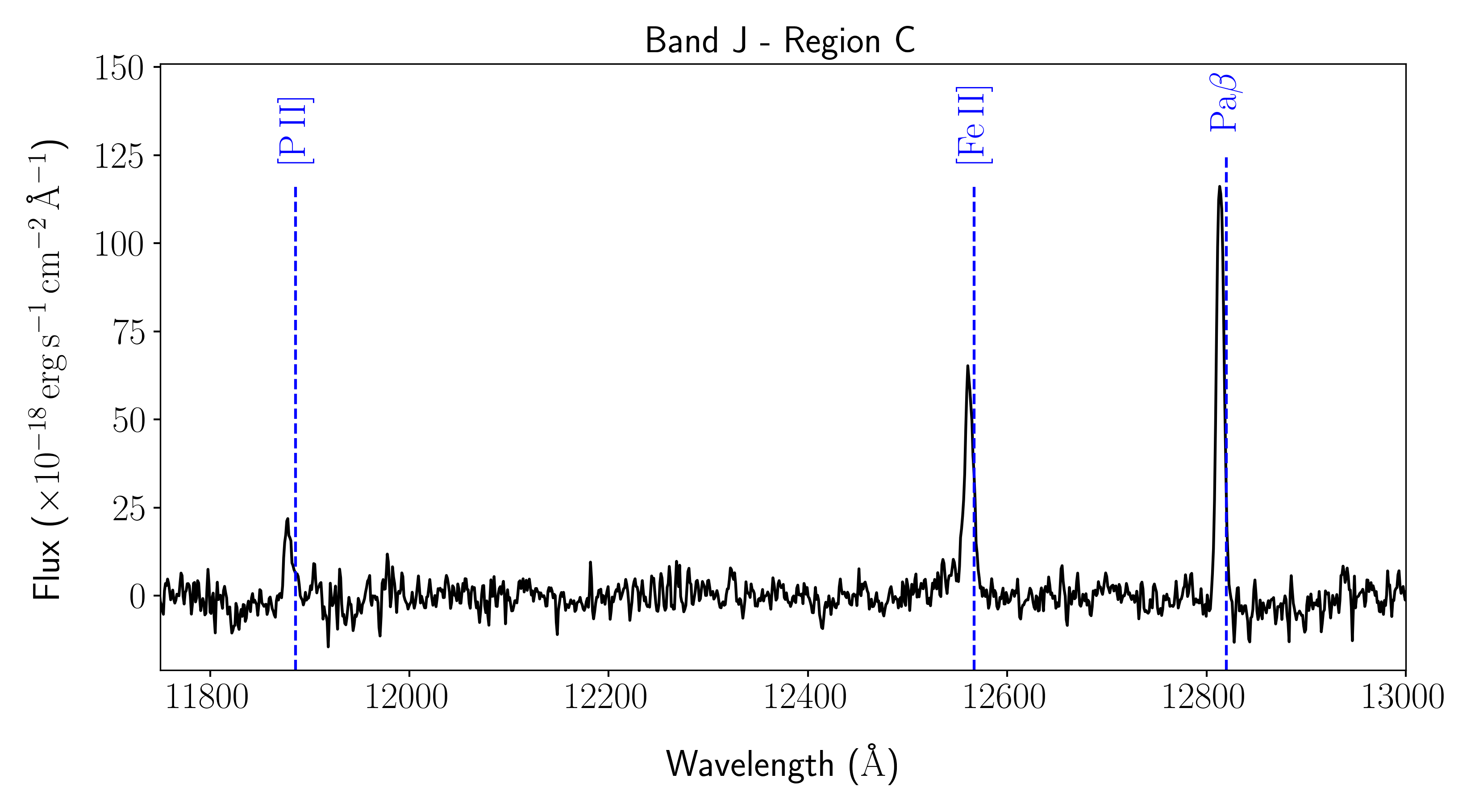}
	\end{minipage}
	\end{minipage}

	\begin{minipage}{\textwidth}
	\centering
	\begin{minipage}{.33\textwidth}
		\includegraphics[width=\textwidth]{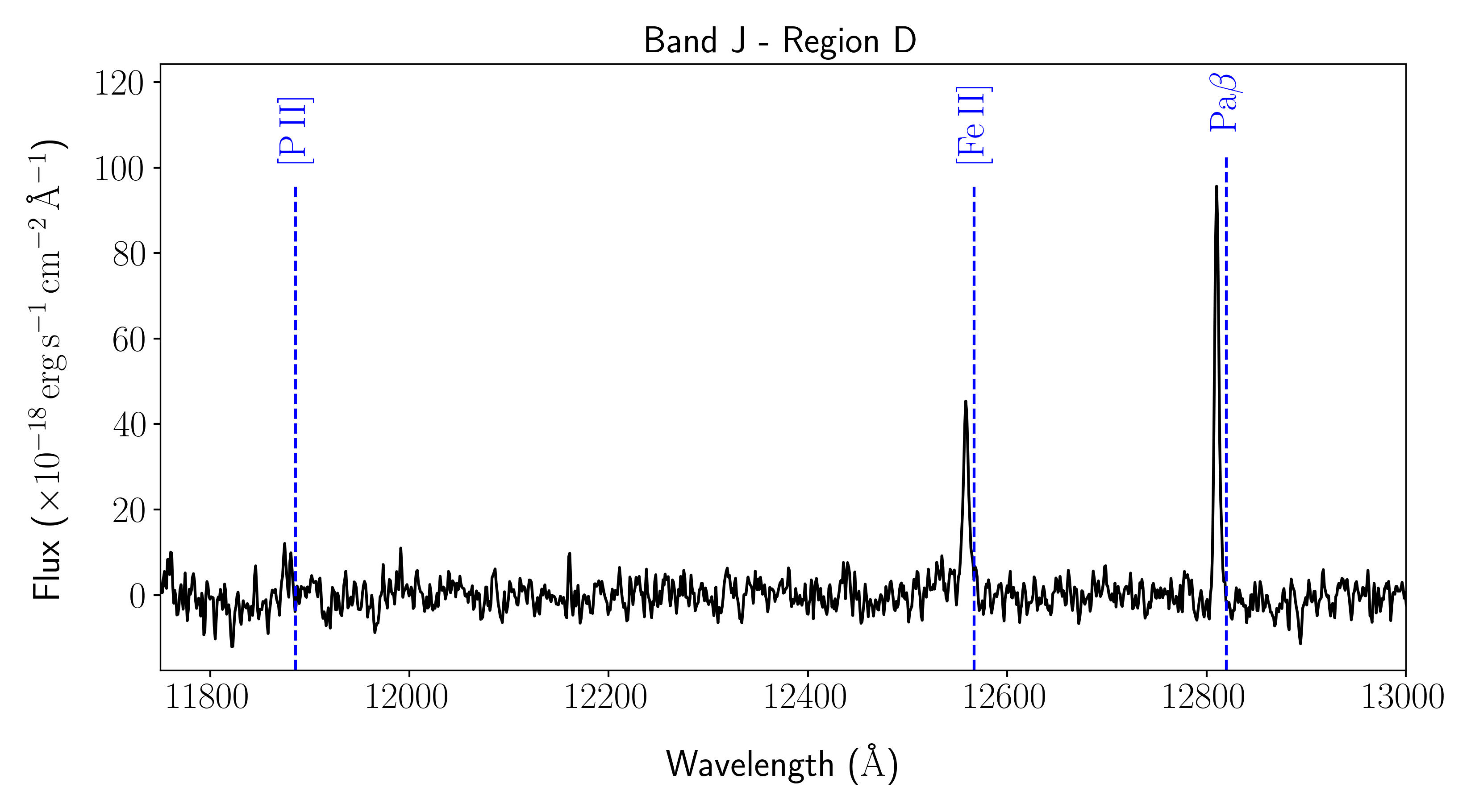}
	\end{minipage}
	\begin{minipage}{.33\textwidth}
		\includegraphics[width=\textwidth]{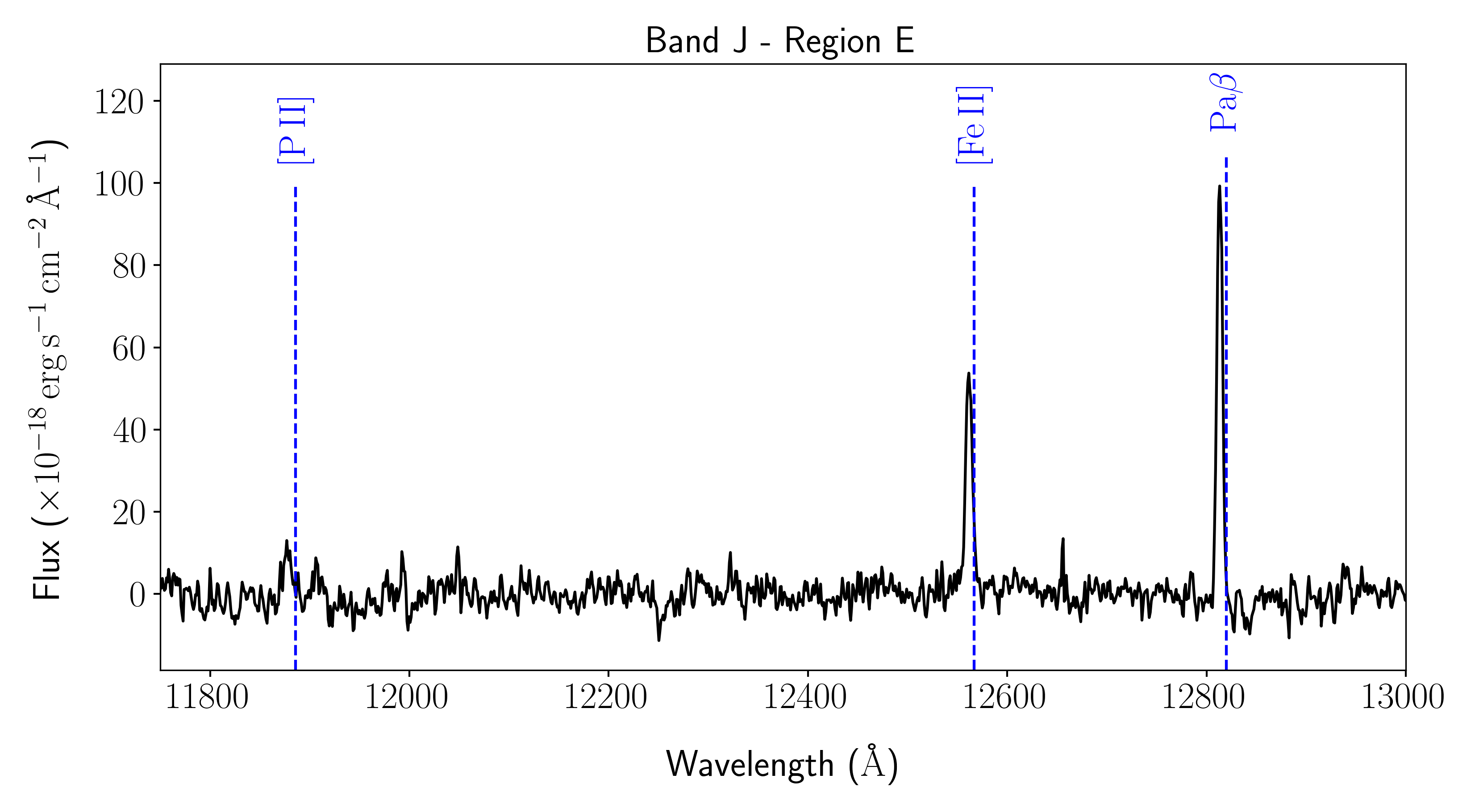}
	\end{minipage}
	\begin{minipage}{.33\textwidth}
		\includegraphics[width=\textwidth]{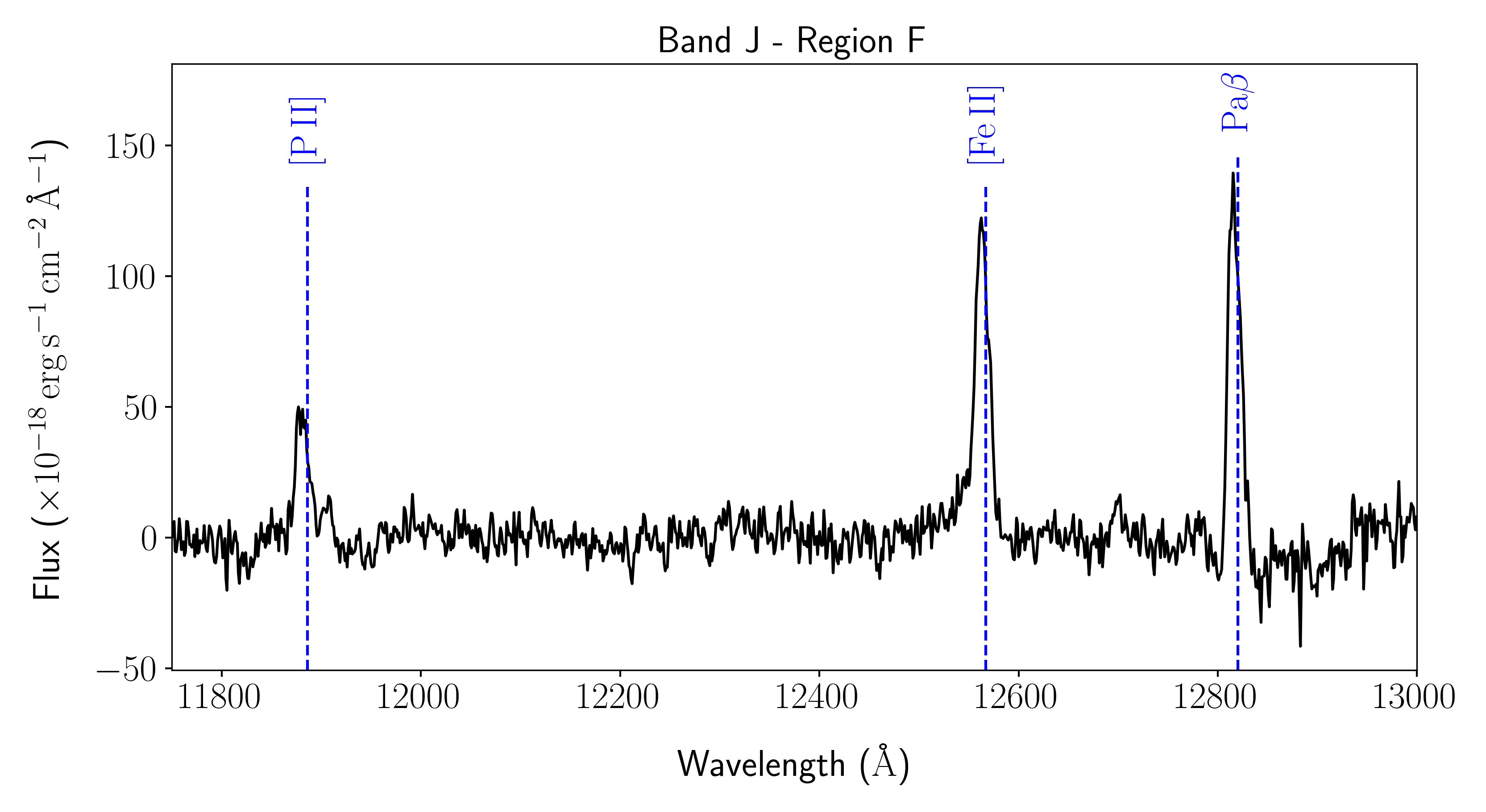}
	\end{minipage}
	\end{minipage}
	
	\begin{minipage}{\textwidth}
	\centering
	\begin{minipage}{.33\textwidth}
		\includegraphics[width=\textwidth]{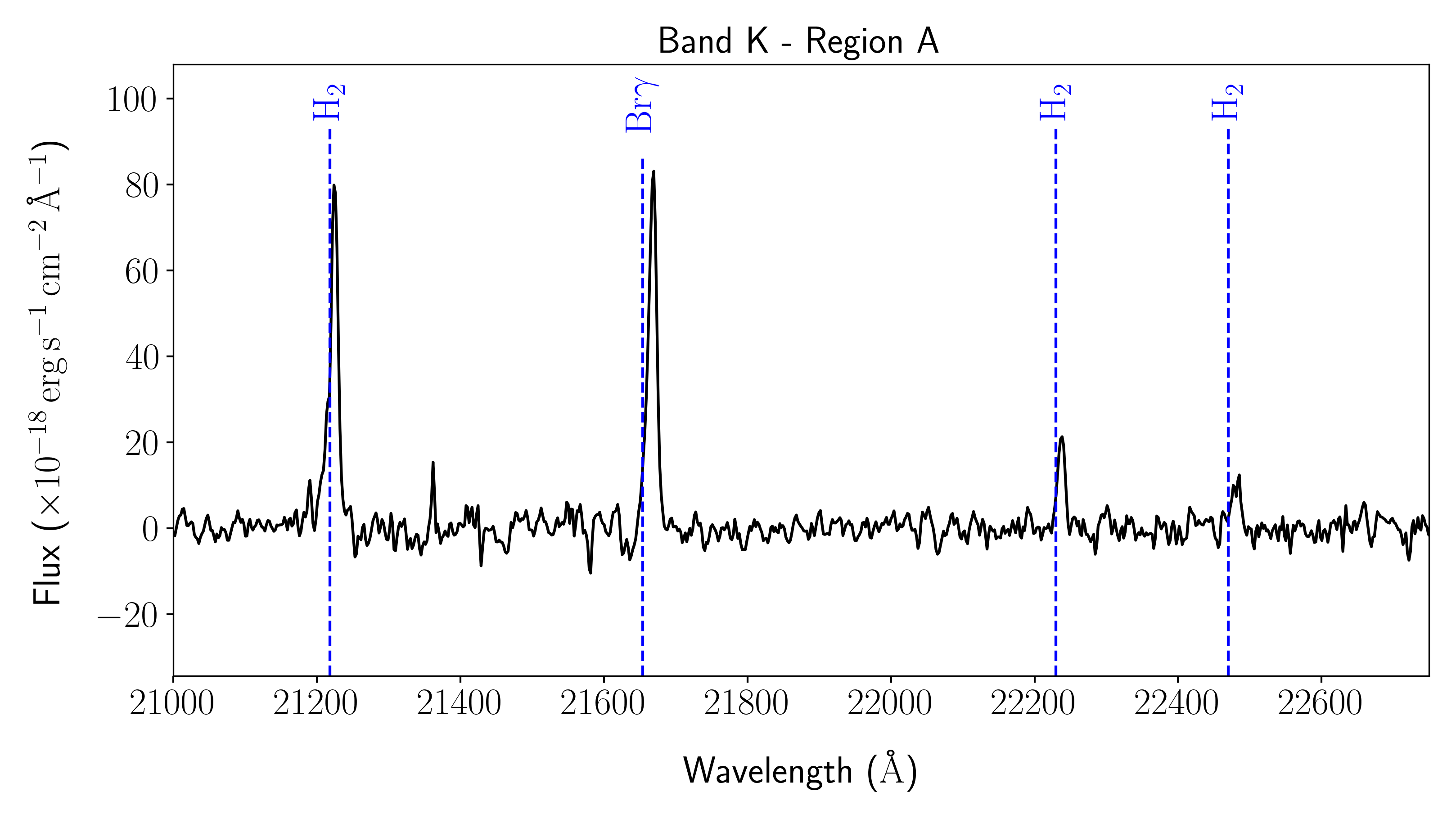}
	\end{minipage}
	\begin{minipage}{.33\textwidth}
		\includegraphics[width=\textwidth]{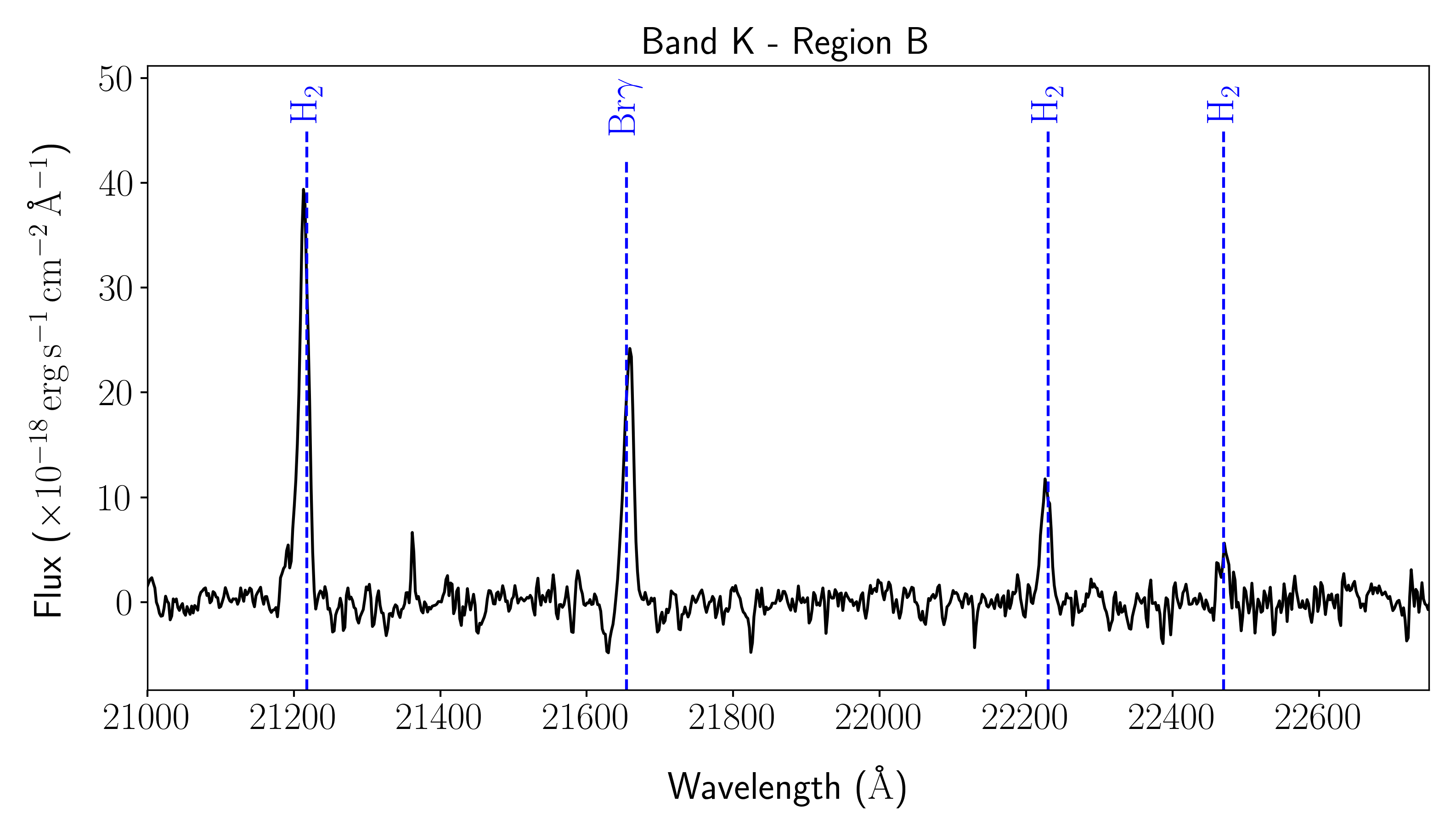}
	\end{minipage}
	\begin{minipage}{.33\textwidth}
		\includegraphics[width=\textwidth]{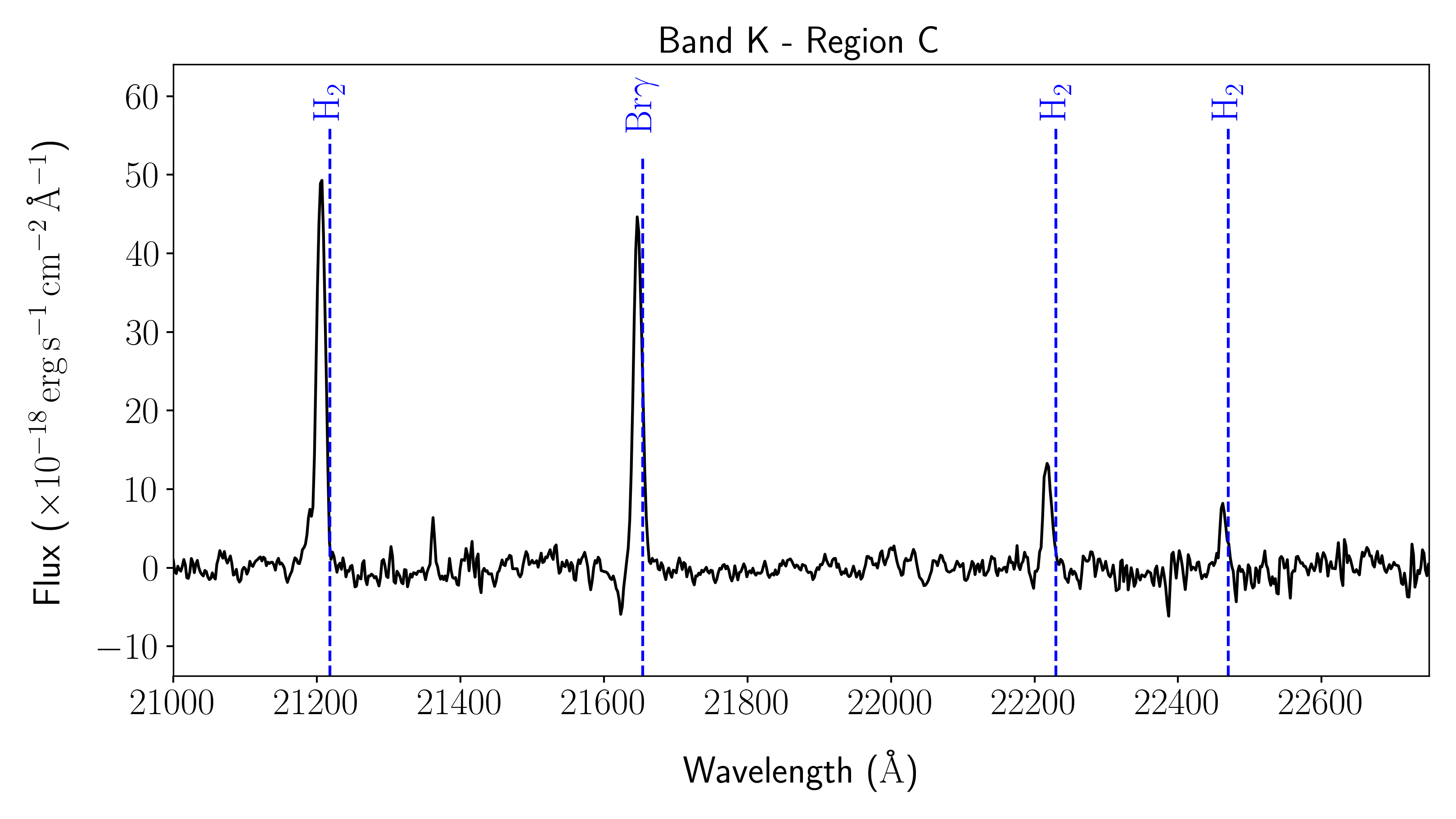}
	\end{minipage}
	\end{minipage}

	\begin{minipage}{\textwidth}
	\centering
	\begin{minipage}{.33\textwidth}
		\includegraphics[width=\textwidth]{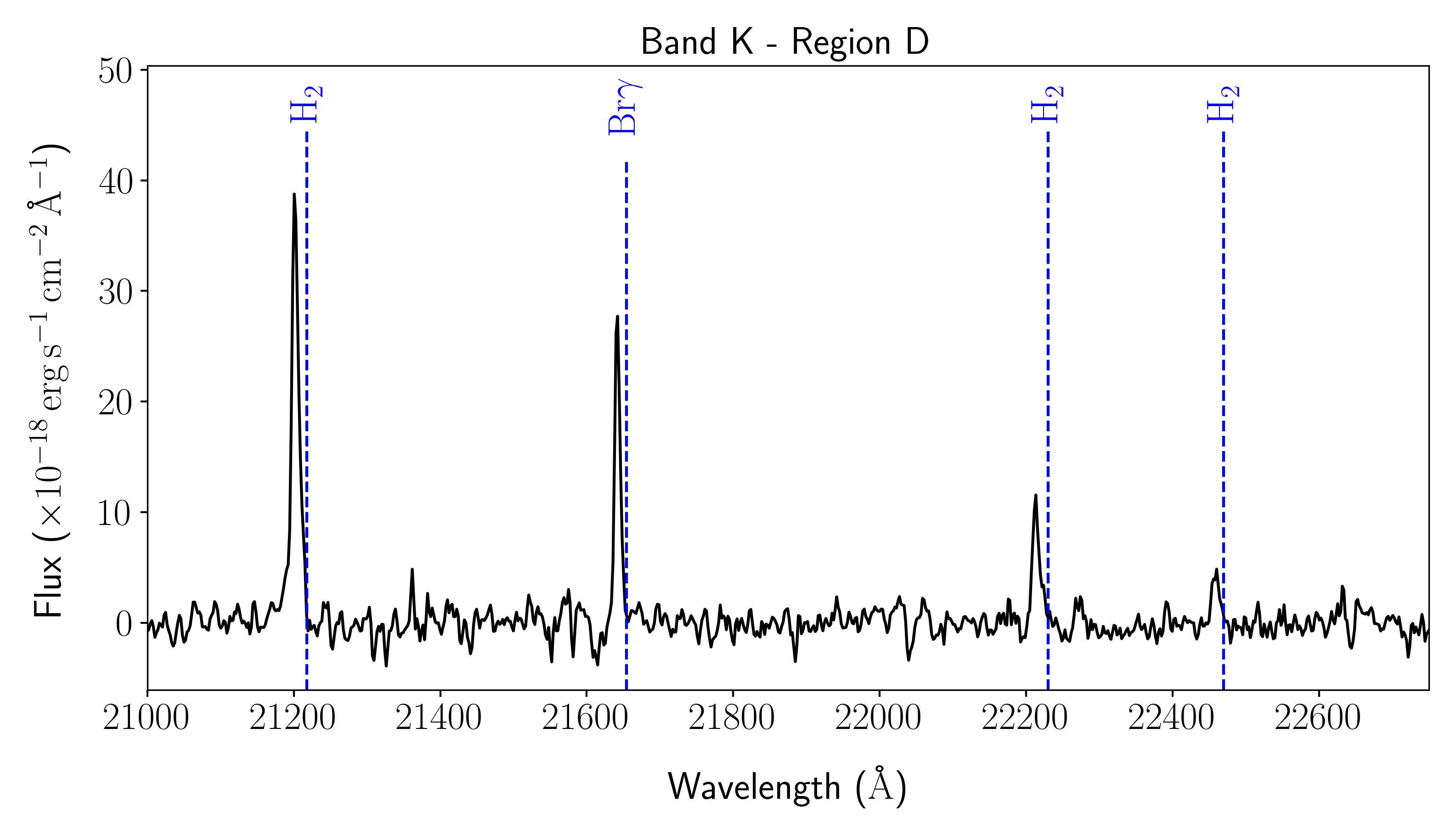}
	\end{minipage}
	\begin{minipage}{.33\textwidth}
		\includegraphics[width=\textwidth]{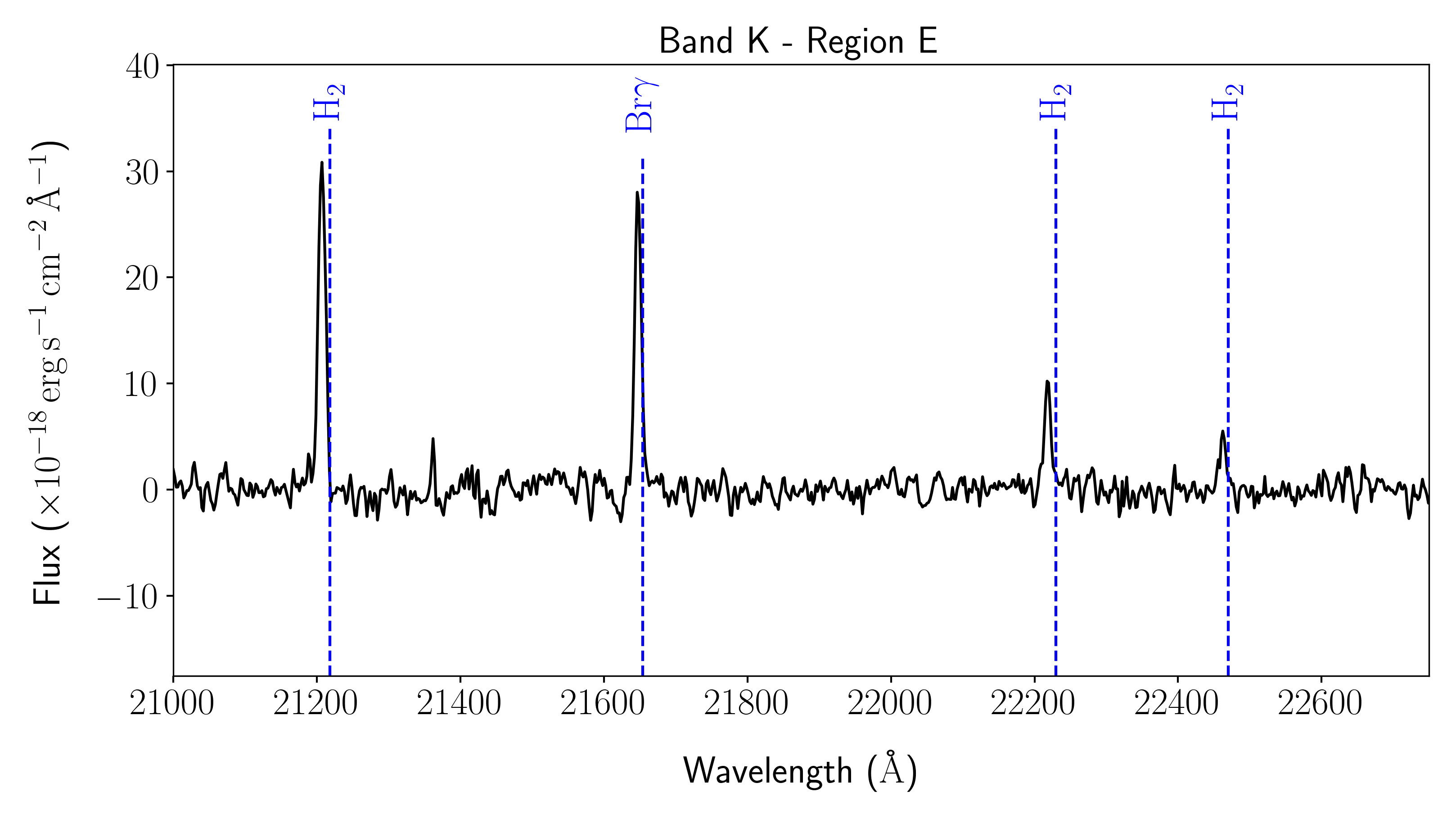}
	\end{minipage}
	\begin{minipage}{.33\textwidth}
		\includegraphics[width=\textwidth]{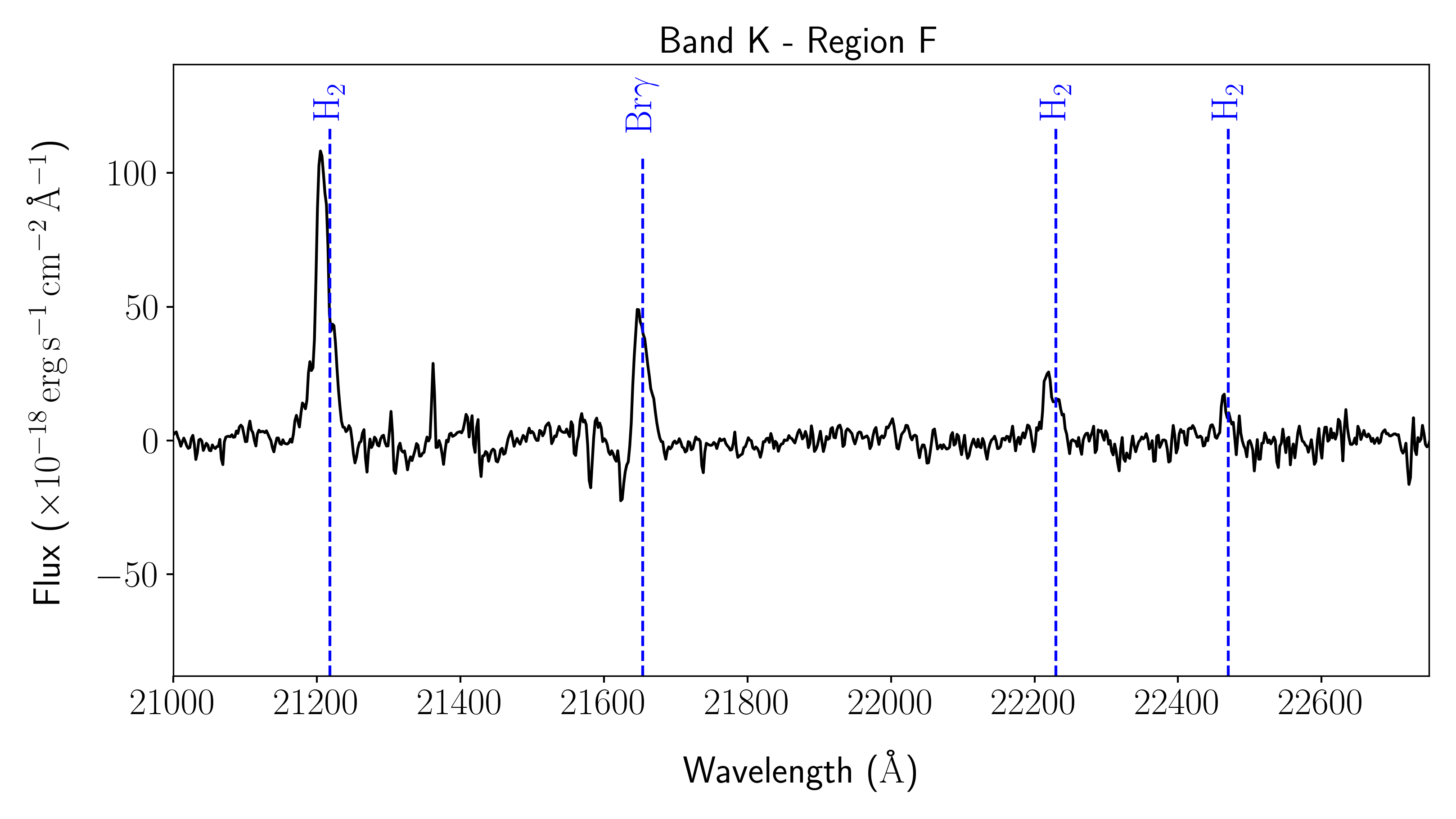}
	\end{minipage}
	\end{minipage}	
	
\caption{Integrated gas spectra of regions A--F corresponding to apertures of 0.2$\arcsec \times$0.2$\arcsec$. \textit{J}-band spectra are shown in the first and second rows. \textit{K}-band spectra are shown in the third and fourth rows. The NIR emission lines are indicated by the blue, dashed, vertical lines. \textit{J}-band emission lines: $\mathrm{[\ion{P}{ii}]\,\lambda 11886}$\AA\ , $\mathrm{[\ion{Fe}{ii}]\,\lambda 12570}$\AA\ and $\mathrm{Pa\beta}$. \textit{K}-band emission lines: $\mathrm{H_2\,\lambda 21218}$\AA\ , $\mathrm{Br\gamma}$, $\mathrm{H_2\,\lambda 22230}$\AA\ and $\mathrm{H_2\,\lambda 22470}$\AA\ .
 }
 \label{fig:SpectraRegions}
\end{center}
\end{figure*}

We show in Fig.~\ref{fig:SpectraRegions} the integrated spectra of regions A--F indicated in Fig.~\ref{fig:LineRatios} corresponding to apertures of 0.2$\arcsec \times$0.2$\arcsec$.

\section{Data from the literature}
\label{appendix:DataLiter}

In Table~\ref{tab:literature_data}, we list the data points collected from the literature that were used to build the NIR diagnostic diagram shown in Fig.~\ref{fig:DiagnosticDiagrams}. The columns are as follows: (1) Object type: Seyfert galaxies, types 1 and 2 (Sy1 and Sy2, respectively), Burst Alert Telescope AGN Spectroscopic Survey (BAT AGN), Star-forming Galaxies (SFGs), Low Ionization Nuclear Emission Regions (LINERs), Supernova remnants (SNRs) and Blue Compact Dwarfs (BCDs). (2) Object name. (3) $\mathrm{H_{2} / Br\gamma}$ emission line ratio. (4) $\mathrm{[\ion{Fe}{ii}] / Pa\beta}$ emission line ratio. (5) Reference.

\onecolumn
\begin{longtable}{llccc}
\caption{Data from the literature used to build the NIR diagnostic diagram shown in Fig.~\ref{fig:DiagnosticDiagrams}. (1) Object type: Seyfert galaxies, types 1 and 2 (Sy1 and Sy2, respectively), Burst Alert Telescope AGN Spectroscopic Survey (BAT AGN), Star-forming Galaxies (SFGs), Low Ionization Nuclear Emission Regions (LINERs), Supernova remnants (SNRs) and Blue Compact Dwarfs (BCDs). (2) Object name. (3) $\mathrm{H_{2} / Br\gamma}$ emission line ratio. (4) $\mathrm{[\ion{Fe}{ii}] / Pa\beta}$ emission line ratio. (5) Reference.} \label{tab:literature_data}\\

\hline 
\multicolumn{1}{c}{Object type} &\multicolumn{1}{c}{Object name} & \multicolumn{1}{c}{$\mathrm{H_{2} / Br\gamma}$}& \multicolumn{1}{c}{$\mathrm{[\ion{Fe}{ii}] / Pa\beta}$} & \multicolumn{1}{c}{Reference} \\
\multicolumn{1}{c}{(1)} &\multicolumn{1}{c}{(2)} & \multicolumn{1}{c}{(3)}& \multicolumn{1}{c}{(4)} & \multicolumn{1}{c}{(5)} \\
\hline 
\endfirsthead

\multicolumn{5}{c}
{ \tablename\ \thetable{} -- Continued} \\
\hline 
\multicolumn{1}{c}{Object type} &\multicolumn{1}{c}{Object name} & \multicolumn{1}{c}{$\mathrm{H_{2} / Br\gamma}$}& \multicolumn{1}{c}{$\mathrm{[\ion{Fe}{ii}] / Pa\beta}$} & \multicolumn{1}{c}{Reference} \\
\multicolumn{1}{c}{(1)} &\multicolumn{1}{c}{(2)} & \multicolumn{1}{c}{(3)}& \multicolumn{1}{c}{(4)} & \multicolumn{1}{c}{(5)} \\
\hline 
\endhead

\hline \multicolumn{5}{l}{{continued}} \\
\endfoot

\endlastfoot

Sy 1 & Mrk 334 & 0.57$\pm$0.06 & 0.57$\pm$0.04 & \citealt{2006AA...457...61R} \\
 & NGC 7469 & 1.93$\pm$0.10 & 0.39$\pm$0.05 & \citealt{2006AA...457...61R} \\
 & NGC 3227 & 2.58$\pm$0.76 & 1.76$\pm$0.14 & \citealt{2004AA...425..457R} \\
 & NGC 4151 & 0.63$\pm$0.05 & 0.57$\pm$0.03 &  \citealt{2004AA...425..457R} \\
 & Mrk 766 & 0.32$\pm$0.04 & 0.28$\pm$0.03 &  \citealt{2004AA...425..457R} \\
 & NGC 4748 & 1.14$\pm$0.26 & 0.84$\pm$0.06 &  \citealt{2004AA...425..457R} \\
 & NGC 5548 & 0.74$\pm$0.15 & 0.45$\pm$0.09 &  \citealt{2004AA...425..457R} \\
 & PG 1612+261 & 1.01$\pm$0.30 & 1.10$\pm$0.10 &  \citealt{2004AA...425..457R} \\
 & NGC 1386 & 1.14$\pm$0.10 & 1.59$\pm$0.14 & \citealt{2002MNRAS.331..154R} \\
 & NGC 6814 & 4.09$\pm$2.52 & 1.31$\pm$0.26 & \citealt{2013MNRAS.430.2002R} \\
 & NGC 3227 & 1.44$\pm$0.25 & 2.71$\pm$0.35 & \citealt{2021MNRAS.tmp..778R} \\
 & NGC 3516 & 4.19$\pm$0.73 & 1.9$\pm$0.24 & \citealt{2021MNRAS.tmp..778R} \\
 & NGC 5506 & 0.22$\pm$0.05 & 0.18$\pm$0.02 & \citealt{2021MNRAS.tmp..778R} \\ 
 \hline
Sy 2 & ESO428-G014 & 0.92$\pm$0.02 & 0.67$\pm$0.02 & \citealt{2005MNRAS.364.1041R} \\
 & NGC 591 & 1.81$\pm$0.08 & 1.14$\pm$0.06 & \citealt{2005MNRAS.364.1041R} \\
 & Mrk 573 & 0.64$\pm$0.09 & 0.61$\pm$0.02 & \citealt{2005MNRAS.364.1041R} \\
 & Mrk 1066 & 0.94$\pm$0.02 & 0.71$\pm$0.01 & \citealt{2005MNRAS.364.1041R} \\
 & NGC 2110 & 3.36$\pm$0.15 & 5.45$\pm$0.14 & \citealt{2005MNRAS.364.1041R} \\
 & NGC 7682 & 5.04$\pm$0.13 & 0.46$\pm$0.11 & \citealt{2005MNRAS.364.1041R} \\
 & NGC 7674 & 0.68$\pm$0.08 & 1.04$\pm$0.10 & \citealt{2005MNRAS.364.1041R} \\
 & NGC 5929 & 2.13$\pm$0.29 & 1.09$\pm$0.05 & \citealt{2005MNRAS.364.1041R} \\
 & Mrk 1210 & 0.33$\pm$0.03 & 0.50$\pm$0.03 &  \citealt{2004AA...425..457R} \\
 & NGC 5728 & 2.97$\pm$0.27 & 0.69$\pm$0.13 &  \citealt{2004AA...425..457R} \\
 & NGC 4945 & 3.13$\pm$0.06 & 0.86$\pm$0.03 & \citealt{2002MNRAS.331..154R} \\
 & NGC 5128 & 2.04$\pm$0.10 & 3.42$\pm$0.16 & \citealt{2002MNRAS.331..154R} \\
 & NGC 4388 & 0.88$\pm$0.08 & 0.40$\pm$0.03 & \citealt{2001AJ....122..764K} \\
 & Mrk 3 & 0.31$\pm$0.04 & 1.24$\pm$0.04 & \citealt{2001AJ....122..764K} \\
 & Mrk 993 & 5.31$\pm$0.00 & 0.34$\pm$0.11 & \citealt{2005MNRAS.364.1041R} \\
 & NGC 5953 & 1.28$\pm$0.00 & 3.50$\pm$0.00 & \citealt{2005MNRAS.364.1041R} \\
 & NGC 1144 & 6.71$\pm$0.00 & 6.50$\pm$0.00 & \citealt{2005MNRAS.364.1041R} \\
 & NGC 788 & 0.81$\pm$0.14 & 0.23$\pm$0.03 & \citealt{2021MNRAS.tmp..778R} \\
 & Mrk 607 & 0.69$\pm$0.13 & 0.26$\pm$0.03 & \citealt{2021MNRAS.tmp..778R} \\
 & NGC 5899 & 2.94$\pm$0.51 & 1.75$\pm$0.23 & \citealt{2021MNRAS.tmp..778R} \\ 
 \hline
BAT AGN & 33 & 1.20$\pm$0.17 & 1.07$\pm$0.02 & \citealt{2017MNRAS.467..540L} \\
 & 308 & 3.95$\pm$0.37 & 5.14$\pm$0.16 & \citealt{2017MNRAS.467..540L} \\
 & 382 & 1.68$\pm$0.18 & 0.46$\pm$0.03 & \citealt{2017MNRAS.467..540L} \\
 & 404 & 1.25$\pm$0.10 & 0.75$\pm$0.01 & \citealt{2017MNRAS.467..540L} \\
 & 517 & 1.24$\pm$0.16 & 0.72$\pm$0.02 & \citealt{2017MNRAS.467..540L} \\
 & 533 & 1.83 $\pm$0.17 & 0.87$\pm$0.04 & \citealt{2017MNRAS.467..540L} \\
 & 585 & 2.62$\pm$0.23 & 0.17$\pm$0.04 & \citealt{2017MNRAS.467..540L}\\
 & 586 & 1.10$\pm$0.15 & 0.42$\pm$0.02 & \citealt{2017MNRAS.467..540L} \\
 & 588 & 3.67$\pm$1.23 & 0.59$\pm$0.07 & \citealt{2017MNRAS.467..540L} \\
 & 590 & 0.68$\pm$0.05 & 0.69$\pm$0.03 & \citealt{2017MNRAS.467..540L} \\
 & 592 & 1.22$\pm$0.82 & 1.54$\pm$0.14 & \citealt{2017MNRAS.467..540L} \\
 & 595 & 0.64$\pm$0.04 & 0.93$\pm$0.01 & \citealt{2017MNRAS.467..540L} \\
 & 608 & 0.28$\pm$0.02 & 0.27$\pm$0.01 & \citealt{2017MNRAS.467..540L} \\
 & 615 & 1.00$\pm$0.02 & 0.38$\pm$0.01 & \citealt{2017MNRAS.467..540L} \\
 & 616 & 0.84$\pm$0.06 & 0.42$\pm$0.01 & \citealt{2017MNRAS.467..540L} \\
 & 635 & 1.09$\pm$0.13 & 0.68$\pm$0.03 & \citealt{2017MNRAS.467..540L} \\
 & 641 & 0.76$\pm$0.15 & 0.73$\pm$0.03 & \citealt{2017MNRAS.467..540L} \\
 & 712 & 0.22$\pm$0.03 & 0.38$\pm$0.01 & \citealt{2017MNRAS.467..540L} \\
 & 717 & 0.75$\pm$0.27 & 0.70$\pm$0.04 & \citealt{2017MNRAS.467..540L} \\
 & 723 & 0.72$\pm$0.03 & 0.53$\pm$0.01 & \citealt{2017MNRAS.467..540L} \\
 & 738 & 0.85$\pm$0.08 & 0.59$\pm$0.00 & \citealt{2017MNRAS.467..540L} \\
 & 739 & 4.75$\pm$2.78 & 0.81$\pm$0.05 & \citealt{2017MNRAS.467..540L} \\
 & 774 & 1.00$\pm$0.60 & 0.55$\pm$0.03 & \citealt{2017MNRAS.467..540L} \\
 & 1117 & 2.50$\pm$0.34 & 0.51$\pm$0.03 & \citealt{2017MNRAS.467..540L} \\
 & 1133 & 2.44$\pm$0.11 & 0.62$\pm$0.01 & \citealt{2017MNRAS.467..540L} \\
 & 1157 & 0.30$\pm$0.11 & 0.14$\pm$0.01 & \citealt{2017MNRAS.467..540L} \\
 & 1158 & 3.29$\pm$0.49 & 0.93$\pm$0.03 & \citealt{2017MNRAS.467..540L} \\
 & 1161 & 0.69$\pm$0.05 & 0.69$\pm$0.02 & \citealt{2017MNRAS.467..540L} \\
 & 1182 & 1.09$\pm$0.13 & 0.39$\pm$0.01 & \citealt{2017MNRAS.467..540L} \\
 & 1198 & 5.41$\pm$0.64 & 0.42$\pm$0.02 & \citealt{2017MNRAS.467..540L} \\
 & 1287 & 3.19$\pm$0.16 & 0.84$\pm$0.01 & \citealt{2017MNRAS.467..540L} \\
\hline
SFGs & NGC 34 & 1.72$\pm$0.14 & 1.08$\pm$0.09 & \citealt{2005MNRAS.364.1041R} \\
 & NGC 7714 & 0.21$\pm$0.02 & 0.29$\pm$0.02 & \citealt{2005MNRAS.364.1041R}  \\
 & NGC 1614 & 0.13$\pm$0.01 & 0.27$\pm$0.02 & \citealt{2005MNRAS.364.1041R} \\
 & NGC 3310 & 0.14$\pm$0.03 & 0.33$\pm$0.03 & \citealt{2005MNRAS.364.1041R} \\
 & NGC 1797 & 0.60$\pm$0.05 & 0.42$\pm$0.02 & \citealt{2013MNRAS.430.2002R} \\
 & NGC 7678 & 0.35$\pm$0.06 & 0.36$\pm$0.06 & \citealt{2013MNRAS.430.2002R} \\ 
 & NGC 6835 & 0.29$\pm$0.05 & 0.42$\pm$0.05 & \citealt{2013MNRAS.430.2002R} \\
 & NGC 1222 & 0.12$\pm$0.02 & 0.22$\pm$0.01 & \citealt{2013MNRAS.430.2002R}  \\
 & NGC 2388 & 0.56$\pm$0 & 0.26$\pm$0 & \citealt{2004ApJ...601..813D} \\
 & NGC 6946 & 0.95$\pm$0 & 0.49$\pm$0 & \citealt{2004ApJ...601..813D} \\
 & M 82 & 0.15$\pm$0 & 0.19$\pm$0 & \citealt{1998ApJS..114...59L} \\
 & II Zw 040 & 0.09$\pm$0 & 0.15$\pm$0 & \citealt{1998ApJS..114...59L} \\
 & NGC 5253 & 0.07$\pm$0 & 0.14$\pm$0 & \citealt{1998ApJS..114...59L} \\
 \hline
LINERs & NGC 5194 & 8$\pm$3 & 2.3$\pm$0.6 & \citealt{1998ApJS..114...59L} \\
 & NGC 7743 & 5$\pm$2 & 1.3$\pm$0.4 & \citealt{1998ApJS..114...59L} \\
 & NGC 660 & 0.42$\pm$0.05 & 0.48$\pm$0.02 & \citealt{2013MNRAS.430.2002R} \\
 & NGC 1204 & 0.61$\pm$0.05 & 0.43$\pm$0.02 & \citealt{2013MNRAS.430.2002R}  \\
 & NGC 1266 & 9.79$\pm$2.28 & 4.21$\pm$0.82 & \citealt{2013MNRAS.430.2002R} \\
 & NGC 7465 & 1.04$\pm$0.20 & 1.23$\pm$0.34 & \citealt{2013MNRAS.430.2002R}  \\
 & NGC 7591 & 1.18$\pm$0.09 & 0.70$\pm$0.03 & \citealt{2013MNRAS.430.2002R}  \\
 & UGC 12150 & 1.39$\pm$0.08 & 0.60$\pm$0.05 & \citealt{2013MNRAS.430.2002R} \\
 & NGC 404 & 5$\pm$0 & 2.7$\pm$0.3 & \citealt{1998ApJS..114...59L} \\
 & NGC 3998 & 2$\pm$0 & 0.6$\pm$0.3 & \citealt{1998ApJS..114...59L} \\
 & NGC 4826 & 1$\pm$0 & 0.7$\pm$0.4 & \citealt{1998ApJS..114...59L} \\
 & NGC 7479 & 11$\pm$4 & 0.5$\pm$0 & \citealt{1998ApJS..114...59L} \\
 & NGC 4736 & 20$\pm$0 & 5.3$\pm$0 & \citealt{1998ApJS..114...59L} \\
 \hline
SNRs & M 2 & 0.3$\pm$0 & 8.5$\pm$0 & \citealt{1998ApJS..114...59L} \\
 & IC 0443 & 11$\pm$0 & 5$\pm$0 & \citealt{1998ApJS..114...59L}  \\
 \hline
BCDs & I Zw 040 & 0.0405$\pm$0.0031 & 0.0226$\pm$0.0018 & \citealt{2011ApJ...734...82I}  \\
 & Mrk 71 &  0.0940$\pm$0.0076 & 0.0055$\pm$0.0005 & \citealt{2011ApJ...734...82I}  \\
 & Mrk 930 & 0.2462$\pm$0.0183 & 0.1633$\pm$0.0076 & \citealt{2011ApJ...734...82I}  \\
 & Mrk 996 & 0.1746$\pm$0.0139 & 0.0760$\pm$0.0065 & \citealt{2011ApJ...734...82I}  \\
 & SbS & 0.0583$\pm$0.0068 & 0.0229$\pm$0.0053 & \citealt{2011ApJ...734...82I}  \\
\hline

\end{longtable}


\bsp	
\label{lastpage}
\end{document}